\documentclass[10pt,aps,pre,amsmath,amsfonts,amssymb,floatfix,superscriptaddress,notitlepage,nofootinbib]{revtex4}
\usepackage{mathrsfs} \usepackage{graphicx,color} \usepackage{bbm} \usepackage{multirow,wrapfig}

\usepackage{enumitem}
\usepackage{bm}
\usepackage{bbm}

\usepackage{xcolor}
\usepackage[hidelinks]{hyperref}

\newcommand{\U}[1]{\underline{#1}}
\newcommand{\Us}{\underline{\sigma}}
\newcommand{\Ut}{\underline{\tau}}
\newcommand{\Hs}{\hat{\sigma}}
\newcommand{\Ht}{\hat{\tau}}
\newcommand{\grad}{\underline{\nabla}}

\newcommand{\MCT}{\text{\tiny MCT}}
\newcommand{\SF}{\text{\tiny SF}}
\newcommand{\K}{\text{\tiny K}}

\newcommand{\IS}{\text{\tiny IS}}
\newcommand{\on}{\text{\tiny onset}}

\let\a=\alpha \let\b=\beta \let\g=\gamma \let\d=\delta
   \let\k=\kappa
\let\l=\lambda \let\m=\mu  \let\x=\xi \let\p=\pi
\let\s=\sigma \let\t=\tau  
   
 \let\Th=\Theta \let\X=\Xi 
   
\let\ee=\varepsilon \let\r=\rho \let\th=\theta \let\io=\infty
\let\om=\omega

\def\LL{{\cal L}}

\def\de{\mathrm d}

\font\msytw=msbm9 scaled\magstep1

\def\to{\rightarrow} \def\la{\left\langle} \def\ra{\right\rangle}

\def\RRR{\hbox{\msytw R}}

\newcommand{\beq}{\begin{equation}} \newcommand{\eeq}{\end{equation}}

\newcommand{\argc}[1]{\left[#1\right]}

\newcommand{\argp}[1]{\left(#1\right)}

\newcommand{\moy}[1]{\left[  #1 \right] }

\begin{document}
	
	\title{Introduction to the dynamics of disordered systems:\\ equilibrium and gradient descent}
	
	\author{Giampaolo Folena}
	\affiliation{Laboratoire de Physique de l'Ecole Normale Sup\'erieure, ENS, Universit\'e PSL, CNRS, Sorbonne Universit\'e, Universit\'e de Paris, F-75005 Paris, France
	}
	\affiliation{James Franck Institute and Department of Physics, University of Chicago, Chicago, IL 60637, U.S.A.}
	
	\author{Alessandro Manacorda}
	\affiliation{Department of Physics and Materials Science, University of Luxembourg, L-1511 Luxembourg}
	\affiliation{Laboratoire de Physique de l'Ecole Normale Sup\'erieure, ENS, Universit\'e PSL, CNRS, Sorbonne Universit\'e, Universit\'e de Paris, F-75005 Paris, France
	}
	\author{Francesco Zamponi}
	\affiliation{Laboratoire de Physique de l'Ecole Normale Sup\'erieure, ENS, Universit\'e PSL, CNRS, Sorbonne Universit\'e, Universit\'e de Paris, F-75005 Paris, France
	}
	
	\begin{abstract}
		This manuscript contains the lecture notes of the short courses given by one of us (F.Z.) at the summer school 
		{\it Fundamental Problems in Statistical Physics XV}, held in Brunico, Italy, in July 2021, and, just before that, 
		at the summer school {\it Glassy Systems and Inter-Disciplinary Applications}, held in Cargese, France, in June 2021.
		The course was a short introductory overview of the dynamics of disordered systems, focused in particular on the equilibrium dynamics (with the associated glass transition),
		and on the simplest case of off-equilibrium dynamics, namely the gradient descent dynamics. A few selected topics (and references) are chosen, based on the authors' own taste and competences, and 
		on pedagogical reasons, without aiming at a complete review of the subject.
	\end{abstract}
	
	\maketitle
	
	\vskip-20pt
	
	\tableofcontents
	
	
	\section{Motivations}	
	\label{sec:intro}
	
	In these notes, we will discuss the dynamics of two simple toy models of disordered systems, 
	that provide a simple description for a broader class of more complex models.
	We will consider (1) the equilibrium dynamics, i.e. the dynamics that starts from the Boltzmann-Gibbs distribution at a fixed temperature and satisfies detailed balance at the same temperature, 
	(2) the simplest case of off-equilibrium dynamics, i.e. a zero-temperature energy minimization via gradient descent from a random initial condition, and (3) a simple combination of the previous two, which provides a simple approximation to the simulated annealing protocol.
	The most important feature of the class of disordered systems we consider here is to have a 
	{\it rough energy landscape}, i.e. an energy function featuring multiple local minima and saddle points that can trap the dynamics for long times; this {\it metastability} phenomenon
	will be one of the central theme of the notes.
	
	While the interest in this problem in physics has been mostly driven by the study of glasses and spin glasses, these phenomena play
	a fundamental role in many other fields of research, hence their study has benefited from a fruitful interaction between different disciplines such as statistical physics,
	probability theory, statistics, machine learning, and computer science. Given the vastity of the field, it is clearly impossible here to give a complete overview,
	so we will rather start the discussion by giving a few examples that will serve as motivations for the rest of the discussion.
	
	\subsection{Glasses and spin-glasses: finding low-energy states}
	\label{sec:IA}
	
\begin{figure}[h]
		\includegraphics[width=0.49\columnwidth]{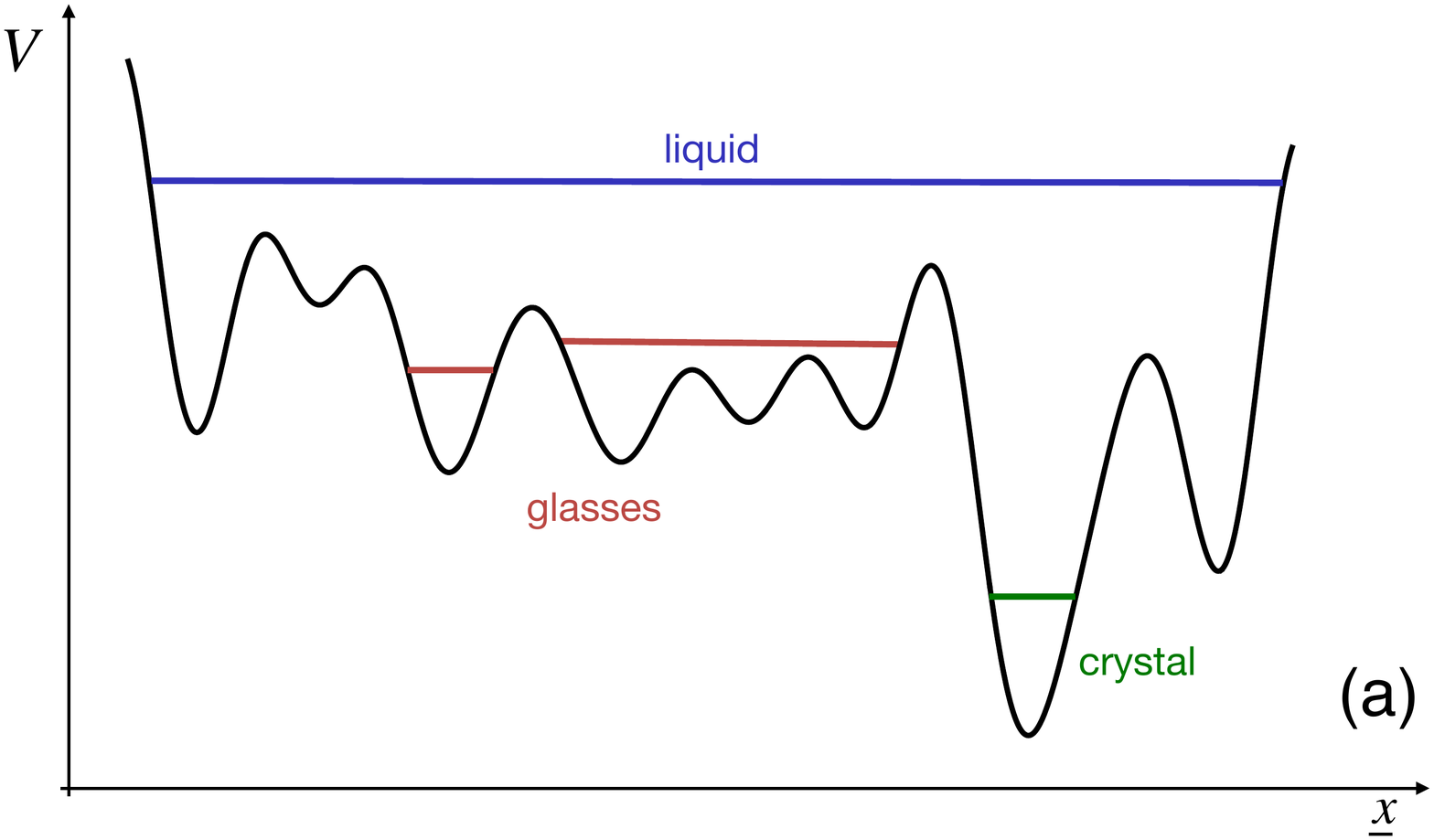}
		\includegraphics[width=0.49\columnwidth]{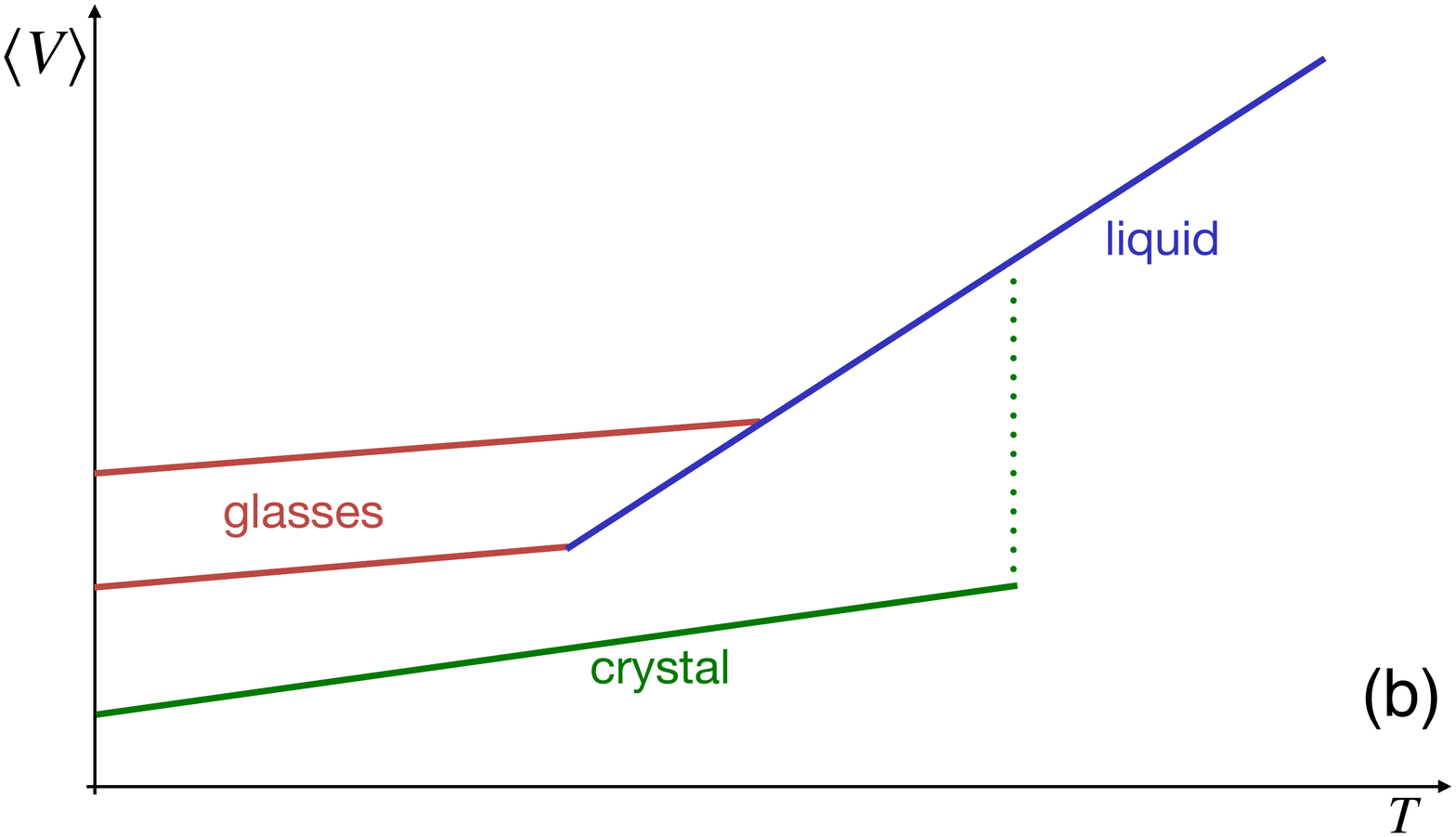}
		\caption{A. Illustration of the potential energy landscape $V(\U{x})$ of a structural glass model. B. Preparation of glassy states by cooling. The 
			glass line with lower energy corresponds to 
			slower cooling.}
		\label{fig:glassPEL}
	\end{figure}

	Structural glasses (such as common window glasses) are microscopically formed by a disordered collection of atoms or molecules. As a simple model, we can consider a system of $N$ classical point 
	particles in three dimensions, described by a set of positions $\U{x}\in \RRR^{3N}$ with total potential energy $V(\U{x})$.
	At low temperatures, the system visits low-energy states. The lowest-energy state is usually a crystal, corresponding to a periodic arrangement of the particles. 
	In glassy systems, the potential energy also features disordered local minima, in which particles' configuration do not display any evident periodicity.
	Many different microscopic arrangements of the particles then give rise to similar potential energy values: at a given value of
	$V(\U{x})$ there are many local minima with similar properties, as illustrated in Fig.~\ref{fig:glassPEL}A.
	
	Glasses are usually prepared by slow cooling from a high-temperature liquid melt. Suppose that the system is initialized in equilibrium at high temperature, in the liquid
	phase. There, it explores a large portion of the phase space, corresponding to the basin of attraction of many local minima, as illustrated in Fig.~\ref{fig:glassPEL}A.
	Upon a slow enough cooling, the system maintains thermodynamic equilibrium and a sharp first-order phase transition to the crystalline state is observed,
	as illustrated in Fig.~\ref{fig:glassPEL}B. However, in glass-forming systems, the nucleation time of the crystal can be astronomically large, and as a result, crystallization is not
	observed~\cite{Ca09}. Upon slow cooling, instead, the system remains 
	confined in the basin
	of attraction of one of the local minima that compose the original liquid state (Fig.~\ref{fig:glassPEL}A). As a result, during the cooling process, the average potential energy
	does not jump discontinuously to the crystal value, but instead displays a smooth crossover from the liquid regime to the glassy one (Fig.~\ref{fig:glassPEL}B).
	Because there are multiple glassy local minima within the liquid basin, different thermal histories (e.g. different temperature cooling rates) can bring the system to slightly 
	different glassy states, hence leading to different low-temperature values of the potential energy. Typically, slower cooling leads to lower values of potential energy (Fig.~\ref{fig:glassPEL}B)~\cite{Ca09}.
	
	Hence, the problem of describing structural glasses corresponds, mathematically, to the problem of characterizing low-energy local minima of the potential energy
	$V(\U{x})$~\cite{sciortino_potential_2005}. Similar problems arise in other physical systems such as spin glasses, which are magnetic systems with impurities. In this case, the system is described
	by a $N$-spin configuration $\U\s$ and an associated magnetic Hamiltonian $H(\U\s)$~\cite{CC05}. Note that,
	both for glasses and spin glasses, the existence of disordered local minima is independent of the presence of explicit randomness in the potential energy function or Hamiltonian~\cite{KT89}.

	\subsection{The jamming transition: colloids, emulsions, granular materials}
	\label{sec:IB}
	\begin{figure}[h]
		\centering
		\vskip-20pt
		\includegraphics[width=0.5\columnwidth]{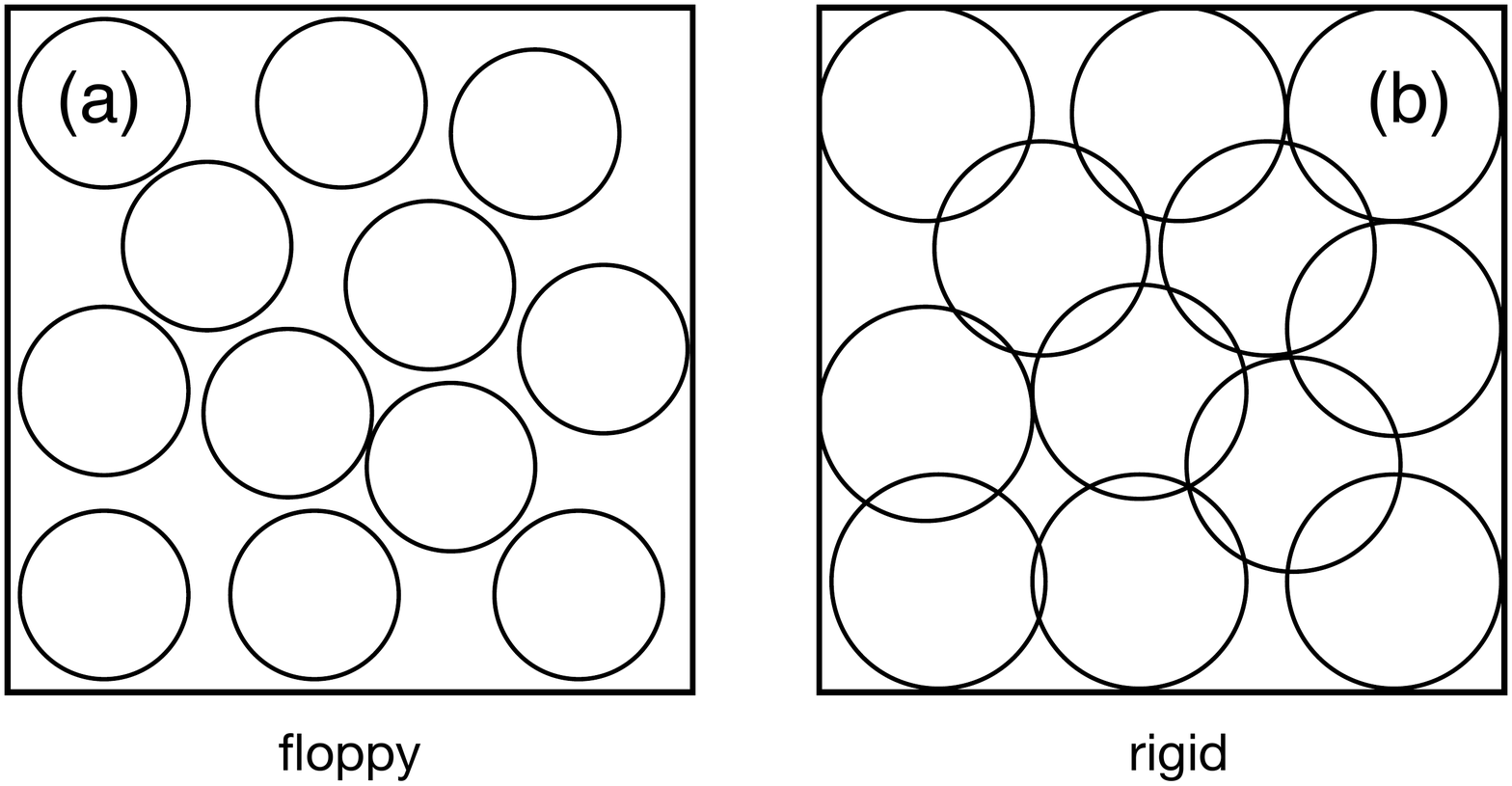}\\
		\hspace{0.4cm}
		\begin{tabular}{r @{\hspace{1cm}}|@{\hspace{1cm}} l}
			$V=0$ & $V>0$ \\
			floppy  & rigid\\
			unjammed & jammed\\
			SAT(isfied) & UNSAT(isfied) \\ 
		\end{tabular}
		\caption{The jamming transition from a floppy (A) to a rigid (B) state.}
		\label{fig:jamming}
	\end{figure}
	
	A special case of interest is that of particle interacting via a finite-range pair potential, i.e. a potential such that there is no interaction 
	when particles are sufficiently well separated.
	A specific common example is that of {\it harmonic soft spheres},
	\beq\label{eq:sspot}
	V(\U{x}) = \sum_{i<j} v(|x_i-x_j|)  \ , \qquad v(r) = \ee (1-r/\ell)^2 \Theta(\ell-r) \ ,
	\eeq
	where $\Theta(x)$ is the Heaviside theta function, and therefore $v(r)=0$ if particles do not overlap, i.e. $|x_i - x_j|>\ell$. Such a potential has multiple local disordered minima, as illustrated in Fig.~\ref{fig:glassPEL}A,
	but on top of that, a new phenomenon emerges due to its finite range. In fact, local minima of the potential (or $T=0$ athermal states) 
	can be separated in two classes. Those with $V=0$ are 
	such that all the overlaps between particles are removed (Fig.~\ref{fig:jamming}A). As a consequence, one can imagine that in the generic
	case the structure is mechanically floppy, because of the absence of interactions. These states are called {\it unjammed}. On the contrary, minima with $V>0$ have at least
	two particles overlapping, and in the generic case there is a full network of particle contacts  (Fig.~\ref{fig:jamming}B) that provides mechanical stability (or rigidity) to the structure.
	These states are called {\it jammed}. Remarkably, a sharp {\it jamming phase transition}~\cite{OLLN02,OSLN03} separates a low-density region where unjammed minima are found with
	high probability (going to one in the thermodynamic limit) from a high-density region where jammed minima are found with high probability.
	
	The model potential in Eq.~\eqref{eq:sspot} has been used as a simple model in soft matter physics, to describe soft colloids, emulsions,
	and soft granular materials~\cite{LN10,LNSW10}. The jamming transition found application to these materials, because it describes the onset of rigidity upon compression.
	
	The problem of finding unjammed minima is also a {\it packing problem}~\cite{ConwaySloane,TS10} that can be formulated as follows:
	is it possible to find $\underline{x}$ such that $V(\underline{x})=0$, i.e. for which there is no overlap between particles? 
	In other words, can a set of $N$ hard spheres of diameter $\ell$, which cannot overlap, be packed in a periodic box $\Omega$ of volume $V$?
	As such, it has a deep connection with many branches of mathematics and coding theory, because the problem of packing spheres (especially
	in high spatial dimension) is connected to the theory of error correcting codes~\cite{MacWilliamsSloane,HuffmanPless}.
	
	Finally, this problem can also be seen as a {\it constraint satisfaction problem}: a configuration space $\U{x} \in \Omega^N$ 
	is given together with a set of constrains, i.e. $|x_i - x_j|>\ell \ ,  \forall i,j$, which have to be satisfied.
	In this language, an unjammed configuration is also called SAT, because all constraints are satisfied, while an unjammed configuration is called UNSAT, because
	some constraints are not satisfied; and the jamming transition correspond to a SAT-UNSAT transition~\cite{AMSZ09}. We will come back to this analogy in Sec.~\ref{sec:ID} below.

	\subsection{Optimization problems: finding low-energy configurations}
	\label{sec:IC}
	
	The problem of finding low-energy configurations in a rough energy landscape, which is a central one in the physics of disordered systems (Sec.~\ref{sec:IA}), 
	is also very important in theoretical computer science. A typical example is the so-called {\it traveling salesman problem}~\cite{ApplegateBixbyChvatalCook,note_test}, which is formulated as follows.
	\begin{wrapfigure}{r}{.2\textwidth}
		\vskip-10pt
		\includegraphics[width=.2\textwidth]{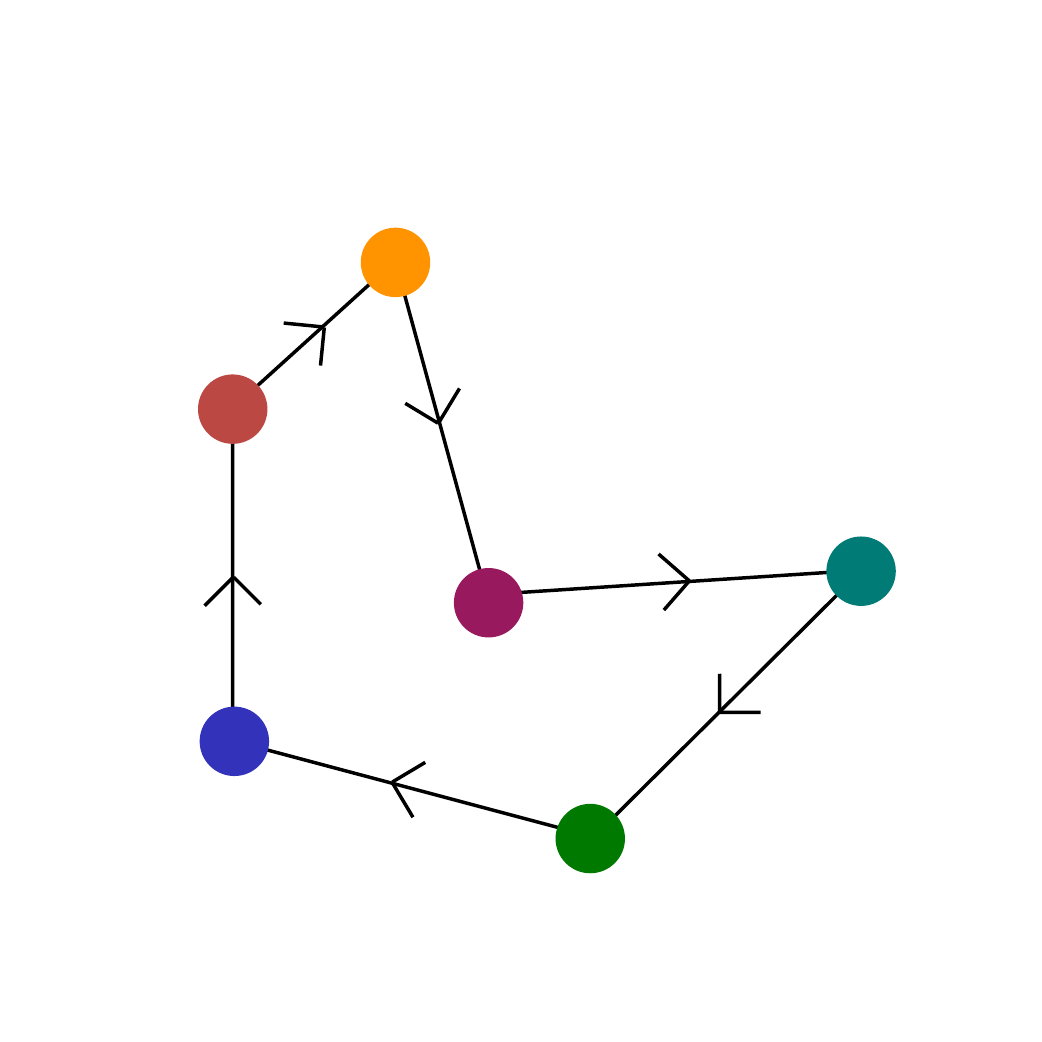}
	\end{wrapfigure}
	One is given a 
	set of cities labeled as $1,\dots,N$, and 
	a cost $w_{ij}$ to go from $i\to j$.
	One is then asked to find a path $i_1,i_2,\dots,i_N$ that goes once and only once through all the cities, 
	and minimizes the total cost of the trip 
	\beq
	E(i_1,\cdots,i_N)=\sum_k w_{i_{k}i_{k+1}} \ .
	\eeq
	This is called an {\it optimization problem},
	i.e. one is given a cost function and has to minimize it over a large space of possibilities.
	Another version, called {\it decision problem}, is to find whether there exist a path such that the cost is below a given threshold, i.e.
	$E(i_1,\cdots,i_N)=\sum_k w_{i_{k}i_{k+1}} < E_0$.
	Note that the cost $w_{ij}$ can represent for example the distance between city $i$ and $j$, the price of the train ticket, or any other positive
	number that is relevant to the problem.
	
	This problem is one of the most intensively studied problems in optimization, and it is obviously relevant for many planning/scheduling problems,
	but it also find applications in biology and astronomy\footnote{See e.g. \url{https://en.wikipedia.org/wiki/Travelling_salesman_problem}.}. Many other optimization problems have been formulated in theoretical computer science,
	and in all cases the problem is to minimize a given cost function over a space of variables, with a number of configurations typically growing 
	exponentially in the number of variables $N$, thus making a brute force search impossible.

	\subsection{Constraint satisfaction problems: finding solutions to a set of constraints}	
	\label{sec:ID}
	
	A special class of optimization problems, akin to the problem formulated in Sec.~\ref{sec:IB}, is obtained when the cost function is a sum of 
	local terms involving a finite number of variables, such that the local cost vanishes for certain assignments of the variables and is positive otherwise.
	An example  is that of graph
	\begin{wrapfigure}{r}{.2\textwidth}
		\includegraphics[width=0.2\columnwidth]{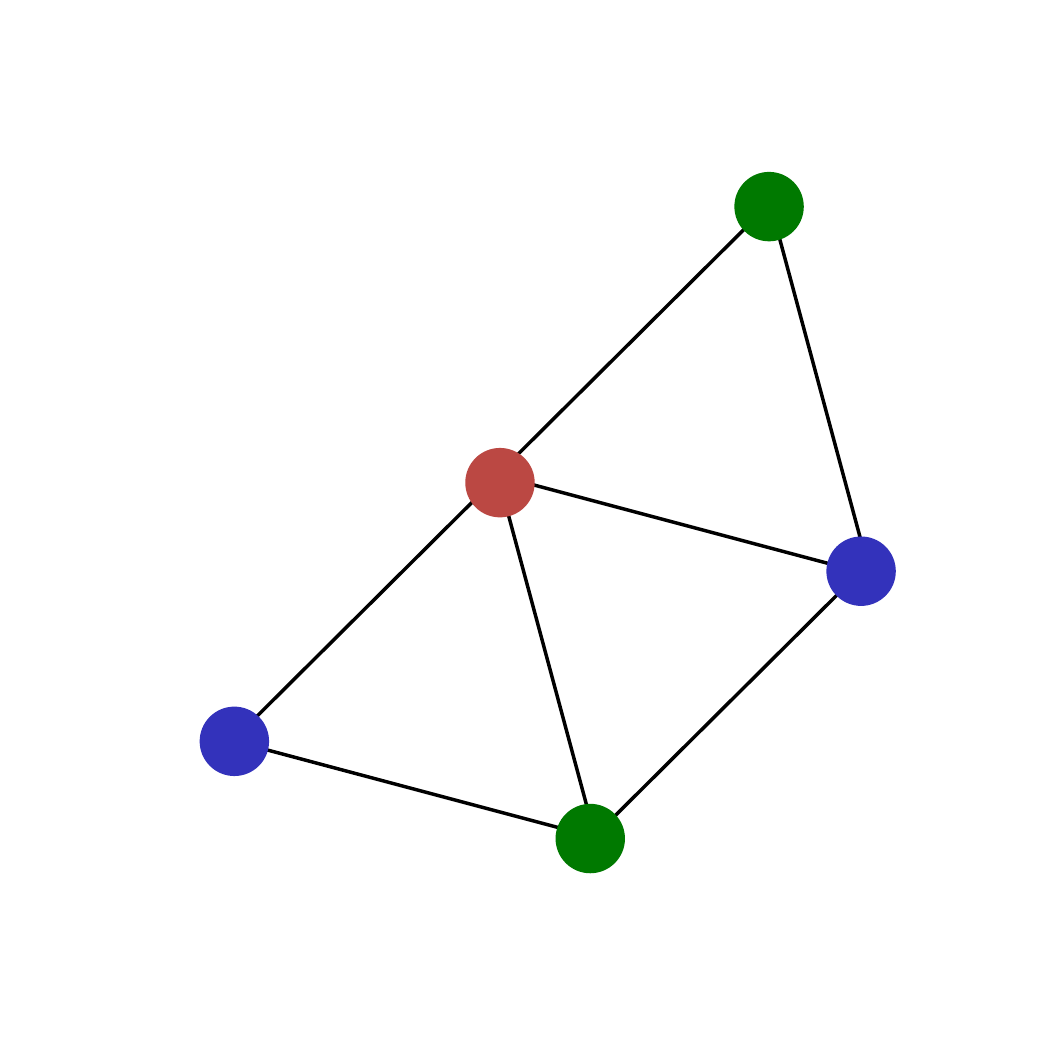}
		\vskip-20pt
	\end{wrapfigure}
	coloring~\cite{JensenToft}. One is given a graph with $N$ nodes and $M$ edges. Each node can be colored by $q$ different colors (e.g. green, red, blue...),
	represented by a Potts spin $\s_i = 1,\cdots, q$.
	The cost function is a sum over edges $<i,j>$,
	\beq\label{eq:COL}
	E(\U{\sigma}) = \sum_{<i,j>} \delta_{\sigma_i,\sigma_j} \ , 
	\eeq
	which counts the number of monochromatic edges connecting two nodes having the same color.
	Clearly, the cost function in Eq.~\eqref{eq:COL} is a sum of local terms, each involving two variables, and vanishing whenever the two variables are different
	and being positive otherwise.
	Similarly, the potential in Eq.~\eqref{eq:sspot} is a sum of pair terms, vanishing when the pair of particles involved is well separated, and being positive otherwise.
	
	One can then formulate two distinct problems. The first is a {\it satisfiability} problem:
	is there an assignment $\U{\sigma}$ of the $N$ variables such that $E(\U{\sigma})=0$, or, in other words, is the problem satisfiable?
	If the answer is yes, the problem is deemed SAT, and UNSAT otherwise. Like in the setting of Sec.~\ref{sec:IB}, for a broad class of cost functions involving disorder,
	a sharp SAT-UNSAT phase transition separates the two phases in the thermodynamic limit, upon varying the number of constraints per variable, i.e. the ratio $\a=M/N$. 
	Furthermore, around the transition,
	there is a critical slowing down of search algorithms that look for solutions~\cite{cheeseman1991really,mitchell1992hard,kirkpatrick1994critical,selman1996critical,monasson1999determining,AMSZ09}.
	
	The second problem is an {\it optimization} (or {\it approximation}) problem, as in Sec.~\ref{sec:IC}: 
	if there is no solution to the set of constraints, is it possible to minimize the number of  unsatisfied constraints, i.e. to find the ground state of $E(\U{\sigma})$?

	\subsection{Machine learning: minimize a loss function}
	
	Similar problems arise in machine learning applications~\cite{EV01,CCCDSTVMZ19rmp}, for example in {\it supervised learning}. An example
	is data classification. Suppose
	one is given a set of
	input data $\U{x}_m$ (e.g. images), and a set of associated labels $y_m \in \{0,1\}$ (e.g. ``dog'' or ``cat''). 
	The problem is to learn, from this set of {\it training} examples, the unknown function $y_m = f(\U{x}_m)$ that relates the input to the output.
	One can then assume a class of functions $y_p=g(\U{x}|\U{\theta}) \in [0,1]$ for $\U{\theta}\in\RRR^P$, such that $y_p$ is the probability of the input $\U{x}$ being labeled as a ``1'',
	which depends on a set of unknown parameters $\U{\theta}$, and define 
	a \textit{loss function} as:
	\beq
	L(\U{\theta}) = \frac{1}{M}\sum_m e\big(y_m,g(\U{x}_m|\U{\theta})\big) \ ,
	\eeq
	where $e(y,y_p)$ is a measure of the error of the predicted probability $y_p$, e.g. the so-called {\it cross-entropy loss}:
	\beq
	e(y,y_p)=-y \log y_p - (1-y) \log (1-y_p) \ .
	\eeq
	For a true label $y=0$, minimizing the cross-entropy is equivalent to minimize $-\log(1-y_p)$, i.e. to minimize $y_p$, and vice versa for $y=1$.
	The aim is then to learn the best value of
	the parameters $\U\theta$ by minimizing the loss function $L(\U\theta)$ calculated on the training examples.
	The training often consists in a gradient descent in the parameter space,
	\beq
	\dot{\U{\theta}} = -\nabla_{\U{\theta}}L(\U{\theta}) \ ,
	\eeq
	possibly in the presence of a little additional noise.
	
	Assuming that the function $f(\U{x})$ is known, together with the statistical distribution of input data,
	the generalization capacity of the trained machine can be evaluated by computing the \textit{generalization error}:
	\beq
	E_g(\U{\theta}) = \mathbb{E}_{\U{x}} \big [ | f(\U{x}) - \Theta[g(\U{x}|\U{\theta})-1/2] | \big ] \ ,
	\eeq
	where $\Theta(x)$ is the Heaviside theta function. In words, we extract a new data point $\U{x}$ at random, we compute its true label $y=f(\U{x})$, and its predicted
	label $y' = \Theta[g(\U{x}|\U{\theta})-1/2]$ (i.e. we assign a label ``1'' if $y_p>1/2$ and ``0'' otherwise), and we then compute the average of $|y-y'|$, which is one in case
	of a mistake and zero otherwise.
	This quantity then provides the probability that the trained machine commits a mistake when classifying a new data point, extracted from the same statistical distribution as the training data, but that was not used in the training.
	
	In order to characterize the local
	minima of $L(\U\theta)$ that are reached by the gradient descent, one can ask several questions:
	\begin{itemize} 
		\item Can we perfectly classify the training data, i.e. reach $L(\U\theta)=0$? The answer to this question
		defines a sharp SAT/UNSAT transition, as a function of the ratio $\alpha=M/P$ between the number of data and the number of parameters when both go to infinity,         
		also called the {\it capacity transition}~\cite{Ga87}. It separates a phase with $L=0$ for low $\a$ from a phase with $L>0$ for high $\a$, as described in 
		Secs.~\ref{sec:IB} and \ref{sec:ID}.
		\item How does the generalization capacity of the machine depend on the dynamics (e.g. by the discretization of the gradient descent, by the initial condition, ...)
		and on the ratio $M/P$~\cite{geiger2020perspective,d2020double}?
		\item What is the role of a little additional noise (e.g. in the so-called stochastic gradient descent)~\cite{MKUZ21,mignacco2021}?
	\end{itemize}
	The analytical and numerical study of the training dynamics, both using simple toy models of data (e.g. random uncorrelated data) and real datasets, has given
	insight on the functioning of the resulting machines.

	\subsection{Inference problems: maximize the likelihood of data}
	
	In inference problems, one is given a set of data, and has to deduce some information about the underlying statistical model from which the data were generated, see e.g.~\cite{schneidman2006weak,CLM09,morcos2011direct,ZK16,cocco2018inverse}. 
	Many inference problems can be framed as {\it unsupervised learning} problems. The data $\{\U{x}_m\}_{m=1,\cdots, M}$ are independently and identically generated
	from an unknown probability distribution $P(\U{x})$.
	One then
	assumes a model probability distribution $Q_{\U{\theta}}(\U{x})$, with unknown parameters $\U{\theta}$. 
	The \textit{likelihood} of the data, i.e. the log-probability that the data were generated by $Q_{\U{\theta}}(\U{x})$, is defined as
	\beq
	\mathcal{L}(\U{\theta}) = \frac{1}{M}\sum_m \log Q_{\U{\theta}}(\U{x}_m) \ .
	\eeq
	Note that the likelihood also coincides with minus the cross-entropy $S_c[P_{\rm emp},Q_{\U{\theta}}]$ 
	of the empirical distribution ${P_{\rm emp}(\U{x}) = M^{-1}\sum_m \delta_{\U{x},\U{x}_m}}$
	and the trial distribution $Q_{\U{\theta}}(\U{x})$.
	One can then seek for the value of $\U{\theta}$ that maximizes the likelihood, or equivalently minimizes the cross-entropy, 
	once again via a gradient descent:
	\beq
	\dot{\U{\theta}} = \nabla_{\U{\theta}}\mathcal{L}(\U{\theta}) = - \nabla_{\U{\theta}}S_c[P_{\rm emp},Q_{\U{\theta}}] \ .
	\eeq
	This intuitive procedure can also be justified more formally, either via the {\it maximum entropy principle}~\cite{schneidman2006weak} or in a Bayesian setting with uniform prior~\cite{cocco2018inverse}.
	
	One can then ask several questions to characterize the quality of the inference, such as:
	\begin{itemize}
		\item If $P(\U{x}) = Q_{\U{\theta}^*}(\U{x})$, can one reconstruct the true $\U{\theta}^*$ via gradient ascent on the likelihood~\cite{ros2019complex}?
		\item Whatever the form of the unknown $P(\U{x})$, are samples from the inferred $Q_{\U{\theta}}(\U{x})$ statistically identical to samples from $P(\U{x})$?
		If the answer if yes, then $Q_{\U{\theta}}(\U{x})$ is called a \textit{generative model}~\cite{trinquier2021efficient}.
		\item Can one use the learned model $Q_{\U{\theta}}(\U{x})$ to infer hidden structure in the data, e.g. to cluster the data or to project them 
		on lower-dimensional manifolds (i.e. lower-dimensional {\it representations} or {\it features})? Can
		one use this knowledge to generate data with desired properties~\cite{cocco2018statistical}?
	\end{itemize}

	\begin{figure}[b]
		\centering
		\includegraphics[width=0.59\columnwidth]{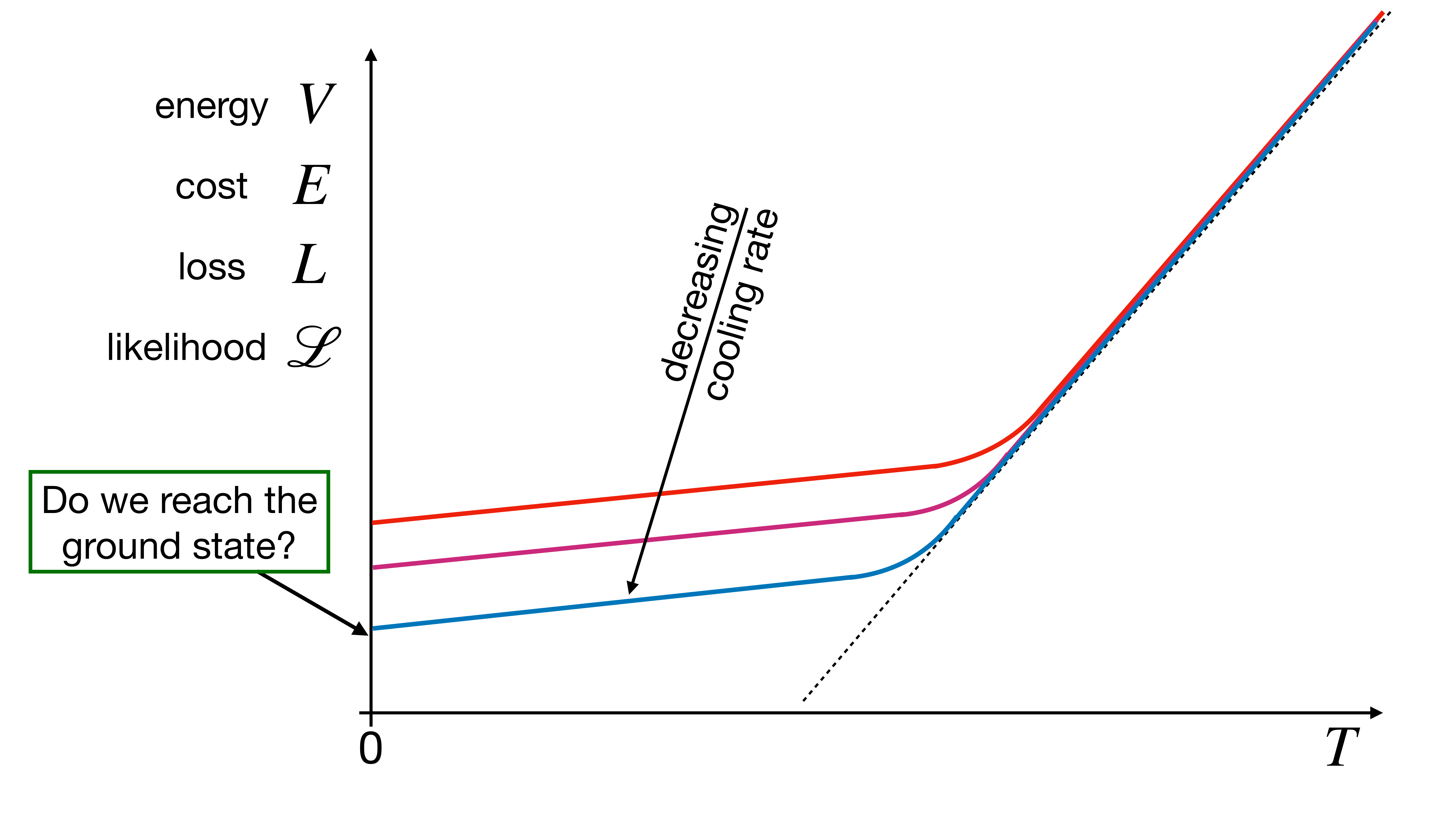}
		\caption{Simulated annealing, i.e. exploring the energy landscape by slowly decreasing the temperature. }\label{fig:SA}
	\end{figure}
	
	\subsection{Simulated annealing: a general algorithm for optimization}
	
	We have seen by this sequence of examples that the optimization of multi-dimensional functions is a highly interdisciplinary problem, which finds application
	in very diverse fields, ranging from soft matter to inference.  A very powerful algorithm to search for a global minimum is simulated annealing~\cite{KGV83}, that was introduced
	by statistical physicists in the context of optimization problems (Sec.~\ref{sec:IC}). In this algorithm,
	a random initial configuration is chosen (corresponding, in statistical physics, to an initial state at temperature $T=\infty$) 
	and the system is then cooled, with different cooling protocols, to zero temperature (Fig.~\ref{fig:SA}). This process mimics in the computer
	the physical cooling process used to created
	structural glasses (Fig.~\ref{fig:glassPEL}B). 
	Simulated annealing is one of the most widely used algorithms for optimization.
	
	Yet, in many of the applications described above, the function to be optimized is {\it rough}, i.e. it displays multiple local minima at different levels,
	separated by saddle points and local maxima, as in Fig.~\ref{fig:glassPEL}A. Because local minima can trap simulated annealing for extremely long times 
	(infinite for gradient descent), and high-energy minima are likely to be more numerous than low-energy ones,
	one might wonder what are the performances of this algorithm in these complex situations. Does the algorithm get stuck in high-energy, bad minima? 
	Is it able to reach the ground state? Or is it able to find a good compromise, i.e. low enough states that provide good approximation to the ground state?
	
	Statistical physics (and other disciplines) have made much progress on this question, see e.g.~\cite{CK93,MR04,ros2019complex,ZK10,KZ13,MKUZ21,mignacco2021,geiger2020perspective,folena2020}.
	In these introductory notes we will consider the two simplest toy models of rough function, that give some insight on the problem:
	\begin{enumerate}
		\item the mixed $p$-spin model;
		\item the perceptron model.
	\end{enumerate}
	Moreover, while in simulated annealing the cooling protocol can be rather arbitrary, we will focus on some limit cases that can be more easily described analytically:
	\begin{enumerate}
		\item full equilibrium (very slow cooling);
		\item gradient descent (very fast cooling);
		\item equilibrium down to some temperature $T$, followed by gradient descent (mixed dynamics).
	\end{enumerate}
	We will see how the performances of these annealing protocols can be studied in the two toy models. The central theme of the notes will be the relation between
	the function geometry (i.e., how the local minima and the saddle points are organized) and the long-time properties of the dynamics.

	\section{Definition and Properties}
	\label{sec:def}
	
	In the following, we will 
	denote by $\Us \in \RRR^N$ the system's configuration, because we will be dealing with continuous spin systems, 
	and by $H(\Us)$ the energy, or cost, or loss, or minus likelihood, function, which we will call {\it Hamiltonian} in the language of physics.
	Let us consider a system in contact with a heat bath at temperature $T=\frac{1}{\beta}$. 
	Given the Hamiltonian $H(\Us)$, its equilibrium Boltzmann-Gibbs distribution is given by
	\begin{equation}
	P_{eq} (\Us) = \frac{e^{-\beta H(\Us)}}{Z} \ , \qquad Z = \sum_{\U\s} e^{-\b H(\U\s)} \ ,
	\end{equation}
	being $Z$ the partition function.
	
	\subsection{Langevin dynamics}
	\label{sec:dyn1}
	
	We consider the simplest model for the dynamics of a system in contact with a heat bath, the {\it over-damped Langevin dynamics}~\cite{Cu02,kurchan2009six}:
	\begin{equation}
	\frac{d \sigma_i}{d t} = -\frac{\partial H}{\partial \sigma_i}  + \xi_i(t) \ , \qquad i=1,\cdots, N \ ,
	\end{equation}
	where $\xi_i(t)$ is a white Gaussian noise that models the thermal bath, with mean and variance given by
	\beq
	\langle \xi_i(t) \rangle =0 \ , \qquad
	\langle \xi_i(t) \, \xi_j(t') \rangle = 2T\delta_{ij}\delta(t-t') \ .
	\eeq
	The evolution of the probability distribution is described by
	the Fokker-Planck equation,
	\begin{equation}
	\frac{dP(\Us,t)}{d t} = \sum_{i}\frac{\partial}{\partial \sigma_i} \Big (\frac{\partial H}{\partial \sigma_i}+T\frac{\partial }{\partial \sigma_i}\Big )  P(\Us,t)= -\mathcal{L}  P(\Us,t) \ .
	\end{equation}
	This equation conserves the total probability, i.e. $\frac{d}{d t}\int d\Us P(\Us,t) =0$, and admits the equilibrium distribution $P_{eq} (\Us)$ as fixed point. It is easy to check
	that
	\begin{equation}
	\qquad 0= -\mathcal{L} P_{eq} (\Us) \ ,
	\end{equation}
	which implies that if at any time the system is in $P(\Us,t) = P_{eq} (\Us)$, then it stays there at all times, because the time derivative of $P(\Us,t)$ then vanishes.
	
	Because $\LL$ is not Hermitian, its left and right eigenmodes associated to an eigenvalue $\l_\a$, respectively $Q_\a$ and $P_\a$, differ.
	For finite $N$ and
	under a few assumptions, physically needed to ensure that the Hamiltonian is confining at infinity and to exclude the existence of disconnected regions of phase space (see Ref.~\cite{kurchan2009six} for details), the operator $\mathcal{L}$ has 
	a unique ground state corresponding to the equilibrium distribution,
	\begin{equation}
	\lambda_0 = 0 \ , \qquad
	P_{0} = P_{eq} \ , \qquad
	Q_{0} = 1 \ ,
	\end{equation}
	and presents a discrete spectrum of strictly positive excited states with eigenvalues $\l_\a>0$. 
	The dynamics starting from a generic initial distribution $P_{in}(\Us)$ can then be decomposed on eigenmodes of $\mathcal{L}$,
	\begin{equation}
	P(\Us,t) = e^{ -\mathcal{L} t} P_{in}(\Us) = \sum_{\alpha}e^{ -\lambda_{\alpha}  t} P_{\alpha}(\Us)\langle Q_{\alpha} | P_{in}\rangle \ .
	\end{equation}
	The $\l_\a$ then physically correspond to decay rates, and the second eigenvalue $\lambda_1$ defines the rate of convergence 
	to equilibrium with a corresponding time scale $\tau_{rel} = 1/\lambda_1$.
	
	\subsection{Observables}
	
\begin{figure}[h]
		\includegraphics[width=0.45\columnwidth]{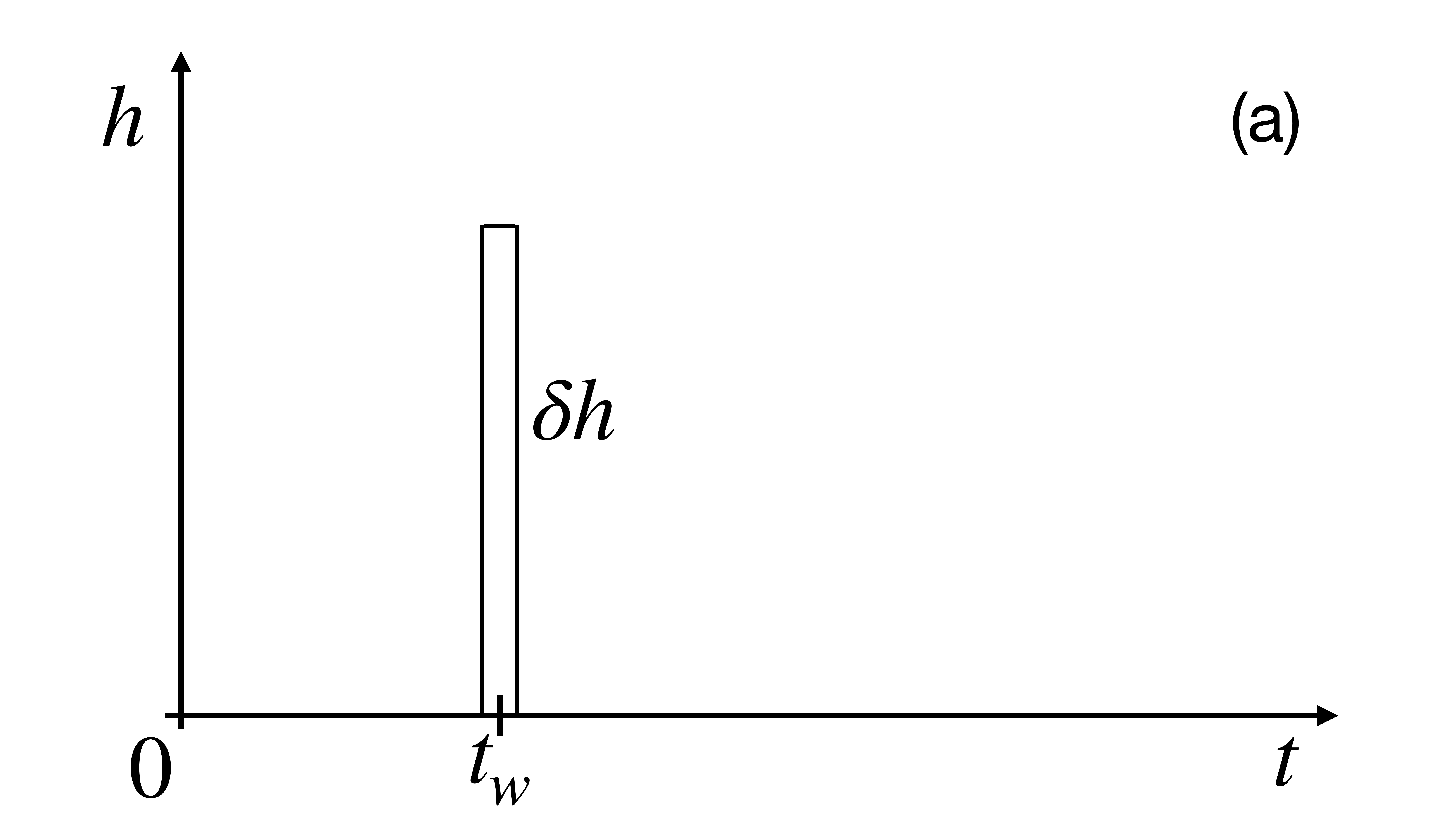}
		\includegraphics[width=0.45\columnwidth]{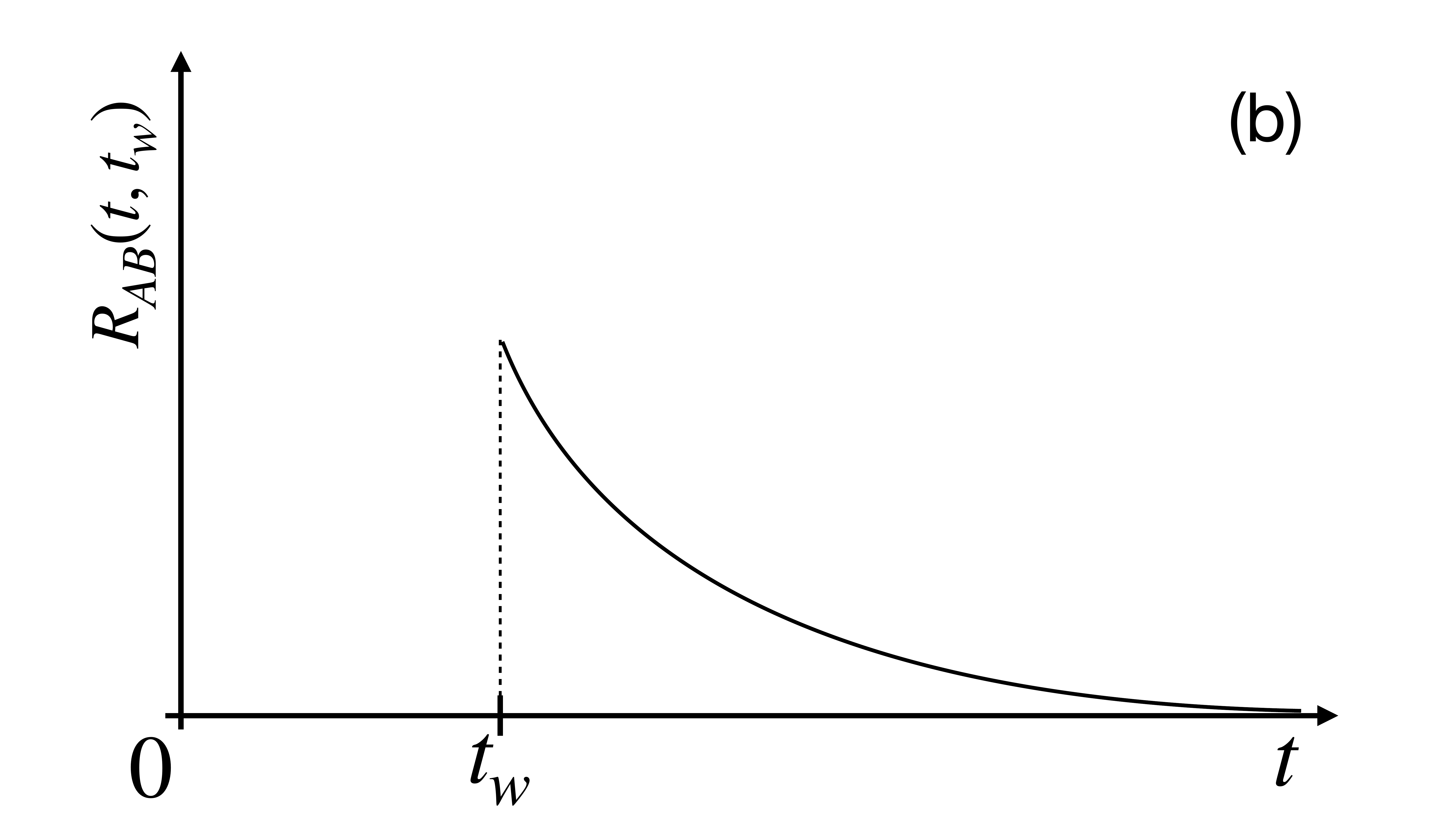}
		\caption{A. Pulse $\delta h$ at time $t_w$. B. Response of the system to the pulse $\delta h$ at time $t_w$.}
		\label{fig:R}
	\end{figure}

	Given generic observables $A(\Us)$ and $B(\Us)$, we define the dynamical average, i.e. the average over the initial condition $P_{in}$ and over 
	different stochastic paths, as
	\begin{equation}
	\langle A(t) \rangle \equiv \langle A(\Us(t)) \rangle = \int \de \Us P(\Us,t)A(\Us) \ ,
	\end{equation}
	and the correlation between observables at two different times as
	\begin{equation}
	C_{AB}(t_w+t,t_w) \equiv \langle A(t_w+t) B(t_w) \rangle \ .
	\end{equation}
	If we perturb the Hamiltonian with an external time-dependent field $h(t)$ conjugated to the observable $B(\Us)$,
	\begin{equation}
	H(\Us) \longrightarrow H(\Us) - h(t) B(\Us) \ ,
	\end{equation}
	we can define the linear response:
	\begin{equation}
	\langle A(t) \rangle_h = \langle A(t) \rangle_0 +\int_{0}^{t}\de sR_{AB}(t,s)h(s) \ .
	\end{equation}
	If the perturbation is a pulse at time $t_{w}$,
	$h(t) = \delta h \, \delta(t-t_w) $,
	then the response of the system precisely corresponds to the response function
	$R_{AB}(t,t_w) = \frac{\delta \langle A(t) \rangle_h} {\delta h}$, as illustrated in Fig.~\ref{fig:R}.
	
	\subsection{Special properties of the equilibrium dynamics}
	\label{sec:TTI}

	Let us consider an initial equilibrium distribution $P_{in}(\Us)=P_{eq}(\Us)$, which
	is then left invariant by the $\mathcal{L}$ operator. Physically, if the system is initialized in equilibrium, then it remains in equilibrium at all subsequent times.
	The invariance of $P_{eq}$ with time has several important implications~\cite{Cu02,kurchan2009six}:
	\begin{itemize}
		\item Time-Translation Invariance (TTI): because the origin of time can be shifted arbitrarily leaving the dynamics invariant, 
		two-times observables depend only on the time differences, i.e.
		\beq\label{eq:TTI}
		C_{AB}(t+t_w,t_w) = C_{AB}(t) \ , \qquad
		R_{AB}(t+t_w,t_w) = R_{AB}(t) \ .
		\eeq
		\item Onsager Reciprocity: combining the time reversal symmetry, which implies $C_{AB}(t) = C_{AB}(-t)$ (because the Langevin dynamics is statistically reversible),
		and TTI, which implies $C_{AB}(t)=\la A(t)B(0) \ra=\la A(0)B(-t) \ra = C_{BA}(-t)$, we have
		\begin{equation}
		C_{AB}(t) = C_{BA}(t) \ .
		\end{equation}	 
		This symmetry has important implications for transport coefficients.
		\item Fluctuation-Dissipation Theorem (FDT): the response and correlation functions are related by
		\begin{equation}
		R_{AB}(t) = -\frac{1}{T} \Theta(t) \frac{d}{dt}C_{AB}(t) \ ,
		\end{equation}	
		where $\Theta(t)$ is the Heaviside function. 
		\item Decorrelation (at finite $N$): because of the discreteness of the spectrum of the evolution operator (Sec.~\ref{sec:dyn1}), the connected  correlation and response functions decay exponentially at long times,
		\beq\label{eq:ergoN}
		\langle A(t) B(0) \rangle_{eq} - \langle A \rangle_{eq}  \langle B \rangle_{eq} \sim e^{-t/\tau_{rel}} \ , \qquad\qquad
		R_{AB}(t) \propto \frac{d}{dt}C_{AB}(t)\sim e^{-t/\tau_{rel}}\ ,
		\eeq
		as illustrated in Fig.~\ref{fig:R}B for the response function. 
	\end{itemize}
	As we will see in Sec.~\ref{sec:eq}, the relaxation time $\tau_{rel}$ can however diverge in the thermodynamic limit, leading to ergodicity breaking at a dynamical phase transition.

	\section{Equilibrium dynamics in the thermodynamic limit:\newline the $p$-spin spherical model}
	\label{sec:eq}
	
	The pure $p$-spin spherical model is defined by the Hamiltonian:
	\begin{equation}
	H_p(\Us) = -\sum_{i_1<i_2<...<i_p}J^{(p)}_{i_1\cdots i_p}\sigma_{i_1}\sigma_{i_2}...\sigma_{i_p} \ ,
	\end{equation}
	where the spins $\s_i\in\mathbb{R}$ with $i = 1,\cdots,N$ are constrained on a sphere $\sum_{i}\s_{i}^2=N$. The quenched disorder is given by couplings $J^{(p)}$ that are i.i.d. Gaussian variables with mean $\overline{J}=0$ variance $\overline{J^2} = \frac{ p!}{2N^{p-1}}$; from now on, an overline denotes the quenched average over the random couplings.
	A mixed $p$-spin is given by a mixture of pure $p$-spins with independent couplings:
	\begin{equation}
	H(\Us) = \sum_{p}\sqrt{\alpha_p} H_p(\Us) \ .
	\end{equation}
	For fixed $\Us$, $H[\Us]$ is a random Gaussian variable with average and covariance given by\\
	\begin{wrapfigure}{r}{.2\textwidth}
		\includegraphics[width=0.19\columnwidth]{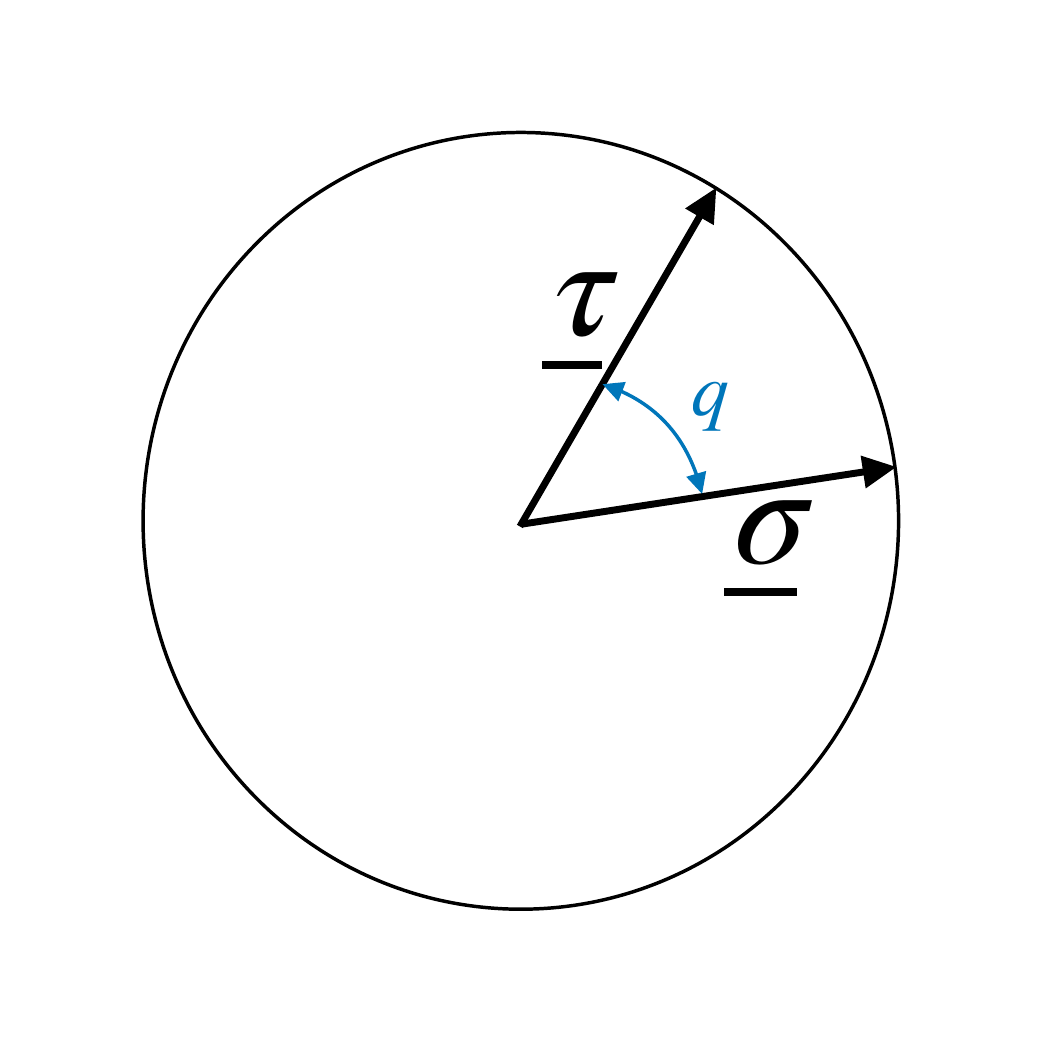}
		\vskip-20pt
	\end{wrapfigure}
	\begin{equation}\label{fluctuations}
	\overline{H[\Us]} = 0 \ , \qquad
	\overline{H[\Us]H[\Ut]} = N f(q_{\Us \; \Ut})\qquad \text{with} \quad f(q) = \frac{1}{2} \sum_p \alpha_p q^p \ ,
	\end{equation}
	where $q_{\Us \; \Ut} = \Us \cdot \Ut / N =  \Hs \cdot \Ht$ is the overlap (or scalar product) between different configurations.
	Energies for different configurations $\Us, \Ut$ are thus correlated, but
	their correlation depends only on the overlap between the configurations.
	The function $f(q)$ is the polynomial that uniquely defines each specific mixed $p$-spin model. 
	By means of $f(q)$ it is possible to define different classes of models, which correspond to different kinds of rough landscapes with 
	ergodicity breaking at low temperature (see Ref.~\cite{folena_mixed_2020} for more details).
	
	As in Sec.~\ref{sec:def},
	we consider the simplest equilibrium dynamics, the overdamped Langevin equation
	\begin{equation}\label{Langevin}
	\partial_t\sigma_i = -\mu \sigma_i -\frac{\partial H}{\partial \sigma_i}+\xi_i \ ,
	\end{equation}
	where $\xi_i$ is the thermal noise with zero mean and 
	white correlation $\langle \xi_i(t) \xi_j(t')\rangle = 2T \delta_{ij}\delta(t-t')$, and
	the term $-\mu \s_i$ is added to enforce the spherical constraint on the spins~\cite{Cu02}. We will discuss later on how the parameter $\mu$ is determined.
	This mixed $p$-spin spherical model with Langevin dynamics 
	is the simplest toy model of rough energy landscape, and it is exactly solvable in the thermodynamic limit $N\to\io$, both for the 
	thermodynamics (via replicas) and for the dynamics (via dynamical mean field theory, or DMFT)~\cite{SZ81,SZ82,KT87b,CS92,CHS93,CK93,Cu02}.
	A very pedagogical review on its solution is Ref.~\cite{CC05}.
	Several results obtained via these methods for the spherical $p$-spin have been rigorously confirmed~\cite{FT06,auffinger2013,auffinger2013b,subag2017,auffinger2018,arous2020}.
	
	\subsection{Equilibrium phase diagram}

	A rather peculiar case is given by the pure $p=2$ spherical model, for which the Hamiltonian $H=-\sum_{i<j}J_{ij}\sigma_i\sigma_j$ is quadratic in the spins $\Us$. This model presents a phenomenology that differs, both thermodynamically and dynamically, from models with $p>2$. This is a result of the fact the GOE matrix $J_{ij}$ can be diagonalized, leading to a decoupling of the degrees of freedom in the diagonal basis:
	\beq
	H=-\sum_{\alpha}\lambda_{\alpha}\sigma_{\alpha}^2 \ .
	\eeq
	Correspondingly, the energy landscape is convex, hence not rough.
	The thermodynamics of the system is characterized by a condensation transition at a critical temperature $T_c$, below which a finite component of the Gibbs measure is concentrated in the lowest eigenvalue $\lambda_0$, and the system enters a ``disordered ferromagnetic'' phase, where a spontaneous magnetization along the lowest eigenvalue of the matrix $J_{ij}$ is present~\cite{KTJ76}.
	The Newtonian dynamics of the system is fully integrable, with $N$ integrals of motion~\cite{barbier2020non}.
	
	\begin{figure}[h]
		\centering
		\includegraphics[width=0.7\columnwidth]{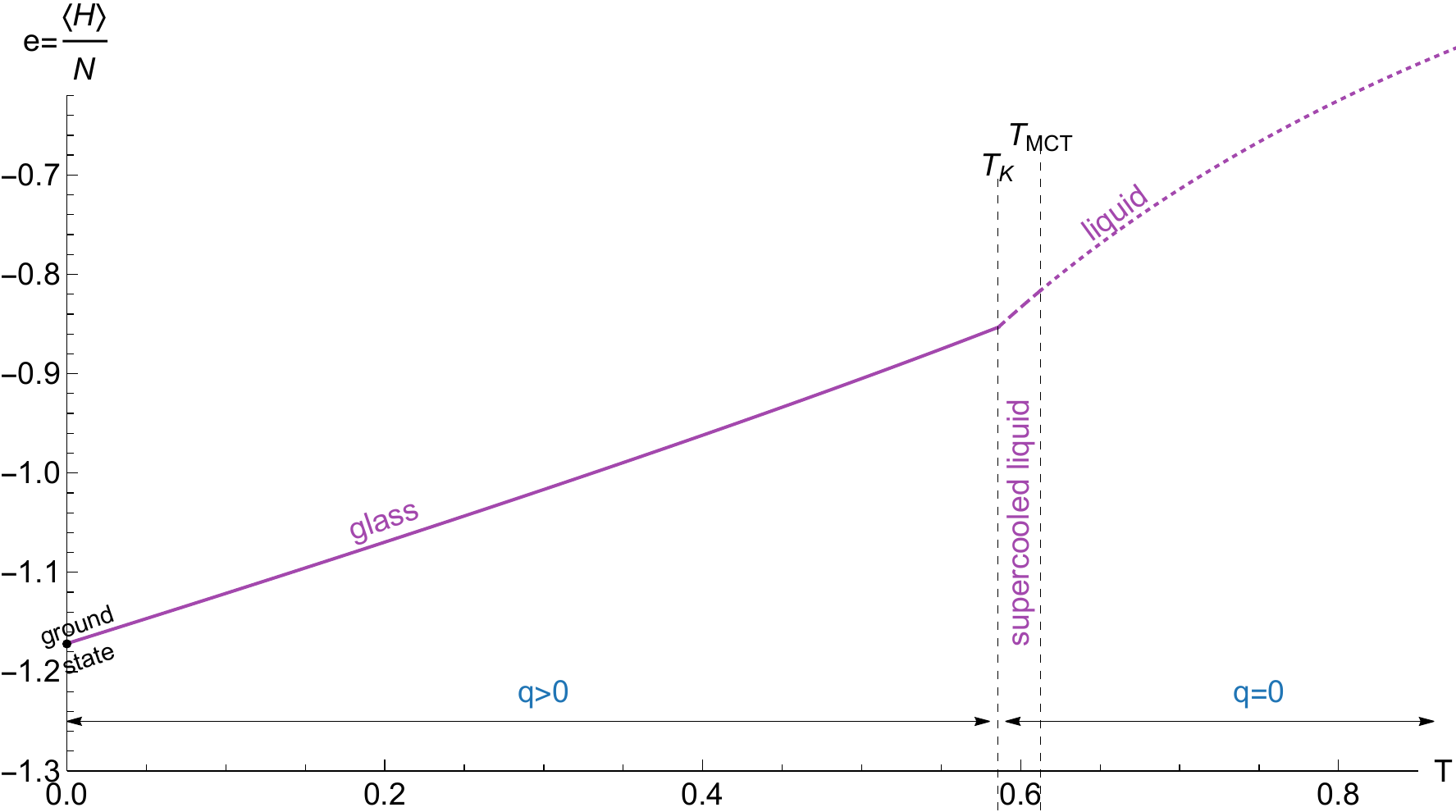}
		\caption{Equilibrium energy in the 3-spin model}
		\label{fig:eTeq}
	\end{figure}
	
	We will thus be mostly interested in the equilibrium behavior that is generically observed in all pure $p$-spin models with $p\geq 3$, and in most cases of mixed models,
	as illustrated in Fig.~\ref{fig:eTeq}.
	A critical temperature $T_{\K}$ (Kauzmann temperature) 
	then separates two different thermodynamic phases.
	\begin{itemize}
		\item \textbf{Paramagnetic phase} ($T>T_{\K}$): The system is paramagnetic, hence the local magnetization is $\la \s_i \ra=0$ for all spins.
		Futhermore, the average overlap of two configurations that are drawn independently from the same Boltzmann-Gibbs measure (i.e. with the same couplings, corresponding to the same physical system) is
		\beq
		q=\overline{\langle \Us \cdot \Ut \rangle}/N = \frac1N\sum_i \overline{\langle\s_i \t_i\rangle}= 
		\frac1N\sum_i \overline{\langle\s_i\rangle\langle \t_i\rangle} =
		\frac1N\sum_i \overline{\langle\s_i\rangle^2} 
		= 0 \ .
		\eeq
		We conclude that two typical equilibrium configurations are orthogonal on the sphere. 
		\item \textbf{Spin glass phase} ($T<T_{\K}$): Few lowest-energy glassy states dominate the Gibbs measure, 
		leading to a spin glass phase with $\la \s_i \ra \neq 0$. Following the previous reasoning, we now obtain $q = \sum_i \overline{\langle\s_i\rangle^2}/N>0$.
	\end{itemize}
	In the language of structural glasses, the paramagnetic phase corresponds to the liquid phase, and the spin glass phase to the glass phase.
	The transition happening at $T_{\K}$ is called a \textit{Random First Order Transition} and it has mixed character~\cite{De81,GM84,Ga85,GKS85,KW87b}: 
	the transition is thermodynamically of second order (e.g. the energy is continuous
	at $T_{\K}$ so there is no latent heat), but 
	$\langle q \rangle$ jumps from zero to a finite value at $T_{\K}$, hence the order parameter is discontinuous as in a first order transition.
	The low-temperature phase for $T<T_{\K}$ is also characterized by replica symmetry breaking~\cite{De81,GM84,Ga85,GKS85,KW87b,MPV87,CC05}.
	

	\subsection{Equilibrium dynamics}
	\label{sec:eqdyn}
	
	We now consider more carefully the equilibrium dynamics of the model. The system is initialized at time $t=0$ in an equilibrium configuration, drawn from the Boltzmann-Gibbs
	measure at temperature $T$, and Langevin dynamics is run at the same temperature $T$, so the system remains in equilibrium at all times and enjoys the special properties
	discussed in Sec.~\ref{sec:TTI}.
	We consider the time-dependent overlap correlation function $C(t,t')$,
	\begin{equation}\label{eq:Cttp}
	C(t,t') = \frac{1}{N} \sum_i \overline{\langle \sigma_i(t)\sigma_i(t') \rangle} = \overline{\langle\Hs(t) \cdot \Hs(t')\rangle} \ ,
	\end{equation} 
	in the limit $N\to\io$ at finite times $t,t'$.
	Because of TTI, we have $C(t+t_w,t_w)=C(t)$, using
	the notation of
	Eq.~\eqref{eq:TTI}.
	This function shows several different regimes upon varying temperature, corresponding to three distinct dynamical phases.
	
	\begin{figure}[t]
		\centering
		\includegraphics[width=0.6\columnwidth]{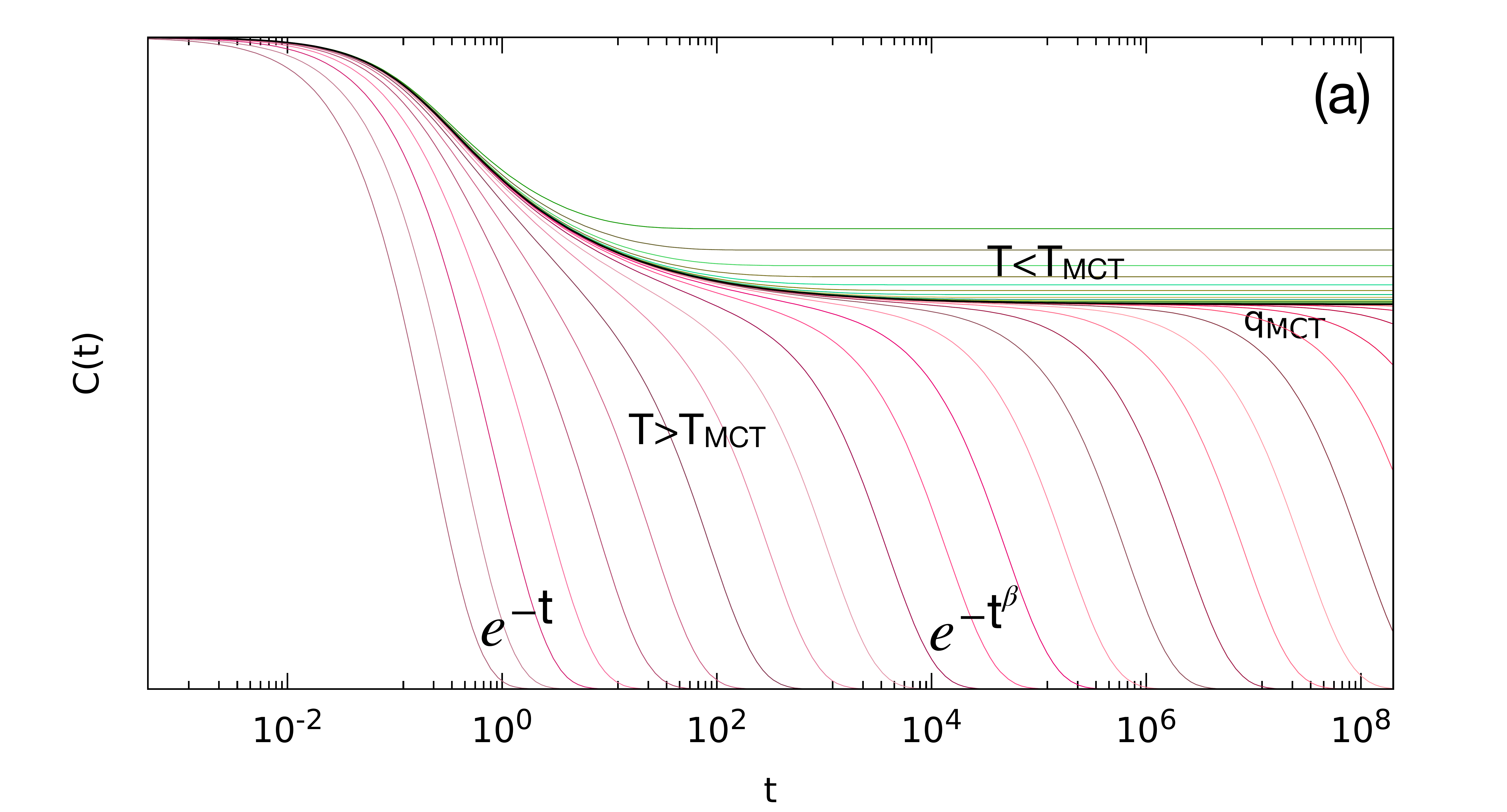}
		\includegraphics[width=0.34\columnwidth]{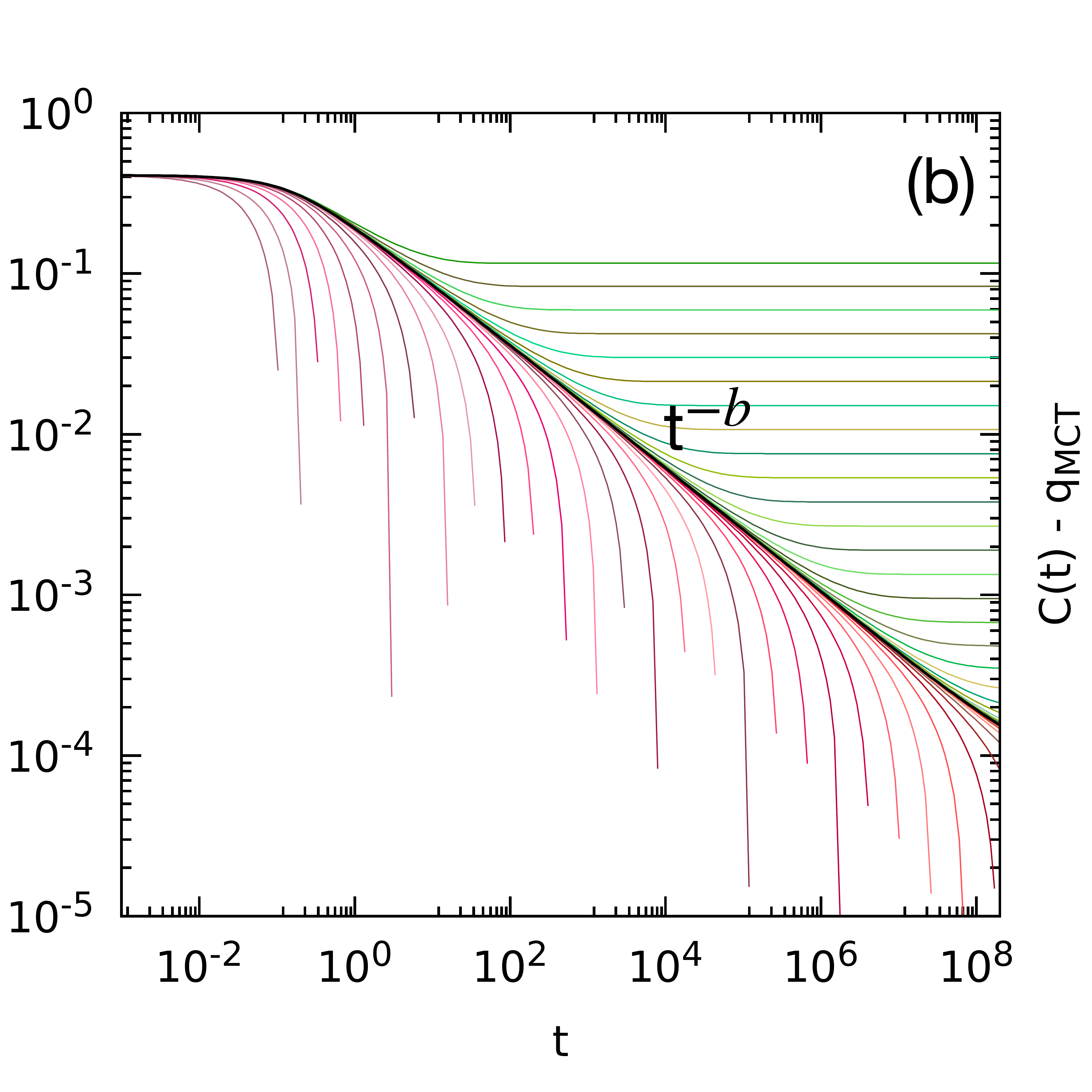}
		\caption{A. Equilibrium overlap correlation function in the (3+4)-model at several temperatures. If $\lim_{t\to\infty}C(t) = 0$ the system is an ergodic liquid (red lines), while for $\lim_{t\to\infty}C(t) = q>0$ it is in a dynamically arrested supercooled liquid phase (green lines). The relaxation time scale $\tau$ diverges upon approaching $T_{\MCT}$ from above.	
			B. A log-log representation of $C(t)-q_{\MCT}$ versus time shows the power-law approach to the plateau from above and below $T_{\MCT}$. Plots adapted from Ref.~\cite{folena_mixed_2020}.
		}
		\label{fig:Corr}
	\end{figure}

	\subsubsection{$T>T_{\MCT}$}

	At high temperature, one observes an exponential decay of 
	$C(t) \sim e^{-(t/\tau)^\beta}$ at long times, with 
	relaxation time $\tau$, and possibly a non-trivial exponent $\beta$ called {\it stretching exponent},
	as illustrated in Fig.~\ref{fig:Corr}A. The stretching exponent decreases by approaching $T_{\MCT}$ from $\beta=1$ to a smaller value $\beta\approx 0.9$; this effect is observed also in numerical simulations of real glasses~\cite{sastry_signatures_1998}.
	Hence, the dynamics is {\it ergodic}. At long times $t\gg\tau$, 
	ergodicity implies $\langle \sigma_i(t)\sigma_i(0) \rangle \sim \langle \s_i\rangle_{eq}\langle \s_i\rangle_{eq}$, see Eq.~\eqref{eq:ergoN}.
	Hence
	the overlap between the state at $t=0$ and that at time $t$
	converges to the typical overlap of two independent equilibrium configurations, i.e. to the thermodynamic value $q=0$.
	Upon cooling, the typical relaxation time $\tau$ increases, and it diverges  as a power-law upon approaching a critical temperature, 
	i.e. $\tau = |T- T_{\MCT} |^{-\gamma}$ when $T\to T_{\MCT}^+$. 
	At $T=T_{\MCT}$, the correlation function does not decay to zero anymore, but it reaches a finite limit, $\lim_{t\to\io} C(t) = q_{\MCT}$. The values of $T_{\MCT}$ and $q_{\MCT}$ are
	obtained by finding the maximal temperature such that the equation
	\beq\label{eq:qeq}
	\b^2 f'(q) = q/(1-q)
	\eeq
	has a solution $q>0$. This equation can be derived following two paths. Either dynamically, by closing the equilibrium DMFT equation in the long-time limit~\cite{CHS93}, or by a replica calculation that evaluates the free energy of the metastable states~\cite{FP95,Mo95}. See Refs.~\cite{CC05,folena_mixed_2020} for pedagogical reviews.
	\begin{wrapfigure}{r}{.2\textwidth}
		\centering
		\includegraphics[width=.2\columnwidth]{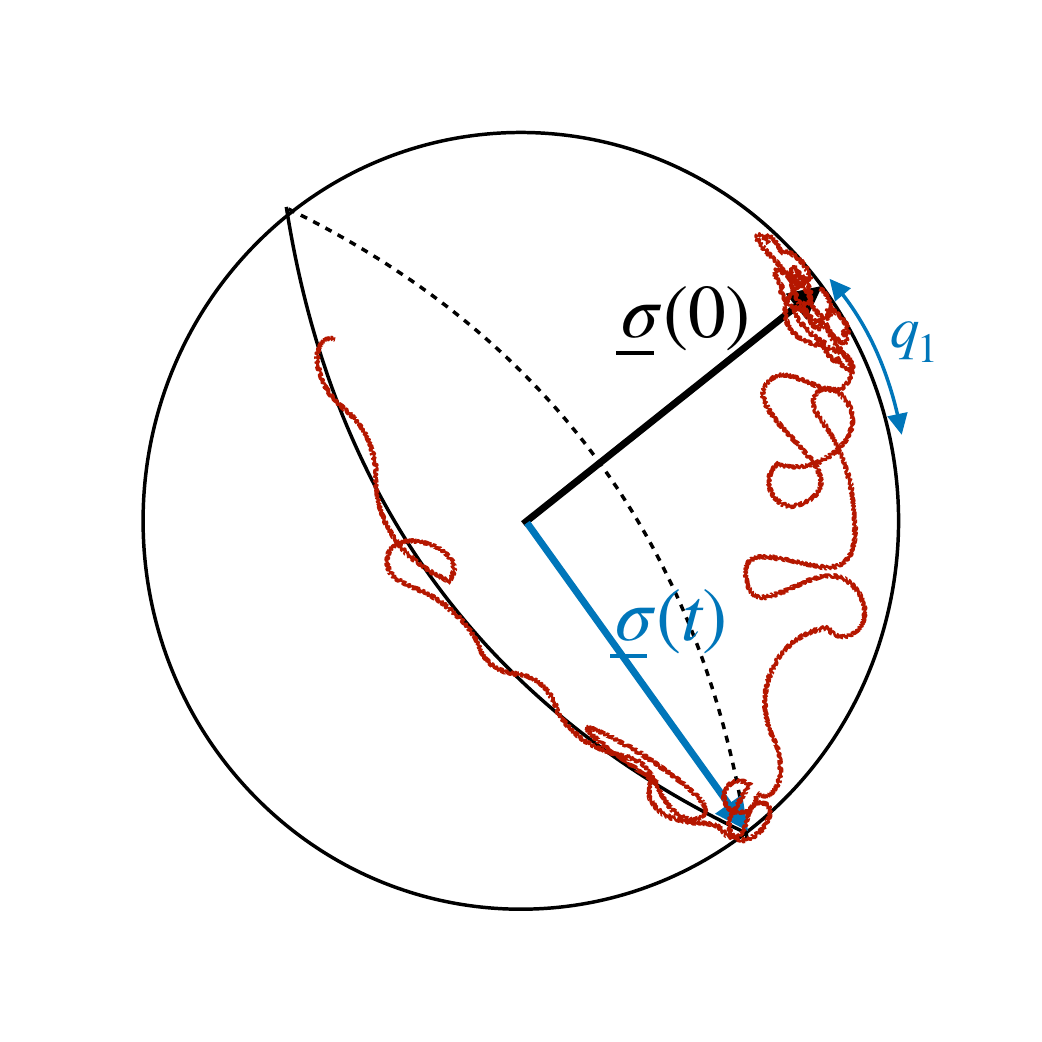}
		\small{\textit{relaxation in the liquid phase}}
		\vskip-40pt
	\end{wrapfigure}
	The critical scaling of $C(t)$ upon approaching
	$T_{\MCT}$ from above 
	is characterized by three time regimes (Fig.~\ref{fig:Corr}B):
	\begin{equation}\begin{matrix}
	&C(t)-q_{\MCT} \propto t^{-b} \ , &\qquad t\ll\tau \ , \\
	&C(t)-q_{\MCT} \propto -t^{a} \ , &\qquad t\lesssim \tau \ , \\
	&C(t) \propto \exp(-t/\tau)  \, &\qquad t\gg \tau \ ,
	\end{matrix}
	\end{equation}
	and the three exponents $\gamma,a,b$ are related by 
	$\lambda_{\MCT} = \frac{\Gamma(1-a)^2}{\Gamma(1-2a)}=\frac{\Gamma(1-b)^2}{\Gamma(1-2b)}$ and $\gamma=\frac{1}{2a}$,
	with a non-universal parameter $\lambda_{\MCT}=\frac{f'''(q_{\MCT})f'(q_{\MCT})}{2q_{\MCT}f''(q_{\MCT})^2}$.
	In the special case of pure models, $f(q) \propto q^p$, one has $q_{\MCT}=(p-2)/(p-1)$, 
	and $\lambda_{\MCT}$ is then identically equal to $0.5$.
	In summary, for $T$ slightly higher than $T_{\MCT}$,
	a typical dynamical trajectory starting from an equilibrium state $\Us(0)$ remains close to the initial state for a long time, of the order of $\t$,
	and then relaxes away from it towards another typical equilibrium state, which is then orthogonal to $\Us(0)$. At $T_{\MCT}$ and below, this relaxation
	process is frozen and the trajectory remains forever confined in the vicinity of the initial state, without reaching equilibrium.
	The critical behavior described above had been previously described within
	Mode-Coupling Theory, an approximate theory of the glass transition~\cite{gotze}. This analogy was crucial to realize that
	the $p$-spin model provides a mean-field theory of the glass transition~\cite{KW87,KT87}.

	\subsubsection{$T_{\K}<T<T_{\MCT}$}
	
	Remarkably, it is found that $T_{\MCT}>T_{\K}$, so the critical divergence of the relaxation
	time happens at a temperature strictly higher than the thermodynamic transition~\cite{KT87,CS92,CHS93}.
	For $T_{\K}<T<T_{\MCT}$, 
	the dynamics is trapped around the initial state for an infinite time: the overlap correlation reaches a finite plateau,
	\begin{wrapfigure}{r}{.2\textwidth}
		\vskip-10pt
		\centering
		\includegraphics[width=.2\columnwidth]{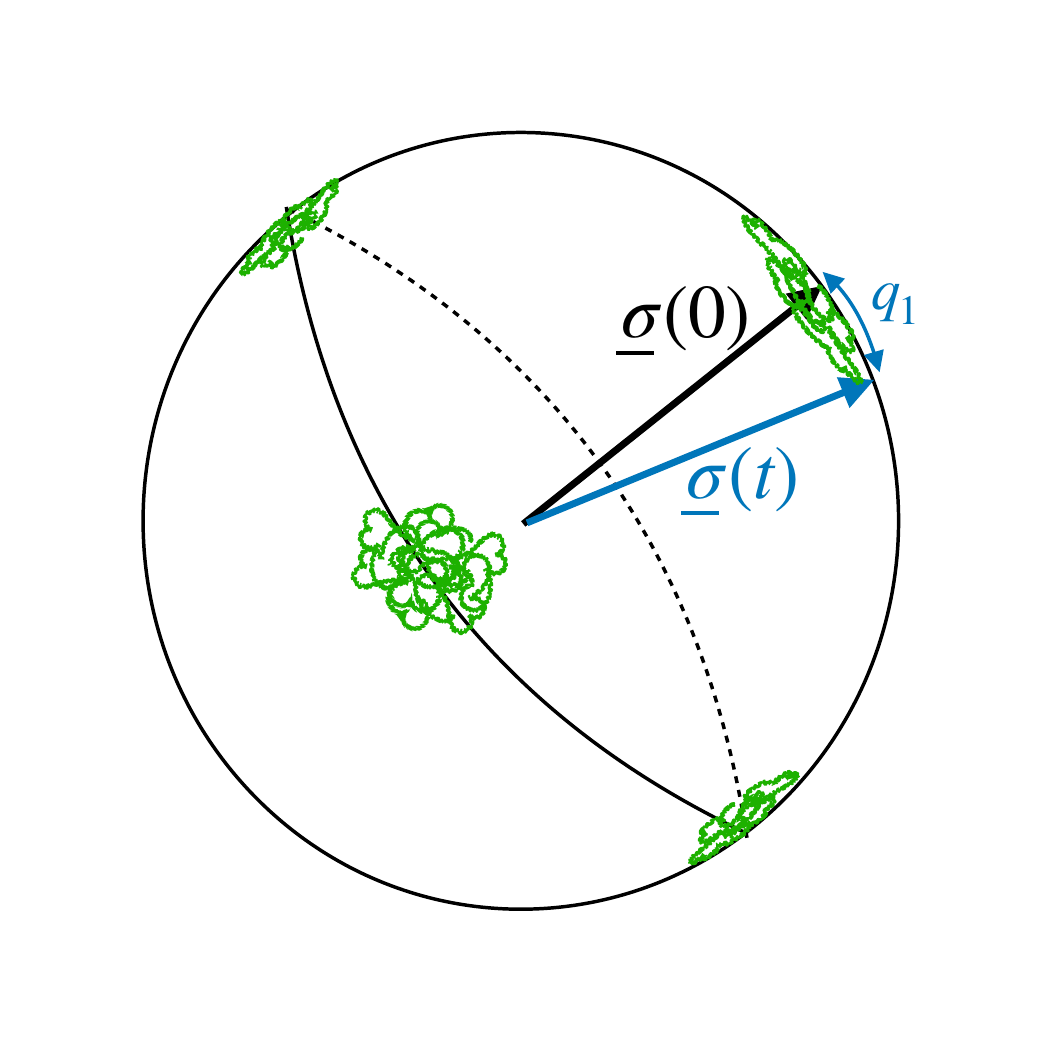}
		\small{\textit{states in the supercooled phase}}
		\vskip-10pt
	\end{wrapfigure}
	$\lim_{t\to\io} C(t)=q_1$, with $q_1$ given by the largest solution of Eq.~\eqref{eq:qeq}. In other words, a typical equilibrium trajectory
	remains confined in a cone of scalar product $\Hs(t) \cdot \Hs(0) \leq q_1$ around the initial condition.
	But yet, we know that the system is thermodynamically a paramagnet, hence typical pairs of equilibrium configurations have overlap $q_0=0$.
	We conclude that each equilibrium configuration is surrounded by a cone of size $\approx 1-q_1$, that is explored by the dynamics starting in that configuration,
	but that there are many such cones, each one corresponding to a distinct independent initial equilibrium state, so that typical pairs of cones are orthogonal
	on the sphere. Each of these cones defines a metastable spin-glass (SG) state. Because of these metastable states, the dynamics is {\it not ergodic}:
	the long-time limit of a typical dynamical trajectories does not reach equilibrium, because it is unable to jump out of the cone defined by the initial state.
	More precisely, the time to escape from the initial metastable state scales exponentially with the system size, $\t \approx e^N$, in this regime.
	Hence, if the limit $N\to\io$ is taken first, no relaxation is observed on finite times; while for finite $N$ systems, some relaxation can be observed, but over extremely
	long time scales (Fig.~\ref{fig:GDT}B).
	
	It is possible to show that in this regime there is an
	exponential number in $N$ of equivalent SG states, each having a typical size $1-q_1>0$ and being orthogonal, i.e. with $q_0=0$, to all other states~\cite{CS95,CC05}.
	The finite limit
	\beq
	\Sigma=\lim_{N\to\io}\frac{1}{N}\log{\#\text{SG}}
	\eeq
	is called  ``complexity'' or entropy of the SG states, and is illustrated in Fig.~\ref{fig:sigma}. 
	In the glass literature, this quantity is also known as configurational entropy~\cite{sciortino_potential_2005}.
	A metastable SG state can be thought, in a first approximation (which is actually exact for the pure $p$-spin model), as a local minimum of the Hamiltonian
	``dressed'' by thermal fluctuations~\cite{CS95,CC05}. Hence, the presence of an exponential number of SG states indicates an extremely rough Hamiltonian at low energies. In Ref.~\cite{folena_mixed_2020} different methods to evaluate the complexity in the $p$-spin model are compared.
	
	Finally, note that while equilibrium configurations are always mathematically well defined, in practice it is impossible
	to sample from the Boltzmann-Gibbs distribution for $T<T_{\MCT}$ by conventional means. Typically, one would take a finite $N$ system,
	initialize it in a random (infinite-temperature) state, and run the dynamics at the target temperature for long enough, until equilibration is reached.
	But because the relaxation time is exponentially large in $N$, the dynamics takes an astronomically large time to equilibrate unless the system is very small, which makes the sampling practically impossible.
	A solution to this problem is the so-called {\it planting trick}, see Refs.~\cite{ZK10,folena_gradient_2021,folena_mixed_2020} for a detailed discussion.

	\subsubsection{$T<T_{\K}$}
	
\begin{wrapfigure}{r}{.4\textwidth}
		\vskip-10pt
		\centering
		\includegraphics[width=.35\textwidth]{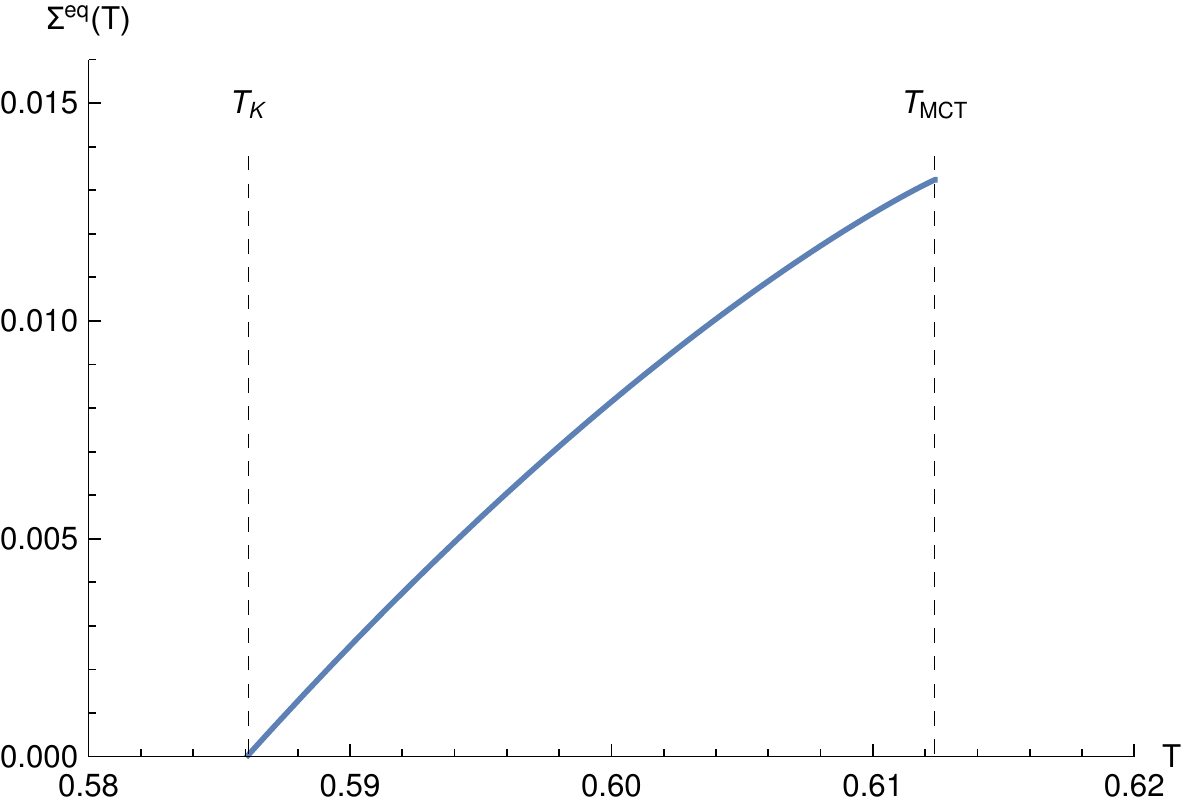}
		\caption{Complexity for the $3$-spin model.}
		\label{fig:sigma}
		\vskip-10pt
	\end{wrapfigure}

	The shape of the equilibrium complexity for a $p$-spin model is illustrated in Fig.~\ref{fig:sigma}.
	The complexity is only defined 
	for $T<T_{\MCT}$, where metastable
	states exist, and it is a decreasing function of temperature. This is consistent with intuition: the lower the temperature, the lower the energy, and it is reasonable
	to expect less local minima of the energy landscape at lower energy.
	The complexity is found to vanish continuously at $T_{\K}$, indicating that at this temperature, the system in equilibrium can only be found in a sub-exponential number
	of SG states (which can be shown to be actually finite). Because the complexity cannot be negative (it is the logarithm of the number of states), the SG states at $T_{\K}$
	are the lowest free energy states, and are thus thermodynamically stable~\cite{CC05}. A true phase transition to an equilibrium spin glass phase then happens at $T_{\K}$.
	For any $T<T_{\K}$, the complexity remains identically zero and the system is found in the lowest free energy states at each $T$, which become,
	when $T\to 0$, the ground states of the Hamiltonian. Throughout this phase, the dynamics is qualitatively similar to that in the regime $T_{\K}<T<T_{\MCT}$: the correlation
	function relaxes to a finite plateau $q_1$, whose calculation now requires replica symmetry breaking~\cite{BFP97}.

	\section{Out-of-equilibrium dynamics}
	\label{sec:noneq}
	
	\begin{figure}[b]
		\centering
		\includegraphics[width=0.63\columnwidth]{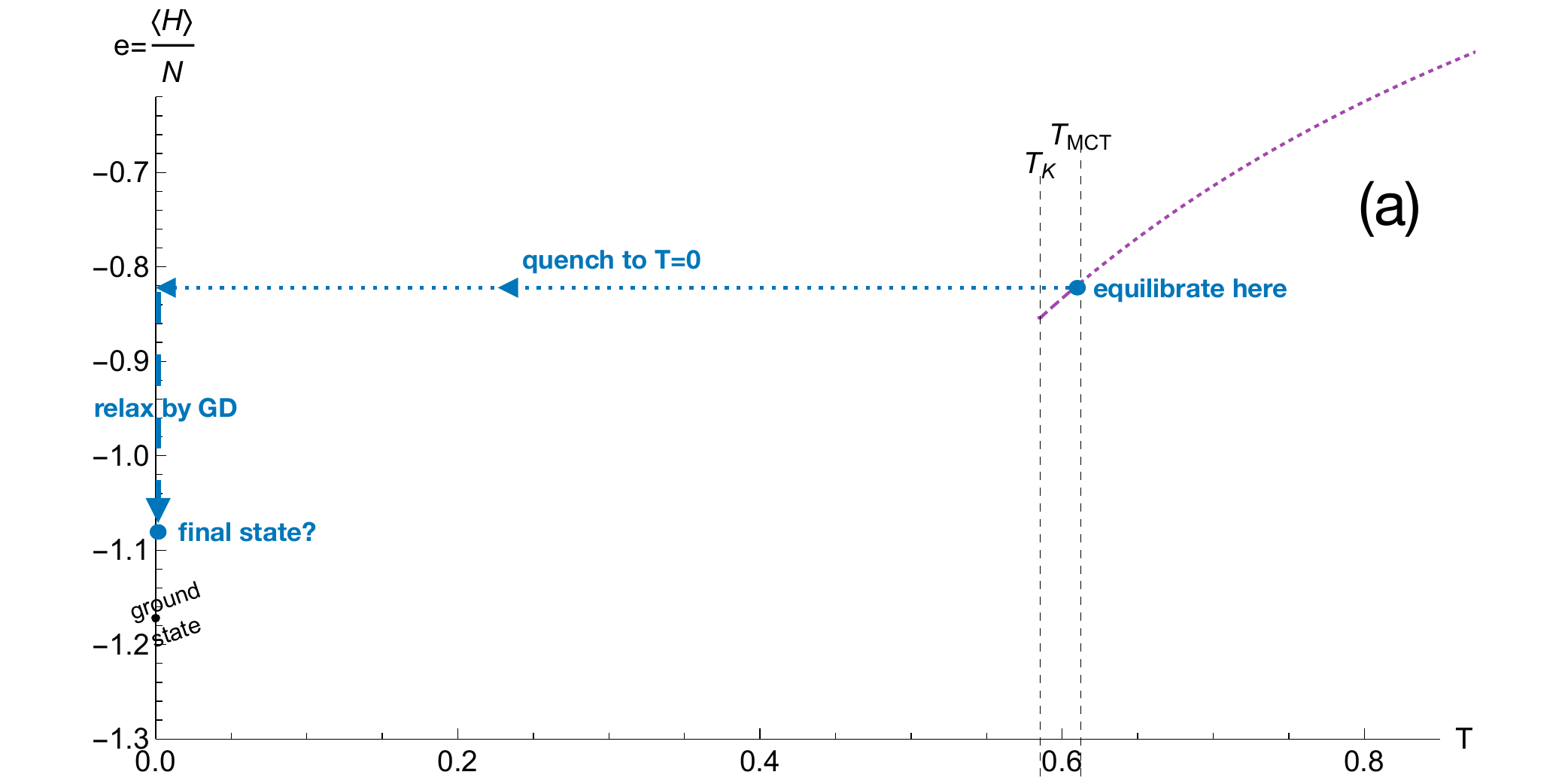}
		\includegraphics[width=0.33\columnwidth,trim=0 0 0 4cm]{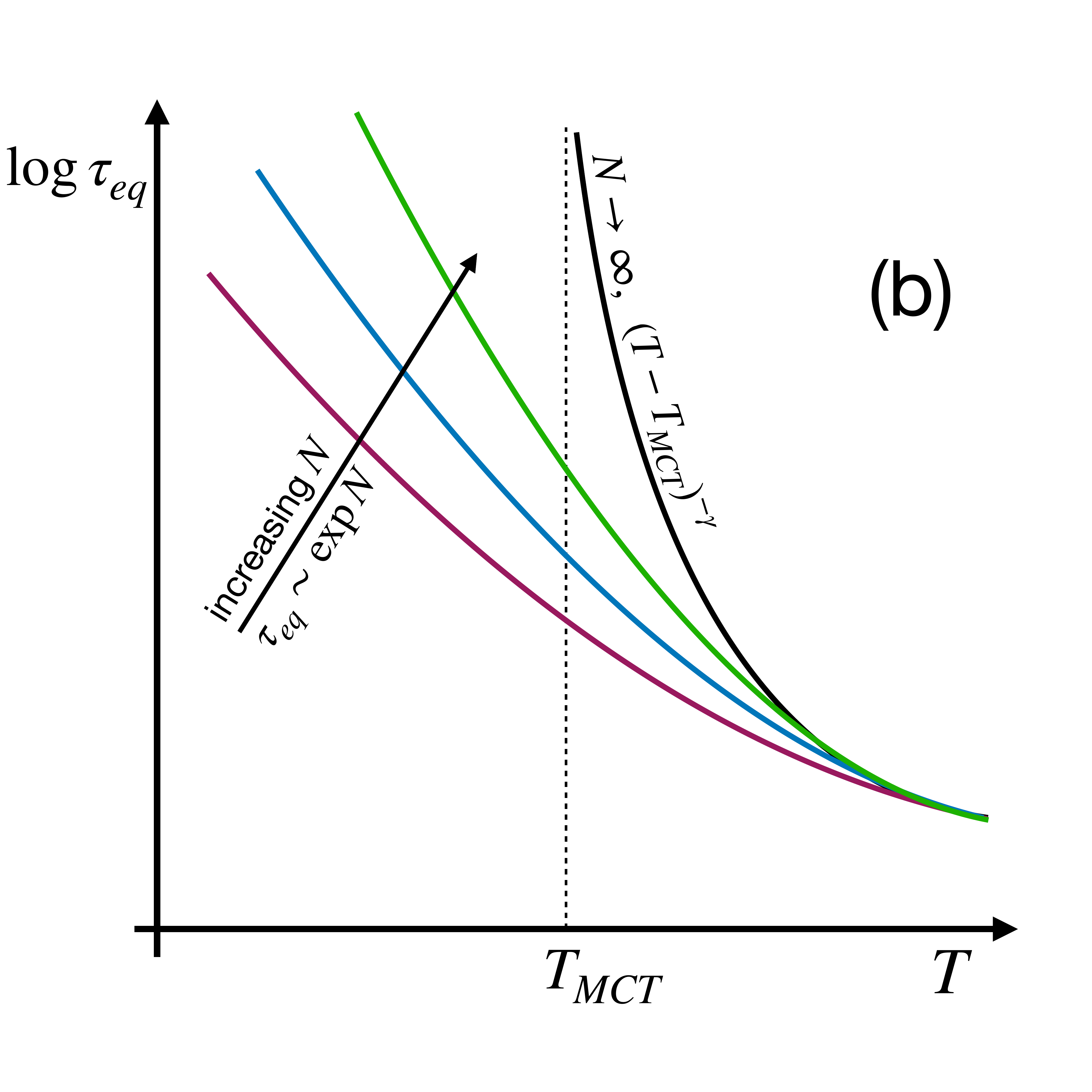}
		\caption{A. Sketch of the gradient descent protocol. B. Sketch of the equilibrium relaxation time as a function of $T$ and its evolution with system size $N$.}
		\label{fig:GDT}
	\end{figure}
	
	\subsection{Gradient descent from an equilibrated configuration}
	
	We will now discuss the 
	simplest case of out-of-equilibrium dynamics, namely 
	gradient descent (GD) dynamics from fixed tempeature $T$. 
	More precisely, the algorithm goes as follows. 
	\begin{enumerate}[label=\alph*)]
		\item Prepare an initial configuration $\Us(0)$ in equilibrium at $T>T_{\K}$, i.e. by sampling it from the Boltzmann-Gibbs distribution 
		\begin{equation}\label{eq:Pin0}
		P_{in}(\Us(0)) = \frac{e^{-\beta H(\Us(0))}}Z \ .
		\end{equation}
		We restrict the initial temperature to $T>T_{\K}$ for simplicity; with this choice, the initial state is in the phase where the system is
		thermodynamically a paramagnet and the Boltzmann-Gibbs distribution can be studied without the need of replica symmetry breaking~\cite{BFP97}, see Sec.~\ref{sec:eq}.
		Based on the discussion of Sec.~\ref{sec:eq}, we know that it is algorithmically hard (i.e. exponentially hard in $N$) 
		to sample from Eq.~\eqref{eq:Pin0} when $T<T_{\MCT}$, but in some models
		one can use the planting trick~\cite{ZK10,folena_gradient_2021}. 
		Even when this is not possible, this prescription to generate $\Us(0)$ is always mathematically well defined and it can be studied by
		analytical techniques 
		in the thermodynamic limit~\cite{FP95,BBM96,BFP97}.
		
		\item Run the gradient dynamics at $T=0$, i.e. the noiseless version of the Langevin Eq.~\eqref{Langevin}, keeping in mind the spherical constraint:
		\begin{equation}\label{GD}
		\frac{\partial\sigma_i}{\partial t}=-\frac{\partial H}{\partial\sigma_i}-\mu\sigma_i \ .
		\end{equation}
		In the $T=0$ limit,
		the Lagrange multiplier can be easily computed by imposing the spherical constraint:
		\beq\label{eq:muT0}
		|\Us|^2 = N 
		\quad\Rightarrow\quad \Us\cdot \frac{\partial \Us}{\partial t}   =- \Us\cdot\grad H-\mu N  =0 
		\quad\Rightarrow\quad \mu =- \frac{1}{N} \Us\cdot\grad H \ .
		\eeq
	\end{enumerate}
	This out-of-equilibrium dynamics, illustrated in Fig.~\ref{fig:GDT}A, does not satisfy TTI and FDT, see Sec.~\ref{sec:TTI}.
	
	The physical motivation for considering this dynamical protocol is the following.  We have seen in Sec.~\ref{sec:eq} that the equilibrium relaxation time of the system grows quickly 
	upon cooling. In the thermodynamic limit, it diverges as a power-law at $T_{\MCT}$. 
	At finite (large enough) $N$, the relaxation time is independent of $N$ for $T>T_{\MCT}$, it grows strongly around $T_{\MCT}$, and for $T<T_{\MCT}$ it is exponential in $N$,
	as illustrated in Fig.~\ref{fig:GDT}B.
	Consider now the simulated annealing protocol illustrated in
	Fig.~\ref{fig:SA}, with a fixed cooling rate $\de T/\de t$. Whenever $T$ is such that $\t_{\rm eq}(T) \de T/\de t \ll T$, the system spends a long time at a given
	temperature before temperature changes, and as a result it can equilibrate easily. When instead $\t_{\rm eq}(T) \de T/\de t \gg T$, the temperature is changing
	so fast compared to the equilibration time that the system is effectively being quenched athermally. 
	Hence, one can approximate the high-$T$ part of a constant cooling schedule by equilibrium dynamics, and the low-$T$ part as a gradient descent dynamics;
	the delicate regime is when $\t_{\rm eq}(T) \de T/\de t \approx T$ and the system is falling out of equilibrium, which happens at some cooling-rate-dependent glass transition
	$T_g$. Because $\t_{\rm eq}(T)$ changes very rapidly with $T$, the crossover regime is a narrow temperature interval around $T_g$. As a result, the idealized description
	where the system is fully equilibrated down to $T_g$ and follows zero-temperature gradient descent dynamics from there, is a good approximation of an actual simulated annealing dynamics,
	but it is much easier to solve analytically. This is why we focus on this protocol in the rest of this section.

	\subsection{Hessian of the final state}
	\label{sec:Hesspspin}
	
	The Hessian of the final state is a fundamental quantity in order to understand the long-time GD dynamics.
	The GD dynamics ends up in a local minimum of the Hamiltonian, with the spherical constraint, hence
	\begin{equation}
	\lim_{t\to\infty}\Us(t)=     \Us_{\infty} \ ,\qquad \text{which is a solution of} \qquad \grad H = -\mu \Us \ .
	\end{equation}
	In the long-time regime we can study the asymptotic relaxation. Defining 
	$\delta\Us(t) = \Us(t)-\Us_{\infty}$, which is small at long times,
	the GD dynamics, from Eq.~\eqref{GD}, can be linearized and gives
	\begin{equation}\label{eq:GDasy}
	\frac{\partial\delta\sigma_i}{\delta t} \sim -\sum_j \frac{\partial^2 H}{\partial \sigma_i\partial\sigma_j}\delta\sigma_j-\mu\delta\sigma_i = -\sum_j M_{ij}(\Us_{\infty})\delta\sigma_j \ ,
	\qquad\qquad
	M_{ij}(\Us_{\infty}) = \left[ \frac{\partial^2 H}{\partial \sigma_i\partial\sigma_j} + \mu \delta_{ij} \right]_{\Us=\Us_\infty} \ ,
	\end{equation}
	where $M_{ij}(\Us_{\infty})$ is the asymptotic Hessian of the GD dynamics. 
	Decomposing it in normal modes, the asymptotic GD dynamics becomes
	\begin{equation}
	M_{ij}=\sum_{\alpha} \l_\a |\U\sigma_\a\rangle \langle \U\sigma_\a |
	\qquad\Rightarrow\qquad
	|\delta\U\sigma(t)\rangle = \sum_{\alpha} e^{-\lambda_{\alpha}t}|\U\sigma_\a\rangle \langle \U\sigma_\a |\delta \U\sigma(0)\rangle \ ,
	\end{equation}
	such that the relaxation is dominated by the lowest eigenvalue $\lambda_{\alpha}$ of the Hessian.
	Note, however, that this is only an asymptotic result that holds at times possibly diverging with $N$, so
	the gradient descent dynamics is more complex than a simple relaxation along eigenmodes.

\begin{wrapfigure}{r}{.4\textwidth}
		\centering
		\includegraphics[width=0.4\columnwidth]{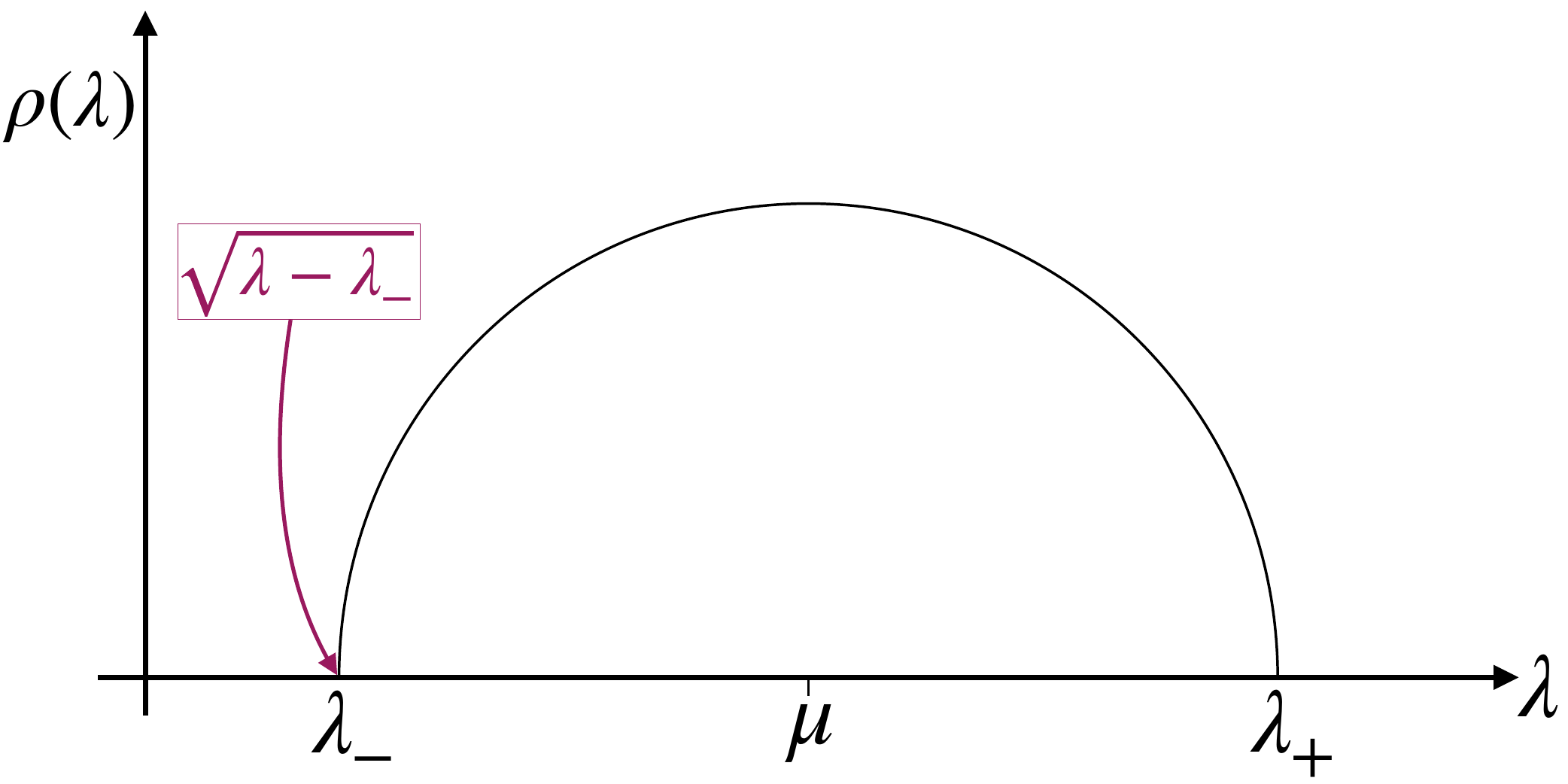}
		\caption{
			Hessian spectrum in the $p$-spin spherical model.
			$\lambda_{-}>0$ stable minimum (gapped spectrum);
			$\lambda_{-}<0$ unstable saddle;
			$\lambda_{-}=0$ marginal minimum.}\label{fig:semicircle}
		\vskip-20pt
	\end{wrapfigure}

	From Eq.~\eqref{eq:GDasy},
	it can be shown that in the $p$-spin model the correlation between the Gaussian couplings and the configuration $\U\s_\io$ can be neglected in the thermodynamic limit. The Hessian of typical stationary points in the energy 
	landscape is thus a shifted GOE matrix (see e.g.~\cite{ros2019complex}), with $\overline{M_{ij}}=\m \d_{ij}$ and $\text{Var}[M_{ij}]= \frac{f''(1)}{N}$. Its eigenvalue spectrum thus has a 
	semicircular shape (Fig.~\ref{fig:semicircle}):
	\begin{equation}
	\rho(\lambda) = \frac{\sqrt{(\lambda-\mu)^2 - 4f''(1)}}{\pi\sqrt{f''(1)}} \ ,
	\end{equation}
	and the two edges of the spectrum are $\lambda_{\pm}=\mu\pm2\sqrt{f''(1)}$.
	In order to evaluate the Hessian matrix, we thus
	need to compute the Lagrange multiplier $\mu$ from Eq.~\eqref{eq:muT0}.
	In the pure model ($H=H_{p}$), it is proportional to the energy:
	\begin{equation}\label{prop}
	\mu_{p} = -\frac{1}{N}\sum_i \sigma_i \sum_{i_2<...<i_p}J_{ii_2\dots i_p}\sigma_{i_2}\cdots\sigma_{i_p}=-\frac{p H_p}{N} \ .
	\end{equation}
	In the mixed model ($H=\sum_p \sqrt{a_p} H_{p}$) the above proportionality does not hold:
	\begin{equation}
	\mu = -\frac{1}{N}\sum_p \sqrt{a_p} p H_{p} \not\propto H \ .
	\end{equation}
	Because (in the thermodynamic limit) the typical spectrum is thus only determined by its Lagrange multiplier $\mu$,
	this difference has important implications in terms of the energy landscape: in the pure $p$-spin model, the spectrum of 
	a typical stationary point is fully determined by its energy, while this is not the case in the mixed $p$-spin. We now discuss 
	the implications of this difference in more details.
	As illustrated in Fig.~\ref{fig:semicircle}, when $\l_->0$ the stationary point is a stable minimum with a gapped spectrum of strictly positive
	eigenvalues; when $\l_-<0$, it is an unstable saddle point with negative eigenvalues; while stationary points with $\l_-0$ are local minima but have
	arbitrarily small eigenvalues, hence they are deemed {\it marginally stable}.

	\subsection{Pure $p$-spin}
	
	As we have already noticed, from Eq.~\eqref{prop}, in pure $p$-spin models, with $f(q)=q^p/2$, there is a direct proportionality between the 
	spectrum-shift $\mu$ and the energy of the minimum $e=H/N$. Recalling that $f''(1)=p(p-1)/2$,
	the marginality condition $\lambda_{-}=\mu-2\sqrt{f''(1)}=0$ fixes a value for $\mu$ and therefore a value for the energy $e$:
	\begin{equation}
	\lambda_{-}=0 \quad \Longrightarrow \quad  \mu_{th} = 2\sqrt{p(p-1)/2} = -p e_{th} \quad \Longrightarrow \quad e_{th}=\sqrt{2(p-1)/p} \ .
	\end{equation}
	The energy corresponding to marginally stable minima is also called the {\it threshold energy}~\cite{CS95,kurchan1996phase,CGP98}. 
	Typical stationary points of the energy landscape are stable minima for $e<e_{th}$, and unstable saddles for $e>e_{th}$.
	
	\begin{figure}[h]
		\centering
		\includegraphics[width=0.51\columnwidth]{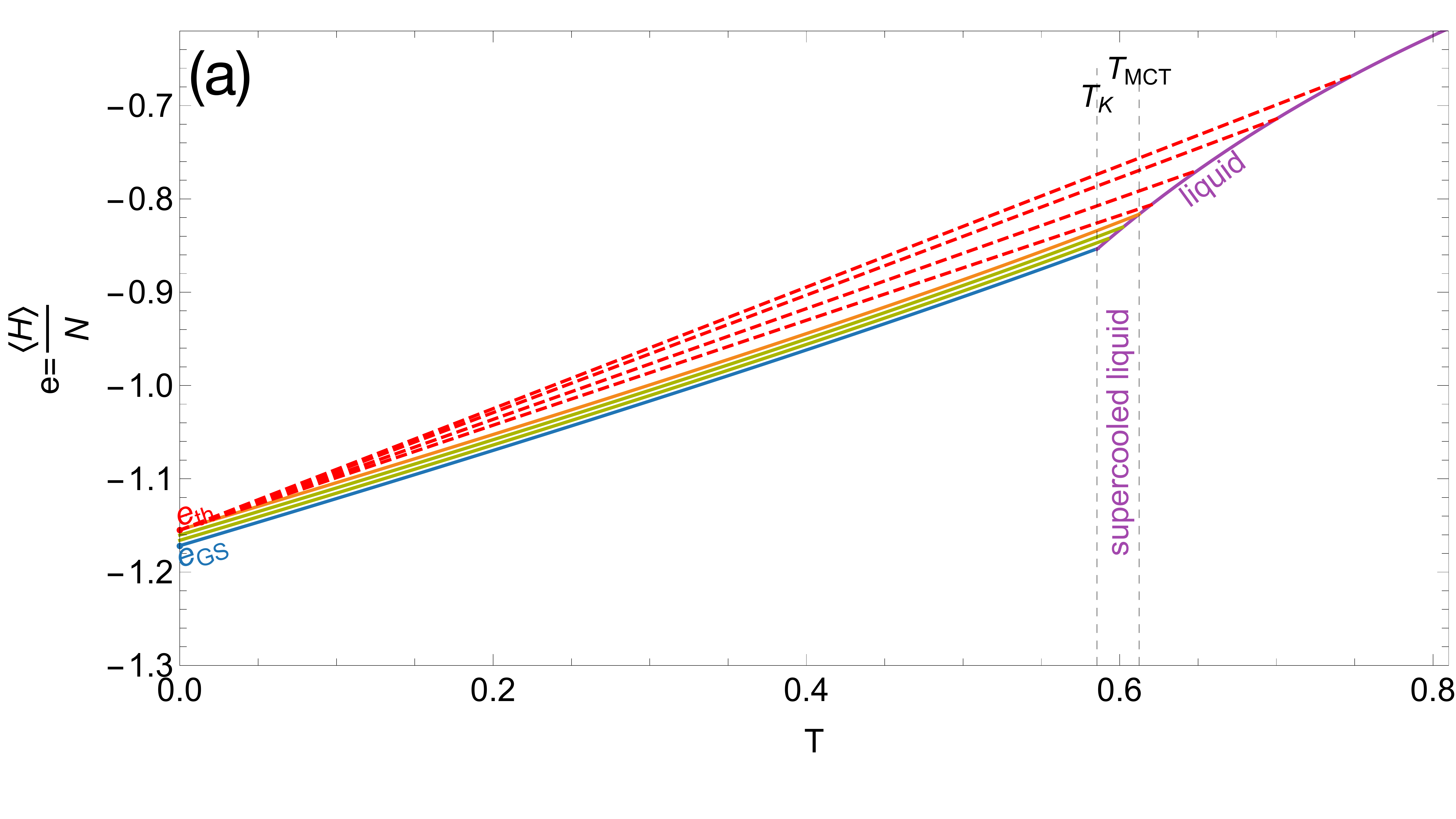}
		\includegraphics[width=0.48\columnwidth]{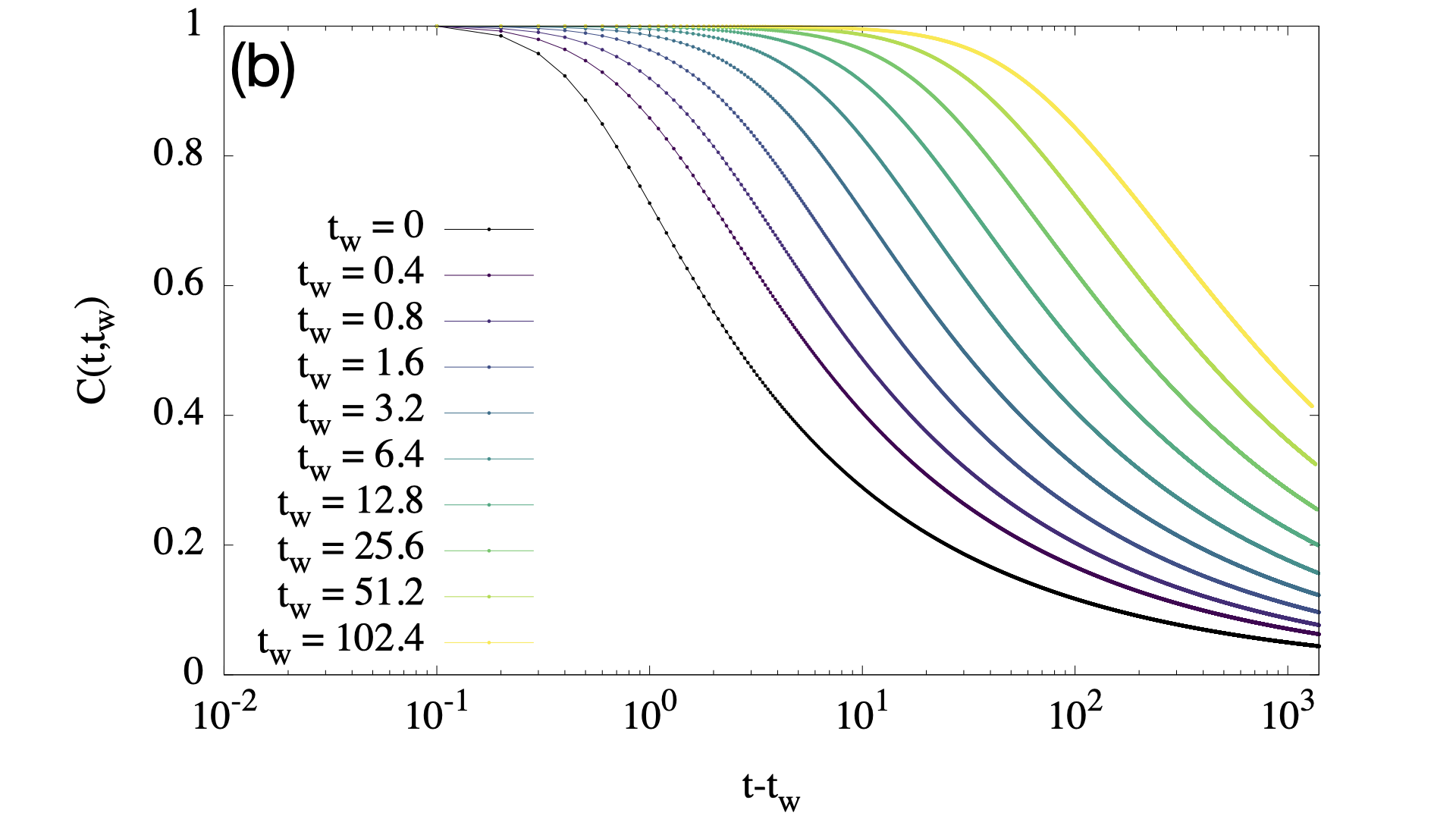}
		\caption{3-spin model. A. Representation of the gradient descent dynamics in the energy-temperature plane. 
			An initial state at temperature $T>T_{\MCT}$ is brought to the threshold level (red dashed lines, only the initial and final point are physical). 
			An initial state at $T<T_{\MCT}$ is brought below the threshold (full colored lines, that represent the evolution of the metastable state from the initial to
			the final temperature).
			B. Aging of the correlation function $C(t+t_w,t_w)$ for gradient descent dynamics from infinite temperature, as a function of $t$ for several $t_w$.}
		\label{fig:3GD}\label{fig:aging}
	\end{figure}
	
	The GD dynamics can be characterized in terms of the time-dependent energy, $e(t)=\langle H(t)\rangle/N$, and the correlation function $C(t,t')$
	given by Eq.~\eqref{eq:Cttp}.
	Two distinct phases, depending on the initial equilibrium temperature $T>T_{\K}$, exist.
	\begin{itemize}
		\item $T<T_{\MCT}$ - If the initial temperature is below $T_{\MCT}$, then one observes~\cite{BBM96}: 
		\begin{enumerate}[label=\alph*)]
			\item
			an exponential relaxation to a nearby local minimum (also called the ``inherent structure'' in the structural glass literature~\cite{sciortino_potential_2005}),
			${e(t)-e_{\IS}(T) \propto e^{-t/\tau}}$, with a temperature-dependent final energy $e_{\IS}(T)$;
			\item
			and a persistent memory of the initial condition, $C(t,0)\to q_{r}(T)$ for $t\to\infty$, i.e. the configuration at time $t$ remains in a cone of width $q_r(T)$ around
			the initial state.
		\end{enumerate}
		In this phase, the dynamics can be mapped onto a {\it restricted thermodynamics} calculation~\cite{FP95}.
		A thermodynamic calculation of $e_{\IS}(T)$ and $q_{r}(T)$ can be achieved by constraining a ``slave'' configuration to
		have fixed overlap with an equilibrium ``master'' configuration, and then finding the overlap that minimizes the free energy of the slave
		configuration, which gives $q_{r}(T)$. This is called a {\it Franz-Parisi}~\cite{FP95,BBM96} or {\it state-following} construction~\cite{ZK10,sun2012}.
		For $T<T_{\K}$ the same mapping is possible but it requires a more complicated replica symmetry breaking scheme~\cite{BFP97}.
		\item $T>T_{\MCT}$: if the initial temperature is above $T_{\MCT}$, then one observes~\cite{CK93,folena2020}:
		\begin{enumerate}[label=\alph*)]
			\item a power-law relaxation to the threshold energy, $e(t)-e_{th}\propto t^{-2/3}$, independently of $T$;
			\item  consistently with the temperature-independence of the final state, memory of the initial condition is lost, i.e. $\lim_{t\to\infty}C(t,0)=0$.  
			Furthermore, $\lim_{t\to\infty}C(t+t_w,t_w)=0$ for any $t_w$, i.e. memory of any finite time configuration is also lost, as illustrated in Fig.~\ref{fig:aging}B. This condition is called
			{\it weak ergodicity breaking}~\cite{CK93}.
		\end{enumerate}
	\end{itemize}

	\begin{figure}[b]
		\centering
		\includegraphics[width=0.4\columnwidth]{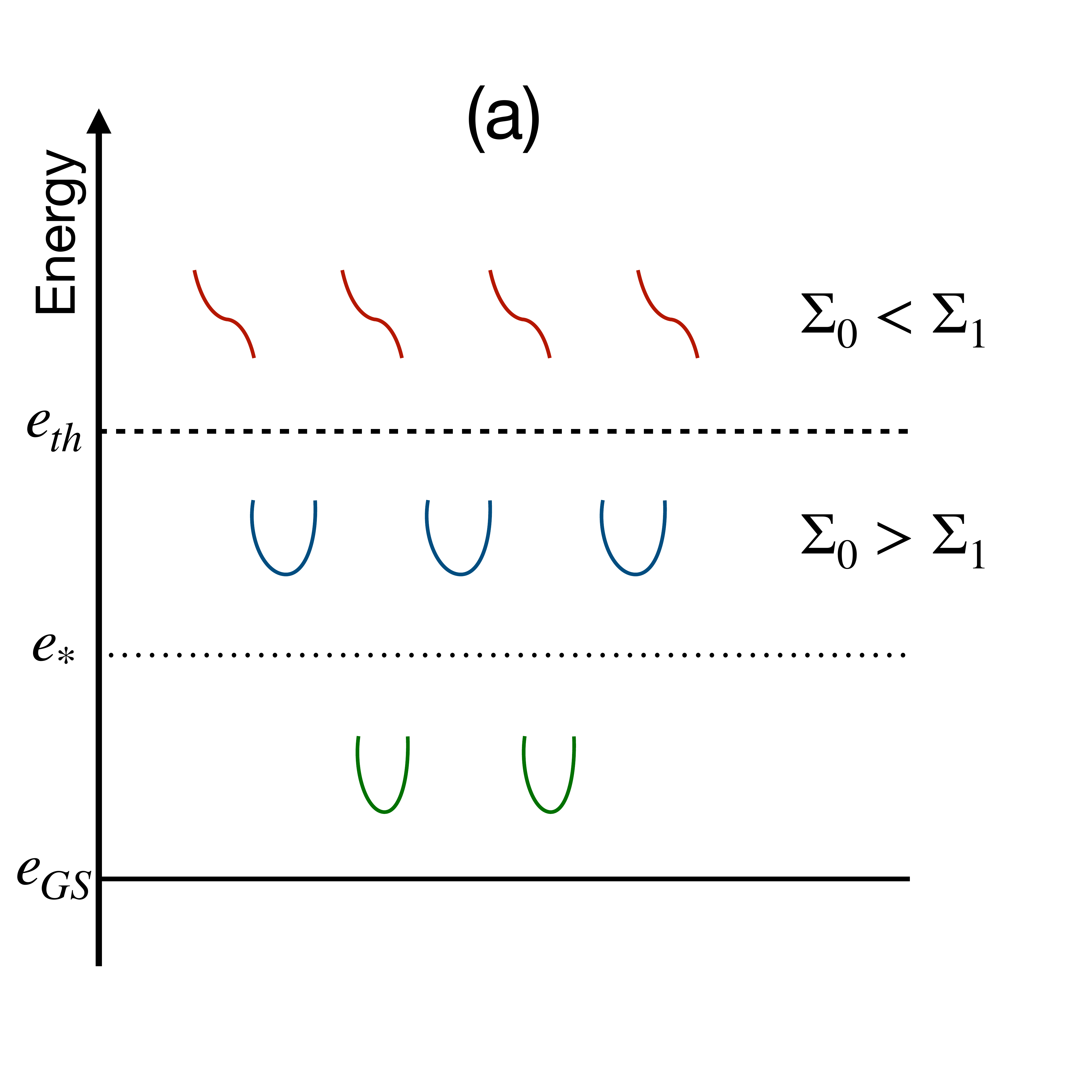}
		\hfill
		\includegraphics[width=0.4\columnwidth]{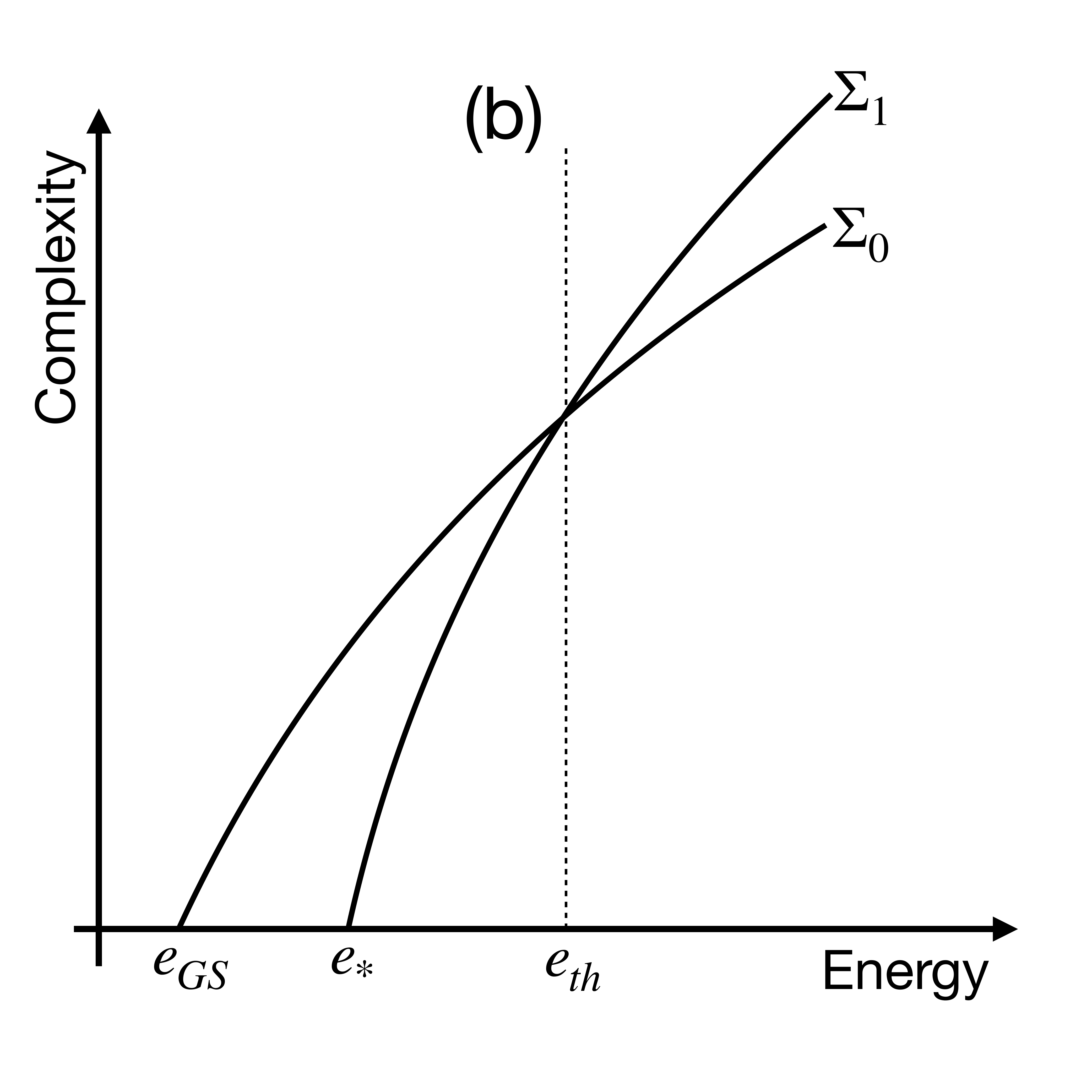}
		\caption{
			A. Schematic illustration of the energy landscape in a pure $p$-spin model.
			B. Complexity of minima $\Sigma_0$ and of saddles of index one $\Sigma_1$ (one negative eigenvalue of the Hessian).}
		\label{fig:geometry}
	\end{figure}
	
	In the pure $p$-spin model it is thus possible to explain the asymptotic (long-time) dynamics in terms of simple geometric properties of the energy landscape~\cite{CS95,kurchan1996phase,CGP98}. 
	At high energies $e>e_{th}$, typical 
	stationary points
	are unstable (they present negative eigenvalues of the Hessian). At the threshold energy $e_{th}$, they become marginally stable minima, and below $e_{th}$ they become stable.
	The threshold manifold is an attractor for the asymptotic dynamics starting from high temperatures ($T>T_{\MCT}$), independently of the used protocol, while for low
	temperatures ($T<T_{\MCT}$) one starts in the basin of attraction of a stable minimum, which is then reached quickly by the gradient descent. This geometric transition
	in the energy landscape is illustrated in Fig.~\ref{fig:geometry}A. A more detailed calculation gives the complexity associated to saddles of order one (or higher), that
	is found to cross that of stable minima precisely at $e_{th}$, see Fig.~\ref{fig:geometry}B. Hence, saddle points also exist below $e_{th}$, but they are exponentially rarer than
	stable minima, while the reverse is true above $e_{th}$.
	
	A simple argument\footnote{Private communication from F.Ricci-Tersenghi.} explains the power-law convergence $\propto t^{-2/3}$ towards the threshold energy,
	and goes as follows. The system, while aging towards the threshold, moves close to saddle points with rarefying number of negative eigenvalues of the Hessian.
	Let us assume that  during this process the dynamics induces random
	\begin{wrapfigure}{r}{.3\textwidth}
		\centering
		\includegraphics[width=0.3\columnwidth]{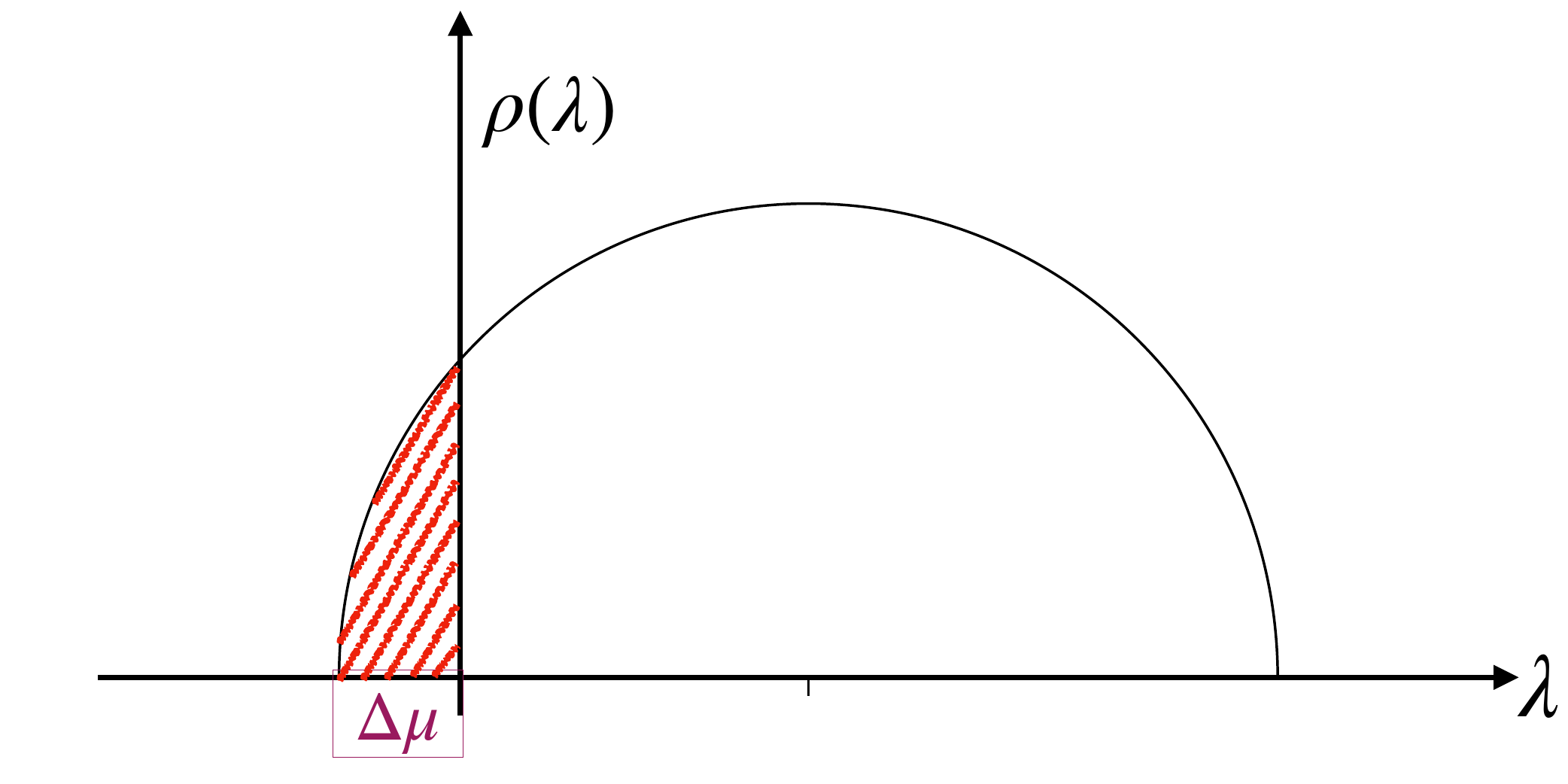}
		\caption{Hessian spectrum during the gradient descent}\label{eq:23arg}
	\end{wrapfigure}
	rotations of the gradient with respect to the local Hessian eigenvectors. 
	The time to escape from the neighborhood of a saddle is thus proportional to the probability that a random rotation brings the gradient in a negative direction. 
	Given that the shift of the spectrum is proportional to the energy (Fig.~\ref{eq:23arg}), 
	\beq
	\l_- = \Delta\mu = \mu - \mu_{th} \propto -(e-e_{th}) = -\Delta e \ ,
	\eeq 
	and
	that $\rho(\lambda)$ vanishes as a square root at the edge (Fig.~\ref{fig:semicircle}),
	the probability of extracting a negative eigenvalue for small $\Delta\mu$ is:
	\begin{equation}
	\rho(\lambda<0) \propto \int_{\Delta\mu}^{0} \de \lambda \sqrt{\lambda-\Delta\mu} \propto |\Delta\mu|^{3/2} \propto \Delta e^{3/2} \ .
	\end{equation}
	Therefore the time to escape is
	$t\propto 1/\rho(\lambda<0) \propto \Delta e^{-3/2}$,
	which justifies the power-law scaling $\Delta e \propto t^{-2/3}$.
	
	\subsection{Mixed $p$-spin}
	
\begin{figure}[h]
		\centering
		\includegraphics[width=0.52\columnwidth]{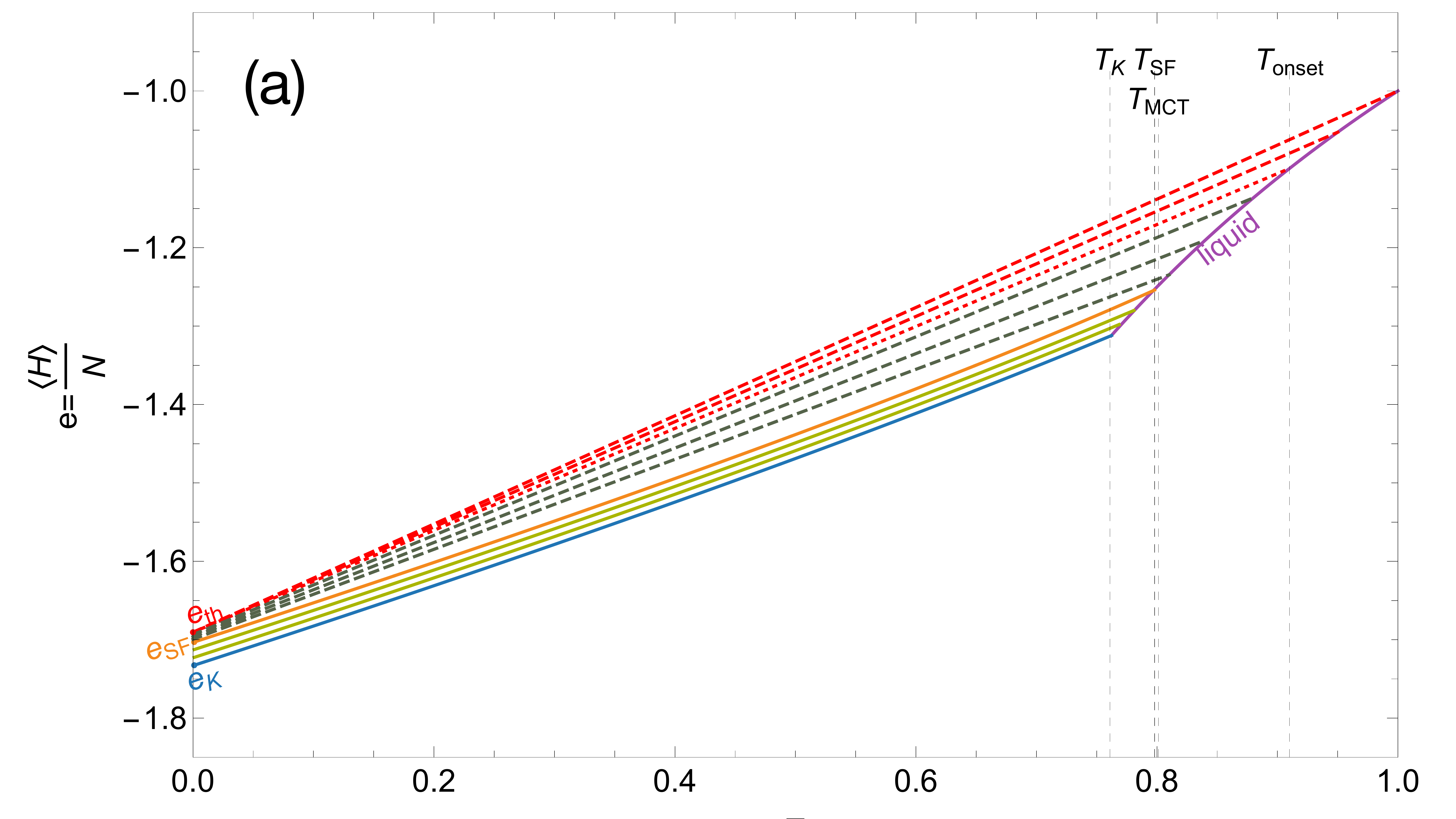}
		\hfill
		\includegraphics[width=0.45\columnwidth]{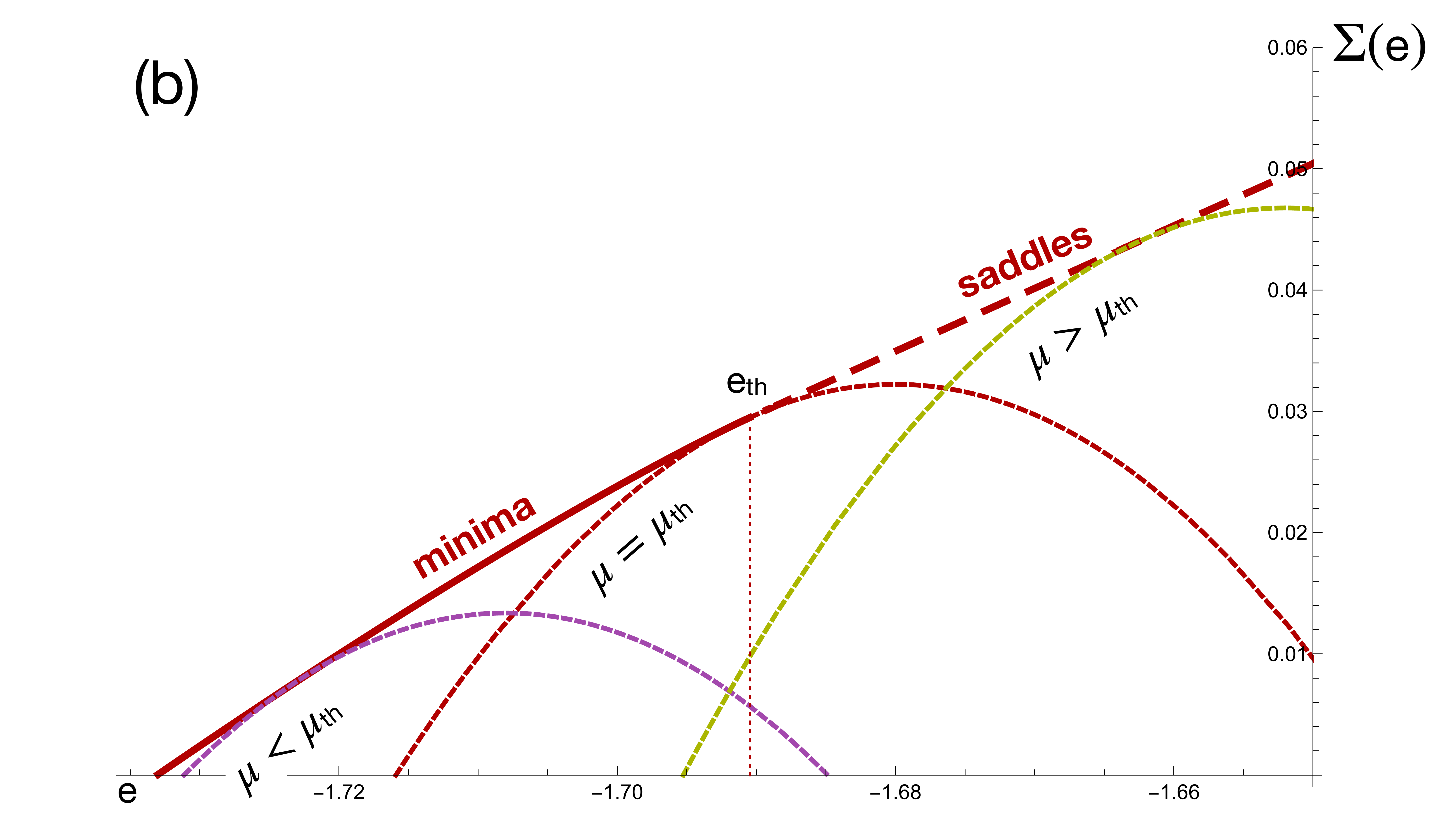}
		\caption{(3+4)-spin model.
			A. Representation of the gradient descent protocol in the energy-temperature plane, as in Fig.~\ref{fig:aging}A.
			B. Complexity as a function of the energy for several values of $\mu$. Dashed lines represent the complexity $\Sigma(e,\mu)$ for fixed $\mu$.
			The envelope is $\Sigma(e)=\max_\mu \Sigma(e,\mu)$, which is represented as a full line when $\mu^*(e) = \text{argmax}_\mu \Sigma(e,\mu) < \mu_{th}$
			(stable minima dominate) and as a dashed line when $\mu^*(e) > \mu_{th}$ (unstable saddles dominate).
		}
		\label{fig:34Comp}\label{fig:34GD}
	\end{figure}

	We next consider a mixed $p$-spin model that presents a RFOT~\cite{sun2012}, e.g. the (3+4)-spin model, with $f(q)=\frac{1}{2}(q^3+q^4)$. 
	The energy and the Lagrange multiplier are:
	\begin{equation}
	e=e_3+e_4 \ ,\qquad
	\mu=-3 e_3-4 e_4 \ ,
	\end{equation}
	where $e_p = \langle H_p \rangle/N$ with $p=3,4$.
	In this case, an exponential number of marginally stable states is found in a finite range of energies. To see this,
	one can compute (an approximation of) the complexity $\Sigma(e,\mu)$, which gives the number of stationary points with energy $e$ and Lagrange parameter $\mu$,
	which controls the shape of the spectrum~\cite{folena2020,folena_gradient_2021}.
	For fixed $\mu$, $\Sigma(e,\mu)$ is a parabola and there exists a family of possible energies $e$, as illustrated in Fig.~\ref{fig:34Comp}B.
	This is true in particular for marginal states with $\mu=\mu_{th}=2\sqrt{f''(1)}$, which are therefore present in a finite range of energies.
	At fixed energy $e$, the dominant states are those with $\mu^*(e) = \text{argmax}_\mu \Sigma(e,\mu)$ (Fig.~\ref{fig:34Comp}B).
	When $\mu^*(e) < \mu_{th}$, the most numerous states (exponentially in $N$) are stable minima, while
	when $\mu^*(e) < \mu_{th}$ the dominant states are unstable saddles. The value $e_{th}$ such that $\mu^*(e_{th}) = \mu_{th}$ thus corresponds to the geometrical
	transition that separates the minima-dominated and saddle-dominated regions of the landscape.
	
	Because of this different structure of the energy landscape,
	in the mixed $p$-spin with a RFOT transition a new phase emerges, which displays both memory and aging~\cite{folena2020}. One thus finds three distinct temperature regimes.
	\begin{itemize}
		\item $T<T_{\SF}$: Memorious exponential dynamics with $C(t,0)\to q_{r}(T)$ for $t\to\infty$ and
		exponential relaxation, $e(t)-e_{\IS}(T)\propto e^{-t/\tau}$ to  the inherent structure. Also in this case, 
		$e_{\IS}(T)$ and $q_{r}(T)$ can be computed via the state following or Franz-Parisi approach. Because relaxation is exponential, transient (aging) effects
		become quickly unobservable.
		\item $T_{\SF}<T<T_{\on}$: Memorious power-law dynamics with $\lim_{t\to\infty}C(t,0)>0$ and
		$e(t)-e_{\IS}(T)\propto t^{-2/3}$ relaxation to a below-threshold energy $e_{\IS}(T) < e_{th}$, which depends on the initial temperature, consistently with the observation
		that memory is preserved. Final states are marginal, i.e. $\mu(t) \to \mu_{th}$, and persistent aging is observable in $C(t+t_w,t_w)$.
		\item $T>T_{\on}$: Memoryless power-law relaxation to the threshold energy,
		with $\lim_{t\to\infty}C(t,0)=0$, $e(t)-e_{th}\propto t^{-2/3}$, and persistent aging with weak ergodicity breaking.
	\end{itemize}
	The high-temperature and low-temperature phases are identical to those of the pure $p$-spin, but in this case there is an intermediate phase.
	Furthermore, in the pure $p$-spin one has $T_{\on} = T_{\SF} = T_{\MCT}$, hence the equilibrium and out-of-equilibrium dynamics both have a phase transition
	at the same temperature, while in the mixed $p$-spin the equilibrium transition at $T_{\MCT}$ is unrelated to the out-of-equilibrium ones at $T_{\SF}$ and $T_{\on}$.

	\begin{figure}[h]
		\centering
		\includegraphics[width=0.69\columnwidth]{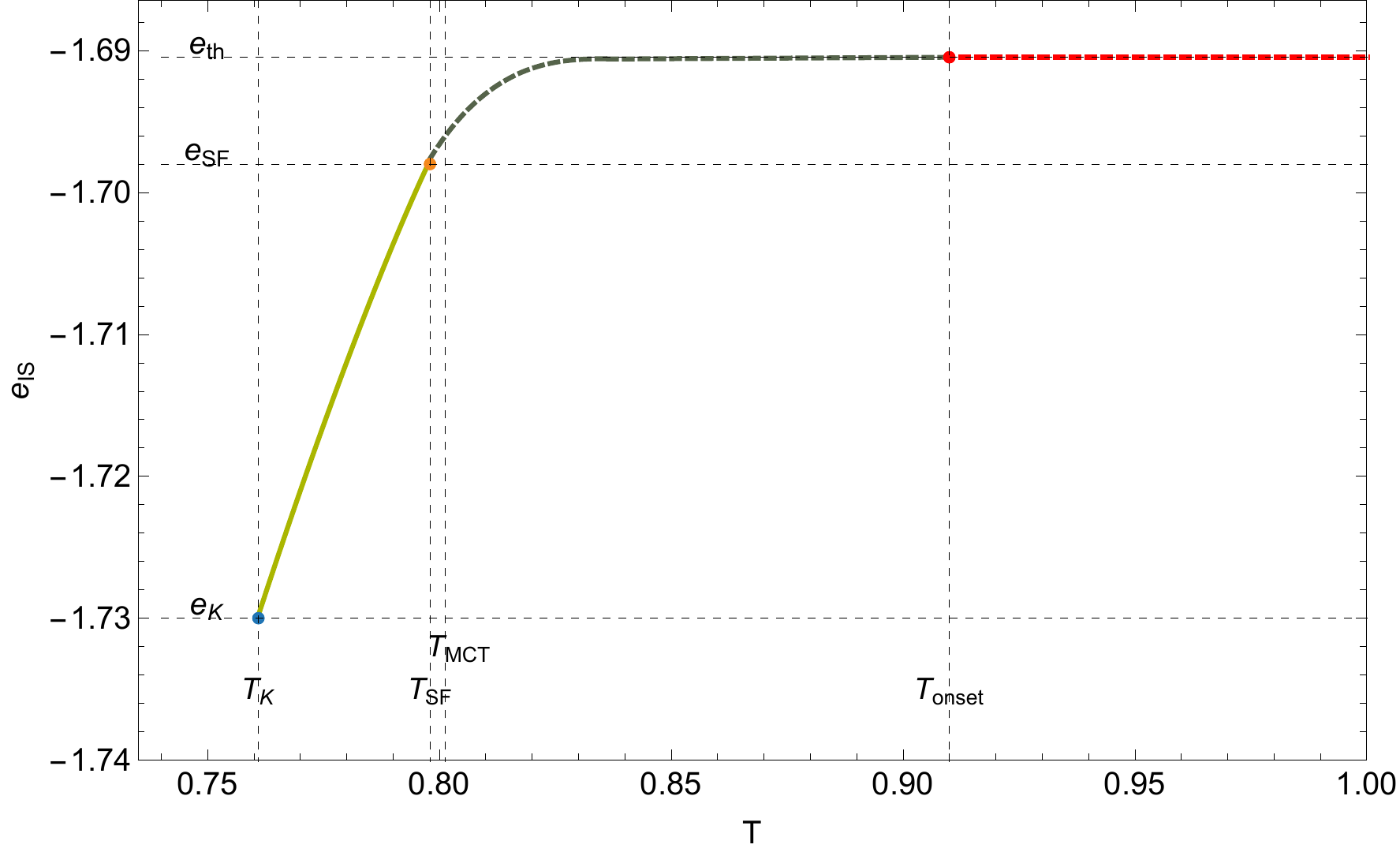}
		\caption{Asymptotic energy vs initial temperature in the (3+4)-spin model.}\label{fig:34GD}
	\end{figure}
	
	The RFOT class of mixed $p$-spin models presents a behavior strongly reminiscent to that of structural glasses~\cite{sastry_signatures_1998,sciortino_potential_2005,ozawa2012jamming,charbonneau2021memory}, as illustrated in Fig.~\ref{fig:34GD}.
	For initial temperatures above $T_{\on}$, gradient descent converges to the $T$-independent energy $e_{th}$ at which marginal states dominate the landscape.
	For $T_{\on}>T>T_{\SF}$, $T$-dependent atypical marginal states, close to the initial condition, are found instead. Finally, for $T<T_{\SF}$, a stable 
	inherent structure close to the initial state is found.
	The geometric interpretation of this relaxation is not fully understood. In particular, the problem of computing the final energy
	$e_{\IS}(T)$ in the intermediate phase without having to solve explicitly the dynamical equations remains open.

	\subsection{Summary}
	
	The study of the out-of-equilibrium dynamics of the spherical $p$-spin model, both pure~\cite{CK93} and mixed~\cite{folena2020}, is extremely useful to get a first insight on the behavior 
	of models with complex (or rough) energy landscapes. In particular:
	\begin{itemize}
		\item A generic scenario for the geometric, thermodynamic and dynamics properties of glassy systems at the mean-field level is obtained.
		\item A simple analytic solution of these models can be obtained using the replica method and dynamical mean field theory (DMFT),
		which gives dynamical equations with fully explicit kernels.
		\item In the high-temperature regime ($T>T_{\on}$), equilibration is easy, and gradient descent converges slowly ($t^{-2/3}$) to the geometric threshold.
		\item In the low-temperature regime ($T<T_{\SF}$), equilibration is hard ($\exp(N)$), but once it is achieved, 
		gradient descent converges fast ($e^{-t/\tau}$) to a stable close minimum.
		\item An intermediate-temperature regime ($T_{\SF}<T<T_{\on}$) exists, where equilibration can be easy ($T>T_{\MCT}$) or hard ($T<T_{\MCT}$), but in any case gradient descent converges slowly ($t^{-2/3}$) to a marginally stable minimum close to the initial condition.
		\item Because of this rich structure, it is exponentially hard in $N$ to find the ground state and low enough energy states by simulated annealing.
	\end{itemize}
	This kind of study can be generalized to other models, different search algorithms, and other cooling protocols, leading to an 
	even richer phenomenology.
	The impossibility of equilibration below $T_{\MCT}$
	is a general property of mean-field systems~\cite{MS06}.

	\section{Constraint satisfaction problem and the jamming transition}
	\label{sec:CSP}
	
	As we discussed in Secs.~\ref{sec:IB} and \ref{sec:ID},
	the \textit{jamming} transition is a characteristic phase transition of particle systems with finite-range interactions, such as granulars and emulsions~\cite{OLLN02,OSLN03,LN10,LNSW10}.
	It occurs when the particle density
	increases so much that a rigid network of contact interactions is formed, and it
	is closely related to the SAT/UNSAT transition
	of constraint satisfaction problems~\cite{AMSZ09}.
	
	The simplest toy model of the jamming transition is the \textit{perceptron}~\cite{FP16}. Its degrees of freedom, like for the $p$-spin model, 
	are a set of $N$ continuous spins $\U{\s}$ with a spherical constraint $\vert \U\s \vert=\sqrt N$. 
	This vectorial spin degree of freedom interacts with $M=\a N$ obstacles $\U{\x}_m$ that act as a quenched disorder. 
	The constraint satisfaction problem is defined by the 
	following Hamiltonian or cost function
	\beq
	H(\U\s) = \sum^M_{m=1} \frac12 h^2_\m \Th(-h_m) \ , \qquad\qquad h_m = 
	\frac{\U\x_m \cdot \U\s}{\sqrt N} - \k \ ,
	\eeq
	being $h_m\geq 0$ the constraint imposed by the $m$-th obstacle, and $\k$ a real parameter representing the diameter of the 
	obstacles. Indeed, one can imagine that an obstacle is located in position $\U{\X}_m=-\sqrt{N} \U\x_m/|\U\x_m|$ on the sphere, and the constraint
	is then equivalent to $\hat \X_m \cdot \hat \sigma < - \k/ |\U\x_m|$, which imposes (for negative $\k$) that the vector $\U\s$ 
	should stay outside of a cone around the obstacle $\U{\X}_m$~\cite{FP16}.
	
	The perceptron with $\k\geq 0$ is instead one of the simplest \textit{classifiers}, i.e. a machine that allows one to separate points in a phase space 
	by assigning them a binary label~\cite{Ga87,GD88}. Imagine that one is given a set of $M$ images of cats and dogs represented by a sequence of $N$ 
	bits encoded in the vectors $\U{x}_m$, $m=1,\ldots,M$. Knowing the labels $y_m\in\{\pm1\}$ corresponding to a cat/dog in 
	the $m$-th image, the goal of supervised learning is to find a vector $\U\s$ such that $y_m = \mathrm{sgn} \left( \U\s \cdot 
	\U{x}_m \right)$, i.e. to correctly classify the training labeled images $\{\U{x}_m,y_m\}$. Calling $\U\x_m = y_m \U{x}_m$, the 
	constraint becomes $\U\s \cdot \U\x_m > 0$ for all $m=1,\ldots,M$. One can show that a stricter constraint, i.e.
	\beq
	\U\s \cdot \U\x_m > \k \sqrt N \ , \qquad \forall \ m = 1,\ldots,M \ , \qquad \k>0 \ ,
	\eeq
	allows one to preserve the correct classification if the images are slightly corrupter by noise~\cite{Ga87,GD88}.
	
	In the simplest setting, the {\it random perceptron}, the input vectors components $x_{i,m}$ are drawn from a white Gaussian distribution with zero mean and unit variance, 
	and a random label $y_m\in\{\pm 1\}$ with probability $1/2$ is assigned to each of them. 
	The obstacles (or patterns) $\U\x_m$ then have the same statistics of the $\U{x}_m$, i.e. they are Gaussian vectors with i.i.d. components of zero mean
	and unit variance.
	In this case, there is nothing to be learned because the labels are random, but one can ask - as a benchmark question - if the machine can learn random noise.
	This problem is related to the bias/variance tradeoff and gives a bound on the generalization error~\cite{abbara2020rademacher}.
	
	We conclude that the random perceptron, whose
	behavior is fully determined by the parameters $\k$ and $\a$, 
	corresponds to a simple random noise classification problem for $\k \geq 0$~\cite{Ga87,GD88},
	and to a particle that has to avoid random obstacles for $\k<0$~\cite{FP16}. 
	In both cases, it is an instance of a continuous-variable constraint satisfaction problem (CCSP)~\cite{FPSUZ17},
	like the sphere packing problem, which consists of finding a configuration of $N$ particle 
	positions $\U{x}$ such that no two particles are overlapping, namely $h_{ij} = \vert x_i - x_j \vert - \ell > 0 \: , \: \forall \: i,j$.
	\begin{wrapfigure}{r}{.3\textwidth}
		\centering
		\includegraphics[width=0.3\textwidth]{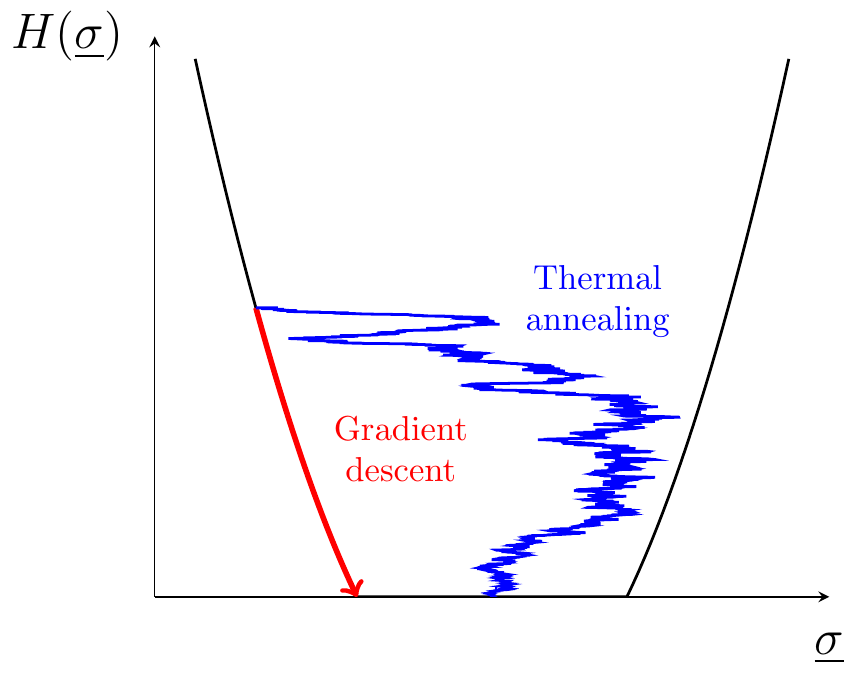}
		\caption{An energy lake in the SAT phase of a CCSP. GD dynamics converges to the shore,
			while thermal annealing can explore the interior of the lake.}
		\label{fig:lake}	\label{fig:thermal_vs_GD}
	\end{wrapfigure}
	In Sec.~\ref{sec:noneq} we analyzed the connection between the out-of-equilibrium dynamics and the energy
	landscape of the $p$-spin model. An important additional ingredient brought in by the perceptron
	(and more generally by CCSPs)
	is that its energy landscape can have ``lakes'', i.e. regions where the energy is identically vanishing. 
	These regions where $H(\U\s)=0$ represent the set of solutions 
	of the CCSP, as shown in Fig.~\ref{fig:lake}.

\begin{figure}[b]
		\includegraphics[width=.45\textwidth]{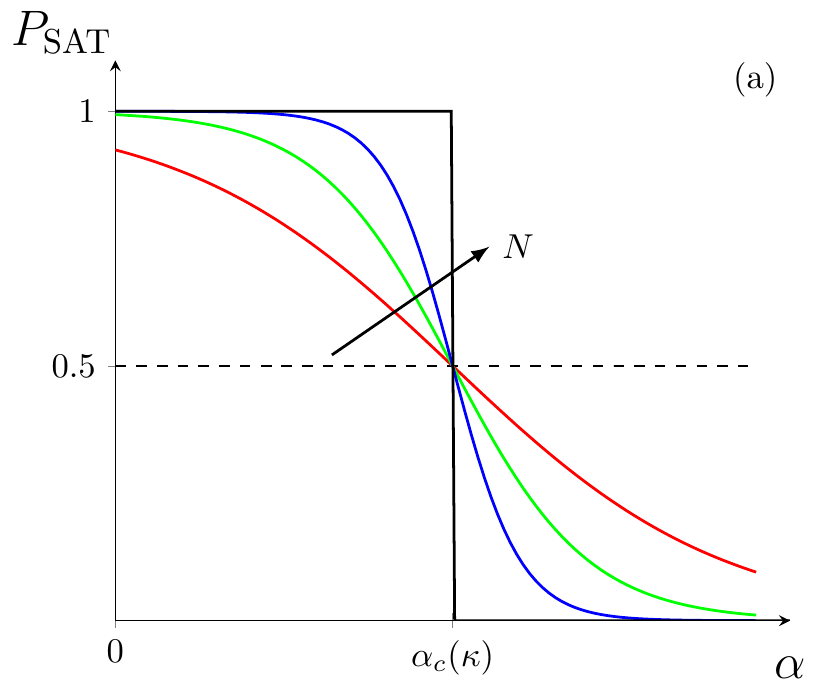}
		\includegraphics[width=.45\textwidth]{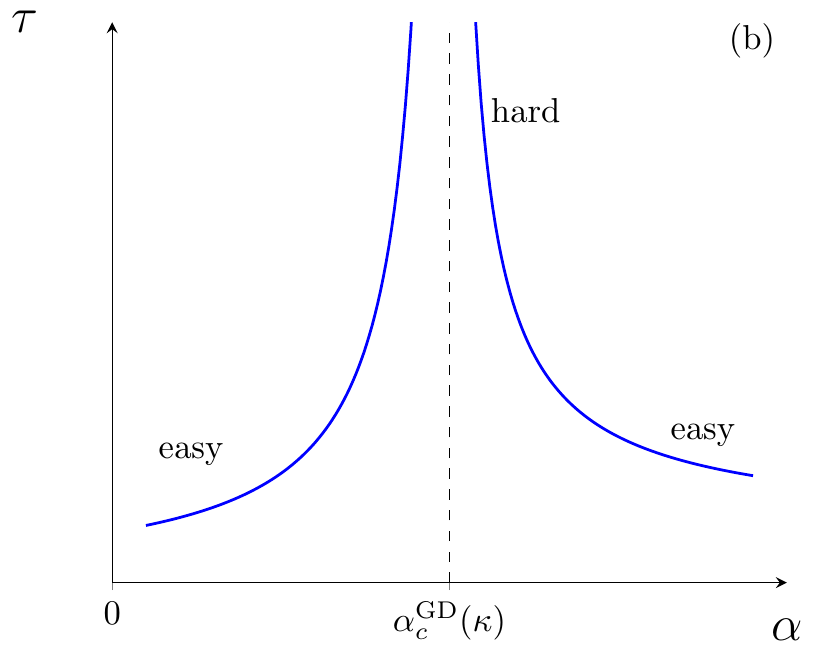}
		\caption{A. Probability of finding a SAT configuration, $\min_{\U\s} H(\U\s)=0$, as a function 
			of the constraint fraction $\a$.
			B.~Characteristic time needed for gradient descent to decide is the system is SAT or UNSAT,
			and its divergence around the jamming transition.
		}        \label{fig:time_SAT}\label{fig:P_SAT}
	\end{figure}

	The jamming transition can be then related to a SAT/UNSAT transition: 
	increasing the fraction of obstacles $\a$ at fixed $\k$, the lakes of the energy 
	landscape shrink until, at the jamming point, their volume 
	vanishes and we are only left with UNSAT minima with $H(\U\s)>0$. At every 
	value of $\a,\k$, one can determine the {\it equilibrium} probability of having a SAT instance,
	$P_{\rm SAT}^{\rm eq}$, as the probability, over the choice of the random obstacles $\U{\x}_m$, that the absolute minimum of 
	$H(\U\s)$ (i.e. the thermodynamic zero-temperature ground state) is at zero energy. 
	The equilibrium SAT probability is almost unity at low obstacle density $\a$ (unjammed, SAT phase) and almost vanishing at high $\a$ (jammed, 
	UNSAT phase), and the transition becomes sharp in the thermodynamic limit around a critical value $\a_c$~\cite{Ga87,GD88,FP16,AMSZ09}, 
	see Fig.~\ref{fig:P_SAT}A. Alternatively, one can estimate a {\it non-equilibrium} $P_{\rm SAT}^{\rm GD}$ 
	as the probability that GD dynamics reaches zero energy starting from a random initial condition, the probability being calculated 
	over the choice of initial condition and the realization of the random obstacles. 
	It is found that $P_{\rm SAT}^{\rm GD}$ has the same behavior of $P_{\rm SAT}^{\rm eq}$ illustrated in Fig.~\ref{fig:P_SAT}A, but with an {\it a priori}
	lower critical value $\a_c^{\rm GD} \leq \a_c$. The two values only coincide for $\k>0$. In the particles literature, the jamming transition is usually
	defined via the gradient descent protocol, hence it coincides with $\a_c^{\rm GD}=\a_J$~\cite{OLLN02,OSLN03}.
	
	Furthermore, the convergence time of GD dynamics diverges 
	at $\a_c^{\rm GD}$, see Fig.~\ref{fig:time_SAT}B, indicating that finding 
	energy minima in the very dilute of very dense case is much easier than
	in the near-critical case~\cite{hwang2020}. This scenario is quite general in search 
	algorithms and applies to several problems (such as the coloring), and it also provides a practical way to 
	construct difficult instances of constraint satisfaction 
	problems~\cite{cheeseman1991really,mitchell1992hard,kirkpatrick1994critical,selman1996critical,monasson1999determining}.

	Unfortunately, the solution of the random perceptron problem is technically more difficult than the $p$-spin.
	Thermodynamic calculations involving the 
	replica~\cite{Ga87,GD88,FP16,FPSUZ17} or cavity~\cite{AFP16} methods are still possible but a bit more involved. 
	Complexity calculations have not been performed yet, and DMFT equations 
	are much more complicated because the memory kernels are given by 
	functionals over a space of trajectories~\cite{ABUZ18,MZ22arxiv}: their
	solution is an open problem at present, for which new numerical algorithms
	are required. Interestingly, the DMFT equations for the perceptron are 
	very similar to those for infinite-dimensional particle systems~\cite{MKZ15,MZ22arxiv}. Therefore, 
	a consistent solution may be easily generalized to both cases.
	From now on, we will focus for simplicity on the GD dynamics starting from equilibrium 
	configurations at infinite temperature, $T=\io$. 
	
	\subsection{Classification of the minima}

	\begin{figure}[b]
		\includegraphics[width=.45\textwidth]{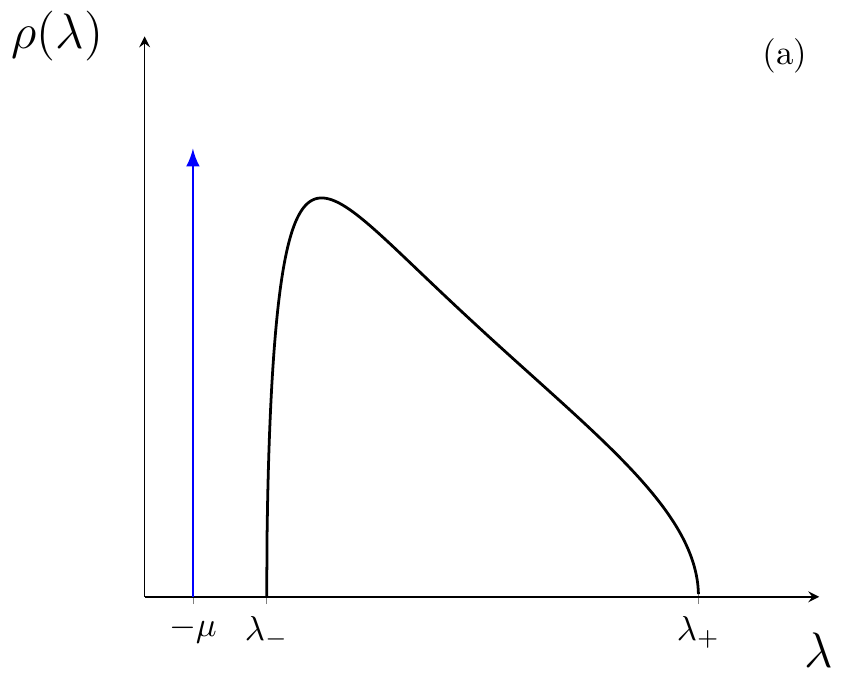}
		\includegraphics[width=.45\textwidth]{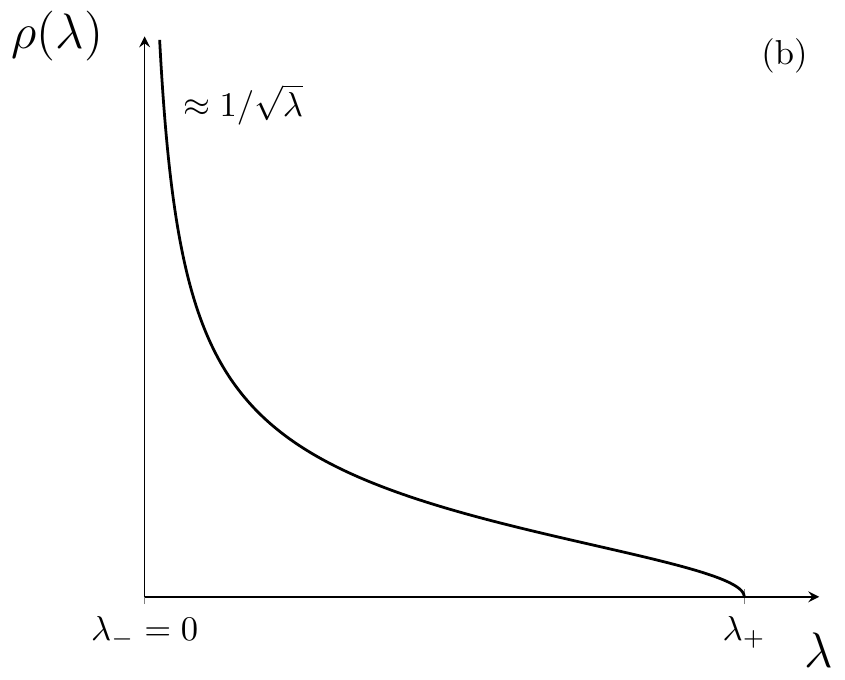}
		\includegraphics[width=.45\textwidth]{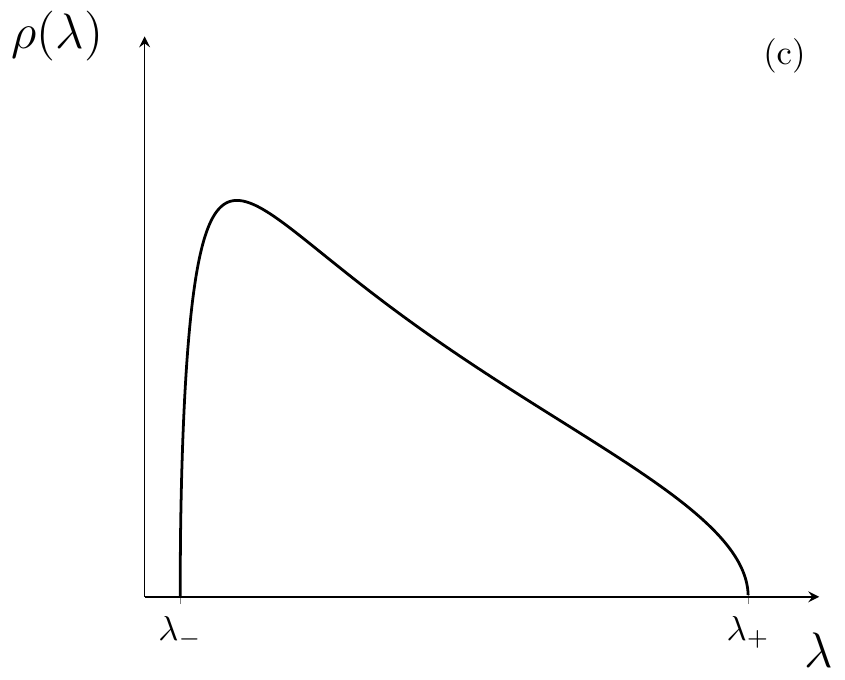}
		\includegraphics[width=.45\textwidth]{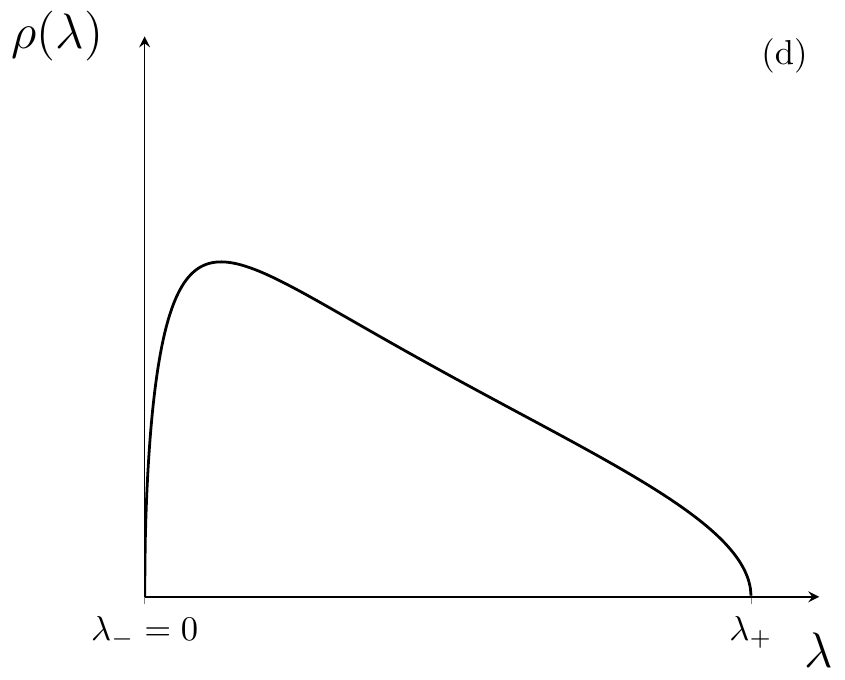}
		\caption{Classification of the minima of the random perceptron. A. Hypostatic case. B. Isostatic case. C. Hyperstatic case, stable. D. Hyperstatic case, marginal.}
		\label{fig:class}        \label{fig:hypostatic}        \label{fig:isostatic}        \label{fig:hyperstatic_s}        \label{fig:hyperstatic_m}
	\end{figure}

	The GD dynamics leads the system towards a local minimum of the energy landscape 
	$H(\U\s)$: we can then repeat the study of Sec.~\ref{sec:Hesspspin} in order to characterize
	the vibrational spectrum of the Hessian matrix in typical stationary points~\cite{FPUZ15}. 
	As in Sec.~\ref{sec:Hesspspin}, taking into account the spherical constraint via a Lagrange multiplier $\m$, 
	stationary points are solutions of
	\beq\label{eq:percmin}
	\frac{\partial H}{\partial \s_i} + \m \s_i = 0 \quad \text{with} 
	\quad \m =- \frac1{N} \U\s \cdot \U\nabla H = \a 
	\argp{ \moy{h^2} + \k\moy{h}} \ ,
	\eeq
	being
	$\moy{f(h)} \equiv \dfrac1{M} \sum_m f(h_m) \th(-h_m)$ the average over 
	the contacts (i.e. violated constraints) for a fixed configuration $\U\s$ and obstacle realization.
	The Hessian matrix is then given by
	\beq\label{eq:Hessian}
	M_{ij} = \frac{\partial^2 H}{\partial \s_i \partial \s_j} + \m \d_{ij} = \frac1{N} 
	\sum^M_{m=1} \x^m_i \x^m_j \th(-h_m) + \m \d_{ij} \ .
	\eeq
	While, in a local minimum, the obstacle positions $\x^m_i$ and gaps $h_m$ are obviously 
	correlated because $\U\s$ is a solution of Eq.~\eqref{eq:percmin}, 
	one can argue that the correlations can be neglected at the 
	leading order in $N$~\cite{FPUZ15}, similarly to the $p$-spin case. The matrix $M$ is then a $N\times N$ matrix given by a sum 
	of a number $\sum_m \th(-h_m)$ of independent random Gaussian projectors, i.e. it is
	a random Wishart 
	matrix with a fraction
	\beq
	c = \a \moy{1} = \frac1N \sum_m \th(-h_m)
	\eeq 
	of uncorrelated 
	patterns, being $c$ the \textit{isostaticity index} indicating 
	the ratio between the number of binding constraints and the degrees of 
	freedom~\cite{FPUZ15}. The eigenvalues of the Hessian matrix therefore 
	follow the Marchenko-Pastur distribution
	\beq
	\begin{split}
		\r(\l) &= (1-c)\Th(1-c) \d(\l+\m) + \frac1{2\p} \frac{\sqrt{(\l-\l_-)(\l_+-\l)}}{\l+\m}
		\mathbbm{1}_{[\l_-,\l_+]}(\l) \ , \\
		\l_\pm &= \argp{\sqrt c \pm 1}^2 - \m \ .
	\end{split}
	\eeq
	Like in the $p$-spin case, the spectrum only depends on global parameters; here, in addition to $\m$ that provides a global shift, it also depends on $c$. 
	The isostaticity index then shows its relevance; it is worth noting that
	the value $c=1$ implies that the number of contacts is equivalent to 
	the degrees of freedom of the system, so for $c\geq 1$ the matrix $\frac{\partial^2 H}{\partial \s_i \partial \s_j}$ 
	has full rank, while for $c<1$ it has a rank equal to $N c$ and it thus has a number
	$N(1-c)$ of zero modes. 
	We can now classify the possible 
	scenarios, all shown in Fig.~\ref{fig:class}.
	\begin{enumerate}
		\item[A.] \textit{Hypostatic case} $c<1$. Here, there are $N(1-c)$ modes with eigenvalue $-\m$, hence we need $\m\leq 0$ for stability.
		For $\m<0$, all the eigenvalues are positive, and the distribution 
		$\r(\l)$ is stable and gapped (Fig.~\ref{fig:class}A).  
		Note that this kind of minima can only exist for $\k>0$, because 
		$\m = \a\argp{\moy{h^2} + \k \moy{h}}$ and $\moy{h}\leq0$ by construction.
		When $\m=0$, the isolated eigenvalue vanishes, generating a finite density
		of zero modes.
		\item[B.] \textit{Isostatic case} $c=1$. Here, when in addition $\m=0$, the system is 
		marginally stable.
		The Marchenko-Pastur distribution diverges at $\l \to 0^+$ as 
		$\r(\l) \sim \l^{-1/2}$, which implies that the density of states
		with vibrational frequency $\om=\sqrt{\l} \sim 0$ (soft modes) is 
		constant at low frequencies, $D(\om \sim 0) \sim \text{const}$ (Fig.~\ref{fig:class}B). The case 
		$c=1$ and $\m>0$, which is in principle possible, is not observed in practice~\cite{FPUZ15},
		because isostaticity is only realized at the jamming transition where $\moy{h}=\moy{h^2}=\mu=0$.
		\item[C, D.] \textit{Hyperstatic case} $c>1$. The isolated eigenvalue is 
		absent and the stability condition $\l_-\geq 0$ requires ${\argp{\sqrt c-1}^2 \geq \m}$. 
		Close to isostaticity, we have $\d c = c-1$ and $\moy{h^2} \ll 
		\vert \moy{h} \vert \sim p$, where $p$ is identified with the mechanical pressure in particle systems~\cite{FPSUZ17}, and the stability criterion thus implies 
		$\d c^2 \geq \text{const} \times p$, as derived in~\cite{WNW05}. The system is therefore
		stable if the criterion above is satisfied ($\l_->0$, Fig.~\ref{fig:class}C), and 
		becomes marginally stable when $\l_- = 0$ (Fig.~\ref{fig:class}D). In the latter case, 
		the distribution goes as $\r(\l)\sim \sqrt\l$ at small $\l$, and 
		the density of soft modes goes as $D(\om\sim0)\sim\om^2$~\cite{DLDLW14}.
	\end{enumerate}

	\subsection{Phase diagram}
	
	\begin{figure}[t]
		\includegraphics[width=.69\textwidth]{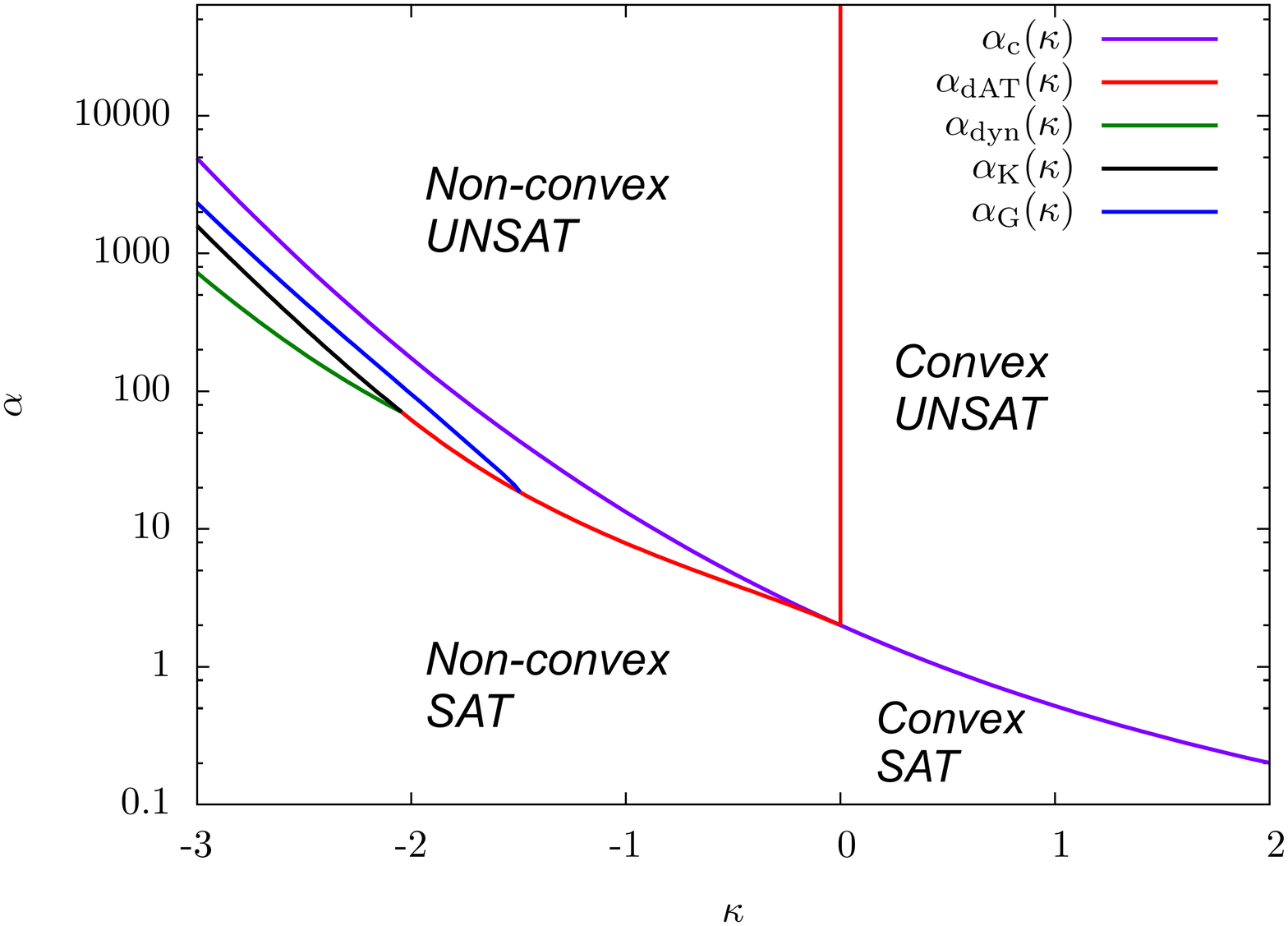}
		\caption{Phase diagram of the random perceptron, adapted from Ref.~\cite{FPSUZ17}.}
		\label{fig:percPD}
	\end{figure}

	The zero-temperature equilibrium phase diagram (i.e. the ground state structure) of the random 
	perceptron has been deeply investigated for both positive~\cite{Ga87,GD88} and negative~\cite{FP16,FPSUZ17} $\k$,
	and it is illustrated in Fig.~\ref{fig:percPD}.
	The control parameters
	are $\k$ and $\a$, and the first one governs the convexity of the 
	solution space, which is convex at $\k>0$ and non-convex at $\k<0$~\cite{FP16}.
	The phase diagram can thus be divided into four main regions: 
	the convex region for $\k>0$ and the non-convex region for $\k<0$ are both separated into a SAT phase at low $\a <\a_c(\k)$ and an UNSAT phase at $\a>\a_c(\k)$. 
	In the convex
	region, only the jamming transition at $\a_c(\k)$ is present; conversely, in the non-convex 
	region the phase diagram is much richer and several phase transitions occur within the SAT phase below
	the jamming line~\cite{FP16,FPSUZ17}.

	\subsubsection{Convex UNSAT phase}
	
	The phase diagram can be obtained from replica computations~\cite{Ga87,GD88,FP16,FPSUZ17}; in the 
	convex region, the replica-symmetric (RS) solution is stable both in the 
	SAT and UNSAT case, and the transition line can be identified as
	\beq\label{eq:alphaJ}
	\a_c(\k) = \argc{ \int^0_{-\io} \frac{\de h}{\sqrt{2\p}} e^{-(h+\k)^2/2} 
		h^2}^{-1} \ .
	\eeq
	The easiest phase to analyze is the convex UNSAT one, 
	$\a > \a_c(\k)$ and $\k>0$~\cite{Ga87,GD88}. The energy minimum is unique
	because of 
	convexity, and the free energy converges to the ground state energy when $T\to 0$, which reads
	\beq
	e_{\rm RS} = \frac12 \argp{\sqrt{\frac{\a}{\a_c(\k)}} -1}^2 \ .
	\eeq
	The GD dynamics therefore converges exponentially to the unique minimum, 
	i.e. $e(t) - e_{\rm RS} \propto e^{-t/\t}$ at long times, being 
	$\t$ the relaxation time~\cite{hwang2020,sclocchi2021,MZ22arxiv}.
	The RS computation also provides the contact number
	\beq
	c = \a \moy{1} = \a \int^{0}_{-\io} \frac{\de h}{\sqrt{2\pi}} e^{-(h+\k)^2/2} \ .
	\eeq
	The perceptron is therefore hypostatic on the convex 
	jamming line, and becomes isostatic only at $\k=0$, i.e. at the boundary of the non-convex phase~\cite{FP16}. The spectrum is 
	gapped in the UNSAT phase at $\a>\a_c(\k)$, and the long-time limit values of $c$, $\m$, 
	$e$ and other observables coincide with those of the unique ground state and can thus be computed exactly both from replica calculations~\cite{Ga87,GD88,FP16,FPSUZ17}
	and from DMFT~\cite{sclocchi2021,MZ22arxiv}. 
	When $\a\to\a_c(\k)^+$, one has $\m\to 0$ with $c<1$ and a finite fraction of zero modes is thus present at jamming (Fig.~\ref{fig:class}A).
	
	\subsubsection{Convex SAT phase}
	
	The convex SAT case is, somehow surprisingly, more complex than the UNSAT
	case, because in the SAT phase the solutions found by thermal annealing 
	and GD dynamics do not coincide. 
	Replica computations~\cite{Ga87,GD88,FP16,FPSUZ17}
	indeed predict the existence of a single convex lake 
	of solutions with $H(\U\s)=0$, surrounded by a convex energy landscape. In this 
	scenario, during thermal annealing the system starts from a high-energy 
	configuration and explores the lake uniformly when temperature is very low. 
	When temperature is finally switched off, the algorithm thus 
	stops in a randomly chosen configuration inside 
	the lake, where $c=e=\m=0$, and all the directions are flat - 
	i.e. $\r(\l)=\d(\l)$, see Fig.~\ref{fig:thermal_vs_GD}.
	Conversely, GD dynamics starts from the same configuration but follows
	the steepest descent path that ends up on the lake ``shore'', implying a final value of 
	$\m=e=0$, as for thermal annealing, but a finite fraction of contacts $0<c<1$
	that have precisely $h=0$~\cite{MZ22arxiv}, as illustrated in Fig.~\ref{fig:nonconvex}A.
	The convergence to the final zero-energy state is exponential as in the UNSAT 
	case. Therefore, the spectrum of the final state of the GD dynamics 
	shows a Marchenko-Pastur distribution
	at finite $\l$ together with a finite fraction of zero modes. 
	Numerical results for the GD dynamics are given 
	in~\cite{hwang2020}.
	The difference between the final states of thermal and athermal dynamics may have important consequences in machine learning applications, see e.g.~\cite{d2020double,mignacco2021}.
	
	The relaxational dynamics of the convex problem also shows features of 
	dynamic criticality: indeed the relaxation time $\t$ of the GD dynamics
	diverges when approaching the jamming transition from both 
	directions~\cite{hwang2020}, see Fig.~\ref{fig:time_SAT}B. It has been numerically shown that 
	$\t \sim 1/\l_1$, being $\l_1$ the first non-zero eigenvalue of the Hessian
	matrix~\cite{hwang2020}. When approaching the transition from the UNSAT phase, $\l_1$ is
	given by the isolated eigenvalue $\l_1 = -\m \sim \vert \moy{h} \vert 
	\sim \d \a$, therefore one has $\t \sim \d \a^{-1}$. On the other hand, 
	from the SAT phase one would naively obtain $\l_1 = \l_-$, which remains finite 
	when $\a \to \a_c(\k)$. However, the relaxation has been observed to be 
	dominated by an isolated, low-frequency eigenmode, which
	is not captured by the density
	$\r(\l)$~\cite{lerner2012toward,IKBSH20,hwang2020,ikeda2020,nishikawa2021}. Computing this isolated mode analytically remains an open problem.
	
	\subsubsection{Non-convex phase}
	
\begin{figure}
		\centering
		\includegraphics{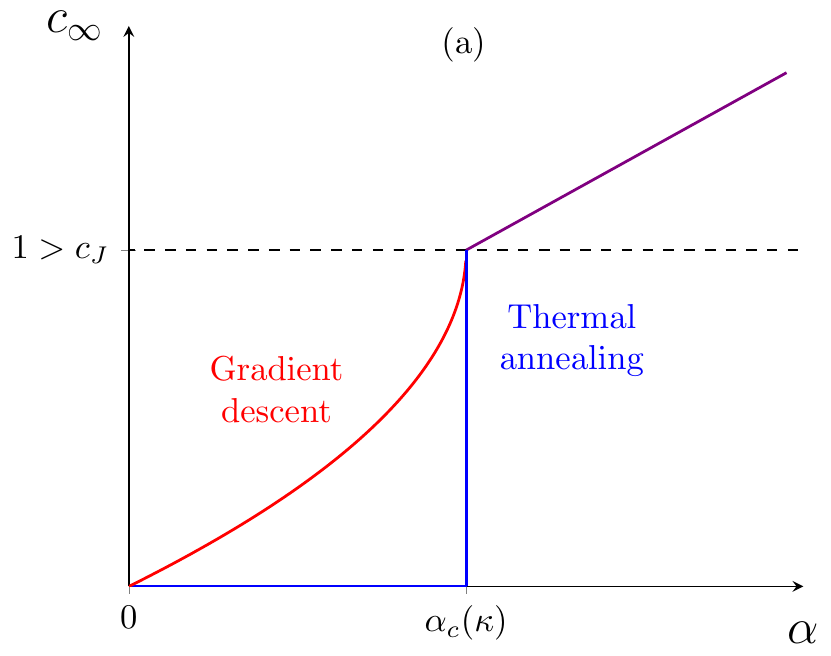}
		~
		\includegraphics{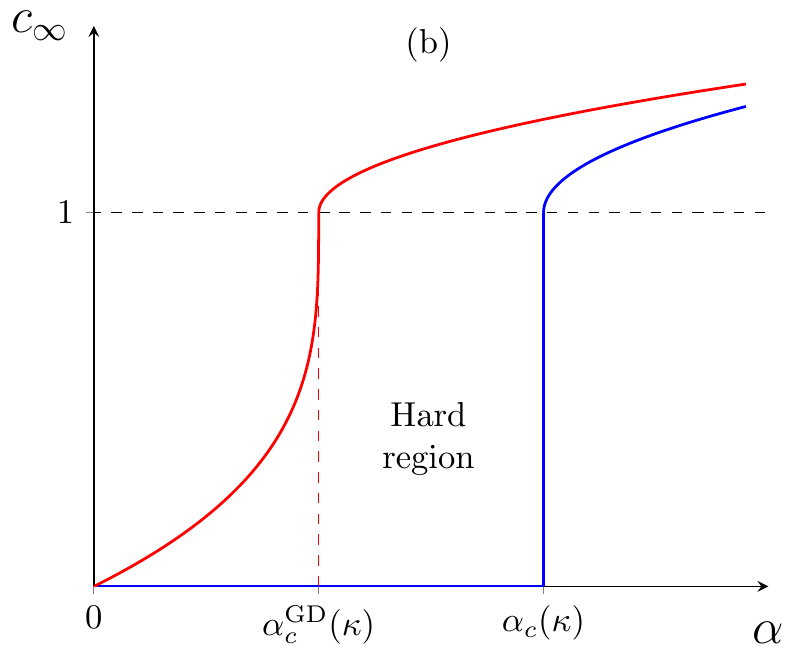}
		\caption{Isostatic index $c$ as computed from GD dynamics (red) and thermodynamics (blue). 
			A. Convex region: the two methods coincide in the UNSAT phase, while in the SAT phase thermodynamics gives $c=0$ while dynamics
			gives $0<c<1$. B. Non-convex region:
			the two methods do not agree and there is a hard region where solutions
			of the CSP exist, but the gradient descent is unable to find them.}
		\label{fig:nonconvex}
	\end{figure}

	When $\k<0$, the energy landscape becomes non-convex and rough, and correspondingly the phase diagram 
	becomes more complex; we refer to Ref.~\cite{FPSUZ17} for details. Concerning the jamming transition, there are two main differences 
	with respect to the convex case:
	\begin{itemize}
		\item the jamming transition is \textit{isostatic}, i.e. $c=1$ identically all along 
		the jamming line, contrarily to the hypostaticity observed in the 
		convex phase;
		\item thermodynamics and GD dynamics \textit{do not agree} on the 
		location of the transition, with $\a_c^{\rm GD}(\k) < \a_c(\k)$ as 
		computed from replica calculations, because the energy landscape is
		developing growing complexity with many local energy minima in which GD dynamics
		gets trapped before the jamming transition occurs at the 
		thermodynamic level - see Fig.~\ref{fig:nonconvex}B. This gap defines a 
		\textit{hard region} $\a_c^{\rm GD}(\k) < \a < \a_c(\k)$, where 
		zero-energy solutions exist but the GD dynamics cannot find them~\cite{ZK16}.
	\end{itemize}
	
	Furthermore, in the UNSAT phase one expects that the energy landscape of the perceptron is very similar to that of the mixed $p$-spin,
	hence GD should display the same power-law relaxation 
	described for the $p$-spin in Sec.~\ref{sec:noneq}, namely $e(t) - e_{th}  \sim t^{-\d}$ with $\d=2/3$, associated to memoryless persistent
	aging with weak ergodicity breaking.
	This has not been carefully checked in the perceptron yet, 
	but it has been investigated in numerical simulations of spherical 
	particles, with $\d \simeq 0.84$ in $d=2$ and 
	$\d \simeq 0.70$ in $d=3$~\cite{chacko2019,nishikawa2021b}. 
	Also, the final state is hyperstatic with $c>1$, and
	similarly to the mixed $p$-spin, we expect it to be marginal, hence $\l_-=0$ and the spectrum goes as $\r(\l) \sim \sqrt \l$,
	see Fig.~\ref{fig:class}D. Note that because the relaxation of the energy is a power-law, the relaxation time $\t$ is formally infinite
	throughout this phase; this is a consequence of the marginal stability of the final state.

	Furthermore, precisely at the non-convex jamming transition,
	the system exhibits other 
	non-trivial critical relations. The distribution of 
	gaps $h$ and forces $f$ exhibit a universal power-law behavior~\cite{Wy12} 
	\beq
	P(h \to 0^+) \sim h^{-\g} \ , \quad P(f\to 0^+) \sim f^\th \ ,
	\eeq
	with $\g\simeq 0.41269\ldots$ and $\th \simeq 0.42311\ldots$. These 
	critical exponents have been obtained analytically by replica calculations~\cite{CKPUZ14}, hence at the thermodynamical jamming point.
	However, they have also been 
	computed by numerical simulations of GD dynamics in
	sphere packings and in the perceptron (hence at the dynamical jamming transition), obtaining perfect agreement within numerical precision,
	despite the different nature of the transition~\cite{CCPZ12,CCPZ15,CKPUZ17,charbonneau2020finite}. 
	Furthermore, scaling relations based on marginal stability~\cite{Wy12,DLBW14,MW15} are also satisfied by the exponents predicted by the replica method.
	These results hints at a strong universality of the jamming transition.

	The non-convex SAT phase, on the other hand, shows more similarities 
	with the convex case. As in the latter, the energy decay follows an 
	exponential law; the relaxation time has been derived from a scaling 
	argument~\cite{lerner2012toward,ikeda2020}, with the result
	\beq
	\t \sim \d c^{-\b} \ , \qquad \b = \frac{4+2\th}{1+\th} \simeq 3.41\ldots \ \ .
	\eeq
	This result is in agreement with some numerical results, but it 
	is very hard to be tested precisely~\cite{nishikawa2021}. Moreover, a derivation of this exponent
	within dynamical mean-field theory is still missing.
	
	Finally, we mention that a modified perceptron can be used to describe a different universality class of non-convex hypostatic jamming, found for example in ellipsoids~\cite{BIUWZ18}.

	\subsection{Summary}
	
	In this section, we discussed the complex behavior of constraint 
	satisfaction problems, based on the paradigmatic perceptron model. 
	The latter is closely related to the sphere packing problem, and their 
	DMFT equations almost coincide in the infinite-dimensional limit. Their energy landscape is even 
	richer than in the $p$-spin case.
	
	We have seen that the jamming transition is associated with a SAT/UNSAT
	transition, and that, in the non-convex case, a hard region exists 
	where GD dynamics and thermodynamics do not give the same result. All
	the results obtained for the \textit{non-convex} perceptron also
	apply to the sphere packing problem of particles in infinite spatial
	dimentions, $d \to \io$.
	
	However, DMFT equations are particularly difficult to solve: the analytical
	understanding of mean-field dynamics is still poor, and the  
	currently known results are limited to the long-time limit of the GD dynamics in the UNSAT 
	phase~\cite{sclocchi2021,MZ22arxiv}. The numerical 
	solution of the DMFT equations has been possible only at very short times for the sphere 
	packing problem~\cite{MZ22arxiv}, while more encouraging results 
	have been found in classification problems~\cite{MKUZ21,mignacco2021}.
	To conclude, we recall some of the main open problems:
	\begin{itemize}
		\item the derivation of the $\b$ exponent of the relaxation 
		time $\t \sim \d c^{-\b}$ in the SAT phase;
		\item the derivation of the $2/3$ exponent in the energy relaxation
		$e-e_{th} \sim e^{-2/3}$ in the UNSAT phase;
		\item the understanding of why jamming criticality is strongly universal, so that mean-field thermodynamic predictions are also consistent with dynamical results in all dimensions;
		\item the effect of finite dimensionality, especially focused on the role of localized 
		modes~\cite{nishikawa2021}.
	\end{itemize}
	While the toy models discussed here provide a good starting point to understand many phenomena that are characteristic of disordered systems with complex landscapes, many problems remain open and a lot of exciting results are surely yet to come.
	
	\acknowledgments
	
	We warmly thank the organizers of the summer school 
	{\it Fundamental Problems in Statistical Physics XV}, held in Brunico, Italy, in July 2021, and of the summer school {\it Glassy Systems and Inter-Disciplinary Applications}, held in Cargese, France, in June 2021, for the invitation to lecture there (F.Z.) and for the invitation to participate in Cargese (A.M., G.F.). We also thank all the participants to those schools for many discussions and for providing a welcoming and stimulating environment.
	
	Our work has received support by the European Research Council (ERC) under the European Union's Horizon 2020 research and innovation programme (grant agreement n. 723955 - GlassUniversality) and by a grant from the Simons Foundation (\#454955, Francesco Zamponi).

	\bibliography{HSmerge,SGmerge}

\end{document}